\documentclass[twocolumn,showpacs,amsfonts,aps,prc,floatfix,superscriptaddress,nofootinbib]{revtex4-1}
\usepackage{float}  
\usepackage{array}
\usepackage{booktabs}
\usepackage[colorlinks,linkcolor=blue,citecolor=blue]{hyperref}
\usepackage{amsmath}
\usepackage{mathrsfs}
\usepackage{amssymb}
\usepackage{graphicx,epsfig,latexsym,overpic,amssymb,color}
\usepackage{multirow}
\usepackage[section]{placeins}
\usepackage{url}
\usepackage{natbib}
\usepackage{threeparttable} 

\preprint{\today}
\begin{document}
 
\title{Revisiting the nuclear island of negative hexadecapole deformations in A$\approx$180 mass region: focusing on moments of inertia and quadrupole-hexadecapole coupling}
\author{Ran Li}
\affiliation{School of Physics, Zhengzhou University, Zhengzhou 450001, China}
\author{Hua-Lei	Wang}
\email{wanghualei@zzu.edu.cn} 
\affiliation{School of Physics, Zhengzhou University, Zhengzhou 450001, China}
\author{Kui Xiao}
\affiliation{School of Physics, Zhengzhou University, Zhengzhou 450001, China}
\author{Zhen-Zhen Zhang}
\affiliation{School of Physics, Zhengzhou University, Zhengzhou 450001, China}
\author{Min-Liang Liu}
\affiliation{Key Laboratory of High Precision Nuclear Spectroscopy, Institute of Modern Physics, Chinese Academy of Sciences, Lanzhou 730000, China} \affiliation{School of Nuclear Science and Technology, University of Chinese Academy of Sciences, Beijing 100049, China}
	
\begin{abstract}
%
For even-even nuclei $^{180-184}$Yb, $^{182-186}$Hf and $^{184-188}$W located on an island of hexadecapole-deformation archipelago, the structure properties, especially under rotation, are reinvestigated by using the Hartree-Fock-Bogliubov-Cranking (HFBC) calculation with a fixed shape (e.g., the ground-state equilibrium shape). The equilibrium deformations, extracted from the potential energy surface, are calculated based on the phenomenological Woods-Saxon mean-field Hamiltonian within the framework of macroscopic-microscopic (MM) model. The impact of different deformation degrees of freedom on, e.g., single-particle levels, total energy, and moment of inertia, is revealed, especially concentrating on the hexadecapole-deformation effects and the quadrupole-hexadecapole coupling. Considering the axially hexadecapole
deformation, the present calculations can well reproduce available experimental data, including the quadrupole deformations and moments of inertia. Interestingly, it is found that the impact of different deformation degrees of freedom on moment of inertia exhibits a similar trend in the HFBC and rigid-body calculations though the latter ignores the pairing effects. Before starting or constructing a complex theory-model, to some extent, such a similarity can provide an alternative way of understanding the effect of, e.g., exotic deformations, on moment of inertia by the calculation of a simple rigid-body approximation. The present findings could offer insights into the static and dynamic effects of hexadecapole deformations, contributing valuable information for the corresponding research in nuclear structure and reaction. \\
		
		
\end{abstract}
\maketitle
	
\section{INTRODUCTION}
\label{introduction}
 

Similar to the periodic table, which provides a systematic way of understanding and predicting the element properties in chemistry, the chart of nuclides is of importance to nuclear physics and has become  a powerful tool for visualizing and studying nuclear properties~\cite{Morfouace2025}. The modern nuclide chart consists of more than 2500 known nuclides (nevertheless, less than one third of theoretical predictions)~\cite{Moller1995,Moller2016,Wang2021}. So far, it has been revealed that the so-called ``island'' (namely, a region of the chart of nuclides where nuclei have prominent property on the nuclear map) or ``archipelago'' of islands may exist on the nulide chart. For instance, some of them have been discovered or predicted, including the archipelago of islands of inversion\cite{Brown2010,Revel2020}, the archipelago of islands of $D_{3h}$ point group symmetry~\cite{Yang2023}, the island of superheavy stability, the island of tetrahedral $T_{d}$ (``pyramid’’) symmetry~\cite{Dudek2006} and the island of asymmetric fission~\cite{Morfouace2025}.  In this project we focus specifically on one island of an archipelago of hexadecapole-deformation nuclei~\cite{Moller2016,Zhang2022}.

As it is well known, the nucleus is a quantum many-body system governed by short-range strong interaction. The density distribution of a nucleus usually decreases to zero exponentially at the border of such a compact system, which helps to introduce the notions of nuclear surface and shape, and further, the one-body mean-field Hamiltonian. Here, using the language of multipole expansion of the nuclear surface with spherical hamornics\cite{Moller2009,Moller1995}, one can describe the nuclear shape with a set of deformation parameters (corresponding to the collective coordinates in the space spanned by orthogonal spherical harmonics).  The multipole deformations, e.g., the quadrupole and hexadecapole ones, are the quantities in this framework. Let us remind that, besides, there are other several parameterization methods of nuclear shape in the literature, such as Cassinian ovals~\cite{Pashkevich1971}, matched quadratic surfaces~\cite{Nix1969} and generalized Lawrence shapes~\cite{Aydin2021,Zhong2014,Lawrence1965}. Needless to say, one has to describe the nuclear shape with other parameters instead of ``multipole'' deformations in those cases.     

It is worth nothing that, comparing with the quadrupole (e.g., prolate, oblate and triaxial) and octupole (e.g., pear-shaped and tetrahedral) deformations, the hexadecapole deformation has received less detailed attention since the effects of the hexadecapole correlations in nuclear low-lying states are often overshadowed by large quadrupole correlation effects~\cite{Lotina2024}. Indeed, the hexadecapole deformation is difficult to extract from experiment with good precision due to the small magnitude and big uncertainty~\cite{Gupta2020, Jia2014}. Nevertheless, it was still pointed out that the hexadecapole correlations are non-negligible in many rare-earth~\cite{Hendrie1968,Erb1972,Wollersheim1977,Ronningen1977} and actinide~\cite{Bemis1973,Hendrie1973,Zamfir1995} nuclei. In the production of superheavy elements, it was illustrated, e.g., by the di-nuclear system model, that the hexadecapole deformation could change significantly the structure of the driving potentials and the fusion probabilities for some reaction channels~\cite{Wang2010}.  In recent years, the developments of related theories and experimental techniques seems to renew the research enthusiasm of hexadecapole deformation in nuclei. For instance, the hexadecapole deformation was revealed to affect the values of the neutrinoless double $\beta$ decay matrix elements of open shell nuclei in Ref.~\cite{Engel2017}. In the hydrodynamic simulation of the relativistic heavy ion collider, it was confirmed the hexadecapole deformation plays an important role~\cite{Ryssens2023}. Based on quasi-elastic scattering, the hexadecapole deformation in the light-mass nucleus $^{24}$Mg was recently determined in experiment~\cite{Gupta2020}. For a recent analysis of in fission calculations, hexadecapole deformation was found to be of importance for near-to-scission configurations~\cite{Han2021}. By the coupled-channel calculations, hexadecapole deformations were determined for $^{74,76}$Kr from inelastic proton scattering cross sections~\cite{Spieker2023}. The impact of hexadecapole deformations on the collective spectra of axially deformed nuclei was investigated using the IBM approach~\cite{Lotina2024}. The emergence and stability of static hexadecapole deformations as well as the impact in the development of dynamic deformation due to collective motion considering quadrupole-hexadecapole coupling are very recently studied for a selected set of radium, thorium, uranium, and plutonium isotopes, using the Gogny Hartree-Fock-Bogoliubov and generator coordinate method frameworks~\cite{Rodriguez-Guzman2025}. 

In our previous study~\cite{Zhang2022}, including the rotational cases, it was found that the nuclear chart has an archipelago of, at least, 7 islands of hexadecapole-deformation nuclei, agreeing with the theoretical calculations, e.g., by M\"oller et al~\cite{Moller1995,Moller2016}, at ground states. Focusing on different nuclear properties (e.g., non-axial deformation degrees of freedom and binding energy), we have formally visited two of them, sitting in the $A \approx 150$ and 230 mass regions with positive hexadecapole-deformation value~\cite{Song2023, Wei2024}. Using the similar theoretical methods, e.g., Potential-Energy-Surface (PES) and Total-Routhian-Surface (TRS) calculations, we have performed numerous calculations on nuclear structure properties, such as triaxial and octupole deformations, fission barriers and pathways, isomers and nuclear mass~\cite{Chai2019,Chai2018b,Guo2025}. In this work, based on MM model~\cite{Meng2022a,Moller1995} and HFBC calculation~\cite{Meng2022b,Yang2016}, we tend to probe nuclear structure properties on the island of nuclei with negative hexadecapole deformations in the A $\approx$ 180 mass region, primarily paying attention to the effects of different quadrupole and hexadecapole deformations on nuclear moment of inertia (MoI) and the comparision with the calculations based on a deformed rigid body. Combining with the recent publications with the inclusion of exotic deformations~\cite{Yang2022, Irene2025}, the prediction is given by a natural extension of the rigid calculations. 
	
This paper is organized as follows: In section II we present theoretical approaches, including the MM model and CHFB calculations. The shape parameterization and nuclear rigid-body approximation are also introduced. Calculated results for, such as single-particle levels, PESs, equilibrium deformations and MoIs, are illustrated and discussed in section III. In particular, the effects of the hexadecapole deformation are presented and analyzed. Finally, a summary and some concluding remarks are given in section IV.

\section{Theoretical method}
\label{method}

In this section, we briefly recall the general procedure of our adopted theoretical methods and provide the necessary references. 

Based on the well-known special functions of applied mathematics, the nuclear shape is defined in terms of the spherical harmonic multipole expansion of nuclear surface $\Sigma$~\cite{Cwiok1987},  
\begin{equation}
\Sigma:
R(\theta,\phi)
=
r_0A^\frac{1}{3}c(\alpha)
\Big[1
+
\sum_{\lambda}
\sum_{\mu=-\lambda}^{+\lambda}
\alpha_{\lambda\mu}
Y_{\lambda\mu}(\theta,\phi)
\Big],
                                                               \label{eqn.01}
\end{equation}
where $R(\theta,\phi)$, sometimes written as $R(\theta,\phi,\alpha)$, indicates the distance from the origin of the coordinate system to the point on the nuclear surface whose position is specified by the angles $\theta$ and $\phi$; the function $c(\alpha)$ ensures the conservation of the nuclear volume and $\alpha$ denotes a set of deformation parameters $\{\alpha_{\lambda\mu}\}$ with non-negative integer indices $\lambda$ and $\mu$ limited by $-\lambda \leq \mu \leq +\lambda$.
Once a set of $\{\alpha_{\lambda\mu}\}$ is given, the nuclear shape will be fixed. If considering the constant density distribution in a nucleus, one can calculate the nuclear MoI based on a rigid-body approximation. For instance, according to the default convension of spherical and Cartesian coordinate systems, the rigid-body (with a fixed shape) MoI $J_{rig.}$ around $x$ axis (here, we just care for the cranking around $x$ axis) reads,
\begin{eqnarray}
	& &\int(y^2+z^2)dm\nonumber\\
	&=&\int[(rsin\theta sin\varphi)^2+(rcos\theta)^2]\rho dv	\nonumber\\
	&=&	\rho \int_{0}^{\pi}d\theta \int_{0}^{2\pi} d\varphi \int_{0}^{R(\theta,\varphi)} (sin\theta - sin^3\theta cos^2\varphi)r^4 dr \nonumber \\
	&=&	\frac{\rho}{5}\int_{0}^{\pi}d\theta \int_{0}^{2\pi}
	R^5(\theta,\varphi)(sin\theta - sin^3\theta cos^2\varphi)d\varphi \nonumber\\
	&=& \int_{0}^{\pi}d\theta  \int_{0}^{2\pi}f(\theta,\varphi) d\varphi 
                                                               \label{eqn.02}
\end{eqnarray}
The above expression can be numerically calculated by using the Gauss-Legendre quadrature, e.g., $\sum_i \omega_i \sum_j \omega_j f(\theta_i, \varphi_j)$, where $\theta_i$, $\varphi_j$ and $\omega_{i,j}$ are nodes and weights, respectively; $f(\theta, \varphi)$ is the integrand. Combining the above two equations, in principle, the rigid-body MoI for a nucleus with arbitrary shape can be obtained, extending the emperical calculation, e.g., in Ref.~\cite{Allmond2017}.    

Generally speaking, most of nuclei prefer to possess three symmetry planes $x-y$, $y-z$ and $x-z$, corresponding to $R(\pi/2 + \theta,\varphi) = R(\pi/2 - \theta,\varphi)$, $R(\theta,\pi/2+\varphi) = R(\theta,\pi/2 - \varphi)$ and $R(\theta,\varphi) = R(\theta, -\varphi)$, respectively. At this case, only the even $\lambda$ and even $\mu$ components are remained in Eq.~(\ref{eqn.01}). Therefore, in the shape parametrization, we just consider the low-order quadrupole and hexadecapole deformation degrees of freedom, including nonaxial deformations [that is, $\alpha$ $\equiv$ ($\alpha_{20}$, $\alpha_{2\pm2}$, $\alpha_{40}$,	$\alpha_{4\pm2}$, $\alpha_{4\pm4}$)]. After requesting the hexadecpole degrees of freedom to be functions of the scalars in the quadrupole tensor $\alpha_{2\mu}$, one can reduce the number of independent coefficients to three, namely, $\beta_2$, $\gamma$ and $\beta_4$, which obey the relationships\cite{Bohr1998,Bhagwat2012},
\begin{equation}
	\left\{
	\begin{array}{lcl}
	\alpha_{20}
	=
	\beta_2\cos\gamma
	\\[2mm]
	\alpha_{22}
	=
	\alpha_{2-2}
	=
	-\frac{1}{\sqrt{2}}\beta_2\sin\gamma
	\\[2mm]
	\alpha_{40}
	=
	\frac{1}{6}\beta_4(5\cos^2\gamma+1)
	\\[2mm]
	\alpha_{42}
	=
	\alpha_{4-2}
	=
	-\frac{1}{12}\sqrt{30}\beta_4\sin2\gamma
	\\[2mm]
	\alpha_{44}
	=
	\alpha_{4-4}
	=
	\frac{1}{12}\sqrt{70}\beta_4\sin^2\gamma.
	\\[2mm]
	\end{array}
	\right.
	                                                             \label{eqn.03}
\end{equation}
Note that, in the deformation space mentioned above, the quadrupole $\beta_2$ and hexadecapole $\beta_4$ deformations satisfy $\beta_2^2 = \alpha_{20}^2 + 2 \alpha_{22}^2$ and  $\beta_4^2 = \alpha_{40}^2 + 2 \alpha_{42}^2 + 2 \alpha_{44}^2$ since the total deformation $\beta_{\lambda}$ at order $\lambda$ is defined by $\beta_{\lambda}^2=\sum_{\mu=-\lambda}^{\mu=+\lambda}\alpha^2_{\lambda\mu}$~\cite{Ryssens2023} and the deformation magnitude $\beta_{\lambda}$ can usually be extracted from the ground-state electric transition rate $B(E\lambda)$ via $\beta_{\lambda}=\frac{4\pi}{(2\lambda+1)ZR_o^{\lambda}}\sqrt{\frac{B(E\lambda)}{e^2}}$, see e.g., Ref.~\cite{Bohr1998}.

For a given nucleus with a fixed shape and taking the realistic Woods-Saxon potential into account, we can write down its model Hamiltonian as follows,
\begin{eqnarray}
	\hat{H}_{\text{WS}}
	=
	\hat{T}
	+
	\hat{V}_{\rm cent}(\vec{r};\hat{\beta})
	+
	\hat{V}_{\rm so}(\vec{r},\vec{p},\vec{s};\hat{\beta}) 
	 +
	\hat{V}_{\rm Coul}(\vec{r},\hat{\beta}), \nonumber \\
                                                   \label{eqn.04}
\end{eqnarray}
where $\hat{T}$ denotes the nucleonic kinetic-energy operator (namely, $-\frac{\hbar^2}{2m}\nabla^2 $). The central Woods-Saxon potential $\hat{V}_{\rm cent}(\vec{r};\hat{\beta})$ reads,
\begin{equation}
	\hat{V}_{\rm cent}(\vec{r},\hat{\beta})
	=
	\frac{V_0[1\pm\kappa(N-Z)/(N+Z)]}
	{1+\exp[\mathrm{dist}_\Sigma(\vec{r},\hat{\beta})/a_0]},
	                                             \label{eqn.05}
\end{equation}
where the plus and minus signs hold for protons and neutrons,
respectively and the parameter $a_0$ denotes the diffuseness of the	nuclear surface. The term $\mathrm{dist}_\Sigma(\vec{r},\hat{\beta})$ represents the distance between the single-particle (nucleon) position point $\vec{r}$ and the nuclear	surface $\Sigma$. The spin-orbit potential $\hat{V}_{so}$ which can strongly affect the level order is defined by,
\begin{eqnarray}
	&&\hat{V}_{so}(\vec{r},\vec{p},\vec{s};\hat{\beta})
	=-\lambda [\frac{\hbar}{2mc}]^2\nonumber\\
	&&\times\Big\{\nabla\frac{V_o[1\pm\kappa(N-Z)/(N+Z)]}
	{1+exp[dist_{\sum_{so}}](\vec{r},\hat{\beta})}\Big\}
	\times\vec{p}\cdot\vec{s},
                                                  \label{eqn.06}
\end{eqnarray}
where $\lambda$ denotes the strength parameter of the effective spin-orbit force acting on individual nucleons. The new surface $\sum_{so}$ is different form the one in Eq. (\ref{eqn.06})	due to the different radius paramter. The Coulumb potential for protons reads  
\begin{eqnarray}
	&&\hat{V}_{Coul}(\vec{r}; \hat{\beta})
	= \frac{Z-1}{4\pi R_0^3/3}e\int \frac{d^3r}{|\vec{r}-\vec{r^\prime}|} . 
                                                  \label{eqn.07}
\end{eqnarray}
The time-independent (stationary) Schr\"{o}dinger Equation $\hat{H}_{WS}\psi_n = e_n \psi_n$, as an eigenvalue equation, is solved numerically since the analytic solutions do not exist. 

To solve the eigenvalue equation, the general computational method is to buid a Hamiltonian matrix by adopting a set of orthonormal basis funcitions and to obtain the eigenvalues and eigenfunctions by diagonalization. The computing efficiency of matrix diagonalization generally decreases rapidly as the number of adopted bases increases.
That is, the computation time to diagonalize a Hamiltonian matrix increases very quickly as the number of bases increases. In principle, the Hilbert space, formed by the orthonormal basis funcitions, on which the Hamiltonian matrix is calculated, usually has an infinite dimension. Though our main goal is to obtain the $N$ lowest eigenstates $\{ \psi_n \}$ and their energy values $\{e_n\}$ of the WS Hamiltonian operator, the number of bases needed to properly describe a nuclear system is usually rather large (e.g., relative to $N$). Efficient implementation of the Hamiltonian matrix diagonalization requires that the matrix elements can be computed as fast as possible and the calculated Hamiltonian matrix has a small number of non-diagonal terms. In general, the former can be simplified by using the time reversal (resulting in the Kramers degeneracy) and spatial symmetries (e.g., the parity and signature are conserved), while the latter can be realized by adopting appropriate basis functions and symmetry properties as well.

In the present project, we compute the WS Hamiltonian matrix in terms of a set of orthogonal complete bases, namely, the eigenfunctions of the axially deformed harmonic oscillator potential in the cylindrical coordinate system (obviously, the spherical basis functions will not be good candidates for a deformed shape). The deformed basis function has the following form~\cite{Cwiok1987},
\begin{equation}
|N n_z \Lambda \Omega \rangle \equiv |n_\rho n_z \Lambda \Sigma \rangle
  =
  \psi_{n_\rho}^\Lambda(\rho)\psi_{n_z}(z)
  \psi_{\Lambda}(\varphi)\chi(\Sigma),
                                                    \label{eqn.08}
\end{equation}
where $N = 2n_\rho + |\Lambda| + n_z$, $\Omega = \Lambda + \Sigma$ and
\begin{equation}
  \left\{
   \begin{array}{lcl}
     \psi_{n_\rho}^\Lambda(\rho)
    &=&
     \frac{\sqrt{n_{\rho}!}}{\sqrt{(n_\rho+|\Lambda|)!}}
     (2m\omega_\rho/\hbar)^{1/2} \\
     &&\times e^{-\frac{\eta}{2}}\eta^\Lambda
     L_{n_\rho}^{|\Lambda|}(\eta),
       \\[2mm]
     \psi_{n_z}(z)
   & =&
     \frac{1}{\sqrt{\sqrt{\pi}2^{n_z}n_z!}}
     (2m\omega_z/\hbar)^{1/4} \\
     &&\times e^{-\frac{\xi^2}{2}}
     H_{n_z}(\xi),
     \\[2mm]
     \psi_{\Lambda}(\varphi)
    & =&
     \frac{1}{\sqrt{2\pi}} e^{i\Lambda \varphi},
       \\[2mm]
   \end{array}
  \right.
                                                                  \label{eqn.09}
\end{equation}
and $\chi(\Sigma)$ represents the spin wave functions, cf. e.g.,
Sec. 3.1 in Ref.~\cite{Cwiok1987} for more details. To obtain the approximate parameters $\omega_\rho$ and $\omega_z$ in Eq.~(\ref{Fig09}), the harmonic oscillator potential is constrained to approach the deformed WS one as much as possible (equivalently, let the harmonic oscillator ellipsoid overlap with the given nuclear shape as much as possible). One can usually obtain them by solving the equation system
\begin{equation}
  \left\{
   \begin{array}{lcl}
     \omega_\rho^2 \omega_z &=& \omega_0^3
     \\[2mm]
     \omega_\rho/ \omega_z &=&\sqrt{<\rho^2>/<z^2>}.
   \end{array}
  \right.
                                                                  \label{eqn.10}
\end{equation}
The average values $<\rho^2>$ and $<z^2>$ can be calculated with the help of the definition of nuclear surface by the integrals (e.g., $<z^2> = \int z^2 dm / \int dm$). The oscillator constant $\omega_0$, which can be determined from the mean square radius of a sphere, depends on the nuclear mass number $A$ and satisfies the relationship, e.g., $\hbar \omega_0 \simeq \frac{5}{4}(\frac{3}{2})^{1/2}\frac{\hbar^2}{mr_0^2}A^{1/3} = 41 A^{1/3}$ MeV, i.e., see Eq.~(2.12) in Ref.~\cite{Ring1980}. By such a treatment, the harmonic ocillator basis functions may satisfy the condition that makes the Hamiltonian matrix to have fewer and smaller off-diagonal elements, greatly improving the computational efficiency.

The model parameters (taken from \cite{Bhagwat2012}) of the Hamiltonian are usually obtained by the inverse problem theory or $\chi^2$-fitting procedure (e.g. fitting the experimental single-particle levels). During our calculation, the eigenfunctions with the principal quantum number $N \leq$ 12 and 14 have been chosen as a basis for protons and neutrons, respectively. It is found that, by such a basis cutoff, the single-particle energies are sufficiently convergent with respect to a basis-space enlargement. In order to evaluate the mixing of wave functions due to the multipole interaction, the Moshinsky bracket, e.g., $\langle Nlj\Omega|Nn_z\Lambda \Omega\rangle$ will be convenient for transferring the WS wave functions expanded by the cylindrical HO basis functions into the spherical HO bases, see the details in Ref.~\cite{Davies1991}. We implant such a transfermation code in this work. Let us remind that the notation $| Nlj\Omega \rangle $ is equivalent to $| nlj \Omega \rangle$ due to the existence of the constraint relation $N=2(n-1)+l$~\cite{Wei2024}.

Once the single-particle levels $\{e_n \}$ are obtained, one can calculate the shell correction $\delta E_{shell}(Z,N,\hat{\beta})$, which is usually the most important part in the MM model. We compute this term by using a phenomenological expression firstly proposed by Strutinsky,
\begin{equation}
\delta E_{shell}(Z,N,\hat{\beta})=\sum 2e_i -\int e\tilde{g}(e)de,
                                                    \label{eqn.11}
\end{equation}
where $e_i$ denotes the calculated single-particle levels and
$\tilde{g}(e)$ is the so-called smooth level density,
which was early defined as,
\begin{equation}
\tilde{g}(e,\gamma)\equiv \frac{1}{\gamma\sqrt{\pi}}
                   \sum_i {\rm exp}[-\frac{(e-e_i)^2}{\gamma^2}],
                                                    \label{eqn.12}
\end{equation}
where $\gamma$ indicates the smoothing parameter without much
physical significance. Later, in order to eliminate any possibly strong
$\gamma$-parameter dependence, the mathematical
form in the right side of Eq.~(\ref{eqn.12}) was optimized
by introducing a phenomenological $p$-order curvature-correction polynomial
$P_p(x)$\cite{Werner1995,Nilsson1969,Strutinsky1975,Ivanyuk1978}.
Then, the $\tilde{g}(e,\gamma, p)$ expression will have the form
\begin{equation}
\tilde{g}(e,\gamma, p) = \frac{1}{\gamma\sqrt{\pi}}
                   \sum_{i=1} P_p(\frac{e-e_i}{\gamma})
                   \times{\rm exp}[-\frac{(e-e_i)^2}{\gamma^2}],
                                                    \label{eqn.13}
\end{equation}
where the corrective polynomial $P_p(x)$ can be expanded in terms of
the Hermite or Laguerre polynomials. The corresponding coefficients
of the expansion can be obtained by using the orthogonality
properties of these polynomials and Strutinsky condition, i.e., cf.
the APPENDIX in Ref.\cite{Pomorski2004}. To a large extent, this method can
be considered standard so far (one can see Ref.\cite{Bolsterli1972} for more details). The smooth density is
calculated with a sixth-order Hermite polynomial and a smoothing
range $\gamma =1.20\hbar \omega _0$, where $\hbar
\omega_0=41/A^{1/3}$ MeV, indicating a satisfactory independence of
the shell correction on the parameters $\gamma$ and
$p$~\cite{Bolsterli1972}.
Besides, there are also some other methods developed for the shell
correction calculations, e.g., the semiclassical Wigner-Kirkwood
expansion method~\cite{Vertse1998} and the Green's
function method~\cite{Kruppa2000}. 

As it is known, the mean field cannot well cover the short-range components of the nucleon-nucleon interactions and the most dominating residual interaction is the pairing force in the mean-field theory. Indeed, during the nuclear structural calculations within the framework of the MM models,  besides the shell correction, another important quantum corrections is the pairing-energy contribution. Up to now, as shown in Ref.~\cite{Gaamouci2021}, there exist various variants of the
pairing-energy contribution in the microscopic-energy calculations. For instance, several kinds of the phenomenological pairing energy expressions are widely adopted in the applications of the
MM approach~\cite{Gaamouci2021}, such as pairing correlation and pairing correction energies employing or not employing the particle number projection technique. In the present work, we calculate the the pairing correlation energy $ E_{pair}(Z,N,\hat{\beta})$ employing the approximately particle number projection technique, namely, the
Lipkin-Nogami (LN) method~\cite{Pradhan1973,Satula1994, Pradhan1973} which aims to minimize the expectation value of the Hamiltonian:
\begin{eqnarray}
	\hat{\mathcal{H}}=\hat{H}_{\text{WS}}+\hat{H}_{\text{pair}}-\lambda_1\hat{N}- \lambda_2\hat{N}^2
	                                                         \label{eqn.14}
\end{eqnarray}
here, $\hat{H}_{\text{pair}}$ indicates the pairing interaction Hamiltonian \cite{Xu2000,Moller1992}. By such a pairing treatment, not only the spurious pairing phase transition but also the particle number fluctuation encountered in the simpler BCS calculation can be avoided. In this method, the LN pairing energy for an even-even nuclear system at ``paired solution'' (pairing gap $\Delta\neq 0$) can be computed by~\cite{Pradhan1973,Moller1995}
\begin{eqnarray}
E_{LN}&=&\sum_{k}2{v_k}^2e_k-\frac{{\Delta}^2}G-G\sum_{k}{v_k}^4
\nonumber \\
     && -4{\lambda}_2\sum_{k}{u_k}^2{v_k}^2,
                                                                  \label{eqn.15}
\end{eqnarray}
where ${v_k}^2$, $e_k$, $\Delta$ and ${\lambda}_2$ represent the
occupation probabilities, single-particle energies, pairing gap and
number-fluctuation constant, respectively. At ``no-pairing solution'' ($\Delta = 0$), one can give the corresponding partner expression as follows
\begin{eqnarray}
E_{LN}(\Delta=0)&=&\sum_{k}2e_k-G\frac{N}{2}.
                                                                  \label{eqn.16}
\end{eqnarray}
Together with Eq.~(\ref{eqn.15}), the pairing correlation $E_{pair}$, defined as the difference between paired solution $E_{LN}$ and no-pairing solution $E_{LN}$($\Delta=0$), will have the expression,
\begin{eqnarray}
E_{pair}&=&\sum_{k}2{v_k}^2e_k-\frac{{\Delta}^2}G-G\sum_{k}{v_k}^4
\nonumber \\
     && -4{\lambda}_2\sum_{k}{u_k}^2{v_k}^2+G\frac{N}{2}-\sum_{k}2e_k.
                                                                  \label{eqn.17}
\end{eqnarray}
At this moment, let us clarify some confusing points for the general readers, in the literature, e.g., cf. Refs.~\cite{Xu1998,Wang2012}, the quantum shell correction and pairing contribution are sometime merged as the ``shell correction'' $\delta E_{shell}$ defined by the difference $E_{LN} - \tilde{E}_{Strut}$ (e.g., see Eq.~(1) in Ref.~\cite{Xu1998}). Here, the definition of $E_{LN}$ includes the term $G\frac{N}{2}$, somewhat different with the expression in Eq.~(\ref{eqn.15}). In fact, the so-called ``shell correction'' $\delta E_{shell}$ e.g., in Ref.~\cite{Xu1998} agrees with the summation $\delta E_{shell}$ and $\delta E_{pair}$ in the present Eqs.~(\ref{eqn.11}) and (\ref{eqn.17}). There is no doubts, one should pay special attention to the basic definiations (such as shell correction, pairing correlation and pairing correction) in the literature and they may be slightly different.

Nuclear collective rotation can be accounted for by the one-dimensional cranking approximation, supposing that the nuclear system is constrained to
rotate around a fixed axis (e.g. the $x-$axis with the largest
moment of inertia) at a given rotational frequency $\omega$~\cite{Voigt1983}. For instance, under rotation, the Hamiltonian, as seen in Eq.~(\ref{eqn.14}), will take the cranking form
\begin{equation}
    \hat{H}^\omega
    =
    \hat{H}_{WS}
    +
    \hat{H}_{pair}
    -
    \omega \hat{j}_x
    -
    \lambda_1 \hat{N}
    -
    \lambda_2\hat{N}^2.
                                                                  \label{eqn.18}
\end{equation}
The resulting cranking LN equation follows the form of the well known
Hartree-Fock-Bogolyubov-like (HFB) equation which can be solved
by using the HFB cranking (HFBC) method, see, e.g., Refs.~\cite{Ring1970,Voigt1983,Satula1994} for a detailed description. The
HFBC equations read,
\begin{equation}
  \left\{
   \begin{array}{lcl}
    \sum_{\beta>0}
    \bigg \{
    \Big [
    (e_\alpha-\lambda)\delta_{\alpha\beta}
    -
    \omega (j_x)_{\alpha\beta}
    -
    G\rho^*_{\bar{\alpha}\bar{\beta}}
    +
    4\lambda_2\rho_{\alpha\beta}
    \Big ]  \\
    \times U_{\beta k}
    -
    \Delta\delta_{\alpha\beta}V_{\bar{\beta} k}
    \bigg \}
    = E_kU_{\alpha k},                                    \\ [5mm]
    \sum_{\beta>0}
    \bigg \{
    \Big [
    (e_\alpha-\lambda)\delta_{\alpha\beta}
    -
    \omega (j_x)_{\alpha\beta}
    -
    G\rho_{\alpha\beta}
    +
    4\lambda_2\rho^*_{\bar{\alpha}\bar{\beta}}
    \Big ]    \\
    \times V_{\bar{\beta} k}
    +
    \Delta^*\delta_{\alpha\beta}U_{\beta k}
    \bigg \}
    =
    E_kV_{\bar{\alpha} k},
       \end{array}
  \right.
                                                                  \label{eqn.19}
\end{equation}
where $\Delta=G\sum_{\alpha>0}\kappa_{\alpha\bar{\alpha}}$,
$\lambda=\lambda_1+ 2\lambda_2(N+1)$, $E_k=\varepsilon_k-\lambda_2$ and $\varepsilon_k$ is the
quasi-particle energy~\cite{Satula1994}. The density matrix and pairing
tensor are denoted by $\rho$ and $\kappa$, respectively. Needless to say, the symmetries of the rotating potential (for example, the signature  and parity are conserved in the present deformation space) can simplify the calculations. After applying an iterative procedure, the HFBC equations can be self-consistently solved. That is, the eigenvectors (matrices $U$ and $V$) and eigenvalues (quasiparticle Routhians $\{ E_k \}$) will be obtained at each frequency $\omega$ and each grid point in the selected deformation space. In principle, one can calculate any physical quantities with the help of the eigenfunctions (namely, matrices $U$ and $V$). For instance, the energy in the
rotating framework can be given by
\begin{eqnarray}
  E^\omega &=& {\rm Tr}(e-\omega j_x)\rho-\frac{\Delta^2}{G}
            - G\sum_{\alpha,\beta >0}
            \rho_{\alpha,\beta}\rho_{\tilde{\alpha},\tilde{\beta}}  \nonumber \\
           && -2\lambda_2 {\rm Tr}\rho(1-\rho),
                                                              \label{eqn.20}
\end{eqnarray}
and the total collective angular momentum can be obtained from,
\begin{eqnarray}
	I_x=\sum_{\alpha,\beta>0}\langle \beta|j_x|\alpha\rangle \rho_{\alpha\beta}
	+\sum_{\alpha,\beta>0}\langle \bar{\beta}|j_x|\bar{\alpha}\rangle 
	\rho_{\bar{\alpha}\bar{\beta}},
\end{eqnarray}
where the signature basis is denoted by $\alpha$ ($\beta$) and $\bar{\alpha}$ ($\bar{\beta}$) for the opposite signature). Then, the energy due to rotation (e.g., used for TRS) and the theoretical MoI $I_x/\omega$ (e.g., usually used for comparing with experimental results) can be calculated. Note that the aligned angular momentum $I_x$ is the total angular momentum $I$ for an even-even nucleus but it is a continuous variable and not conserved since the operator $\hat{j}_x$ does not commute with the Hamiltonian in the present cranking model. Concerning this point, one can also understand from the classic mechanics that, in the Lagrangian, the single-particle potential function includes the generilized coordinates $\theta$ and $\phi$ (which are not ignorable or cyclic coordinates) and so that the angular momentum is not conserved.    
	
So far, the MM calculations generally include several standard steps, i.e., as summarized in Refs.~\cite{Meng2022a, Meng2022b,Chai2018}, by which one can calculate the total energy or Routhian at each sampling deformation point. Then, the smoothly energy/Routhian surface can further be given with the help of the interpolation techniques, e.g., by a spline interpolation. Finally, some physical quantities and/or processes, such as the equilibrium deformations, shape co-existence, fission path, can be studied. In the present project, we focus on the MoIs and energy surfaces affected by the deformations, especially the hexadecapole one.

\section{Results and Discussion}
\label{Results}
\begin{figure}[htbp]
\centering
\includegraphics[width=0.22\textwidth]{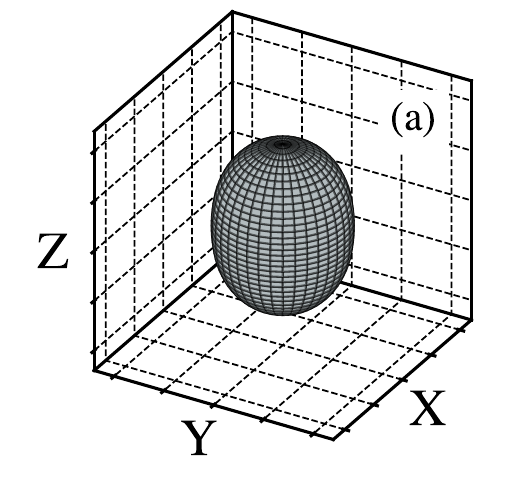}
\includegraphics[width=0.22\textwidth]{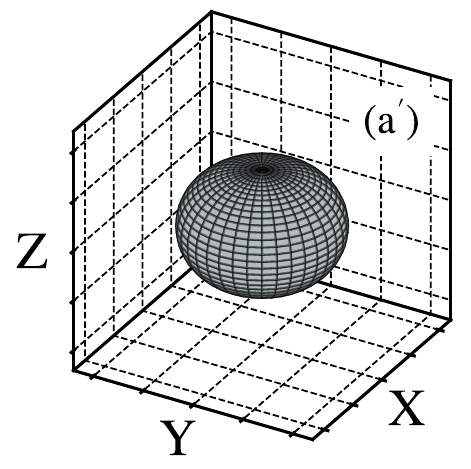}\\
\includegraphics[width=0.22\textwidth]{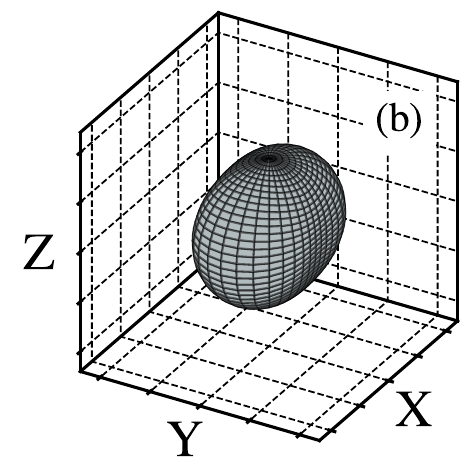}
\includegraphics[width=0.22\textwidth]{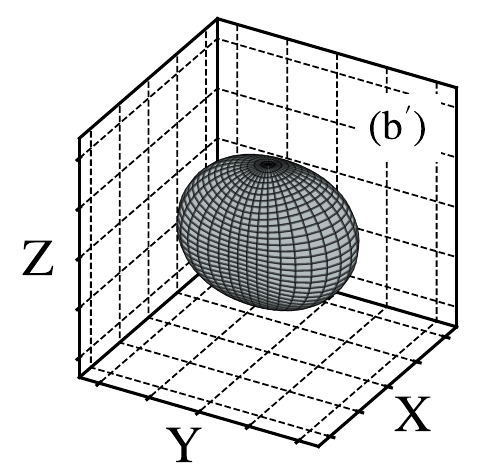}
\caption{Illustrations of nuclear surfaces, defined by Eq. (\ref{eqn.07}), for single deformation parameter $\alpha_{20}= +0.3$ (a), $\alpha_{20}= -0.3$ (a$^\prime$), $\alpha_{22} = +0.3$ (b), and $\alpha_{22}= -0.3$ (b$^\prime$).}
                                                                \label{Fig01}
\includegraphics[width=0.22\textwidth]{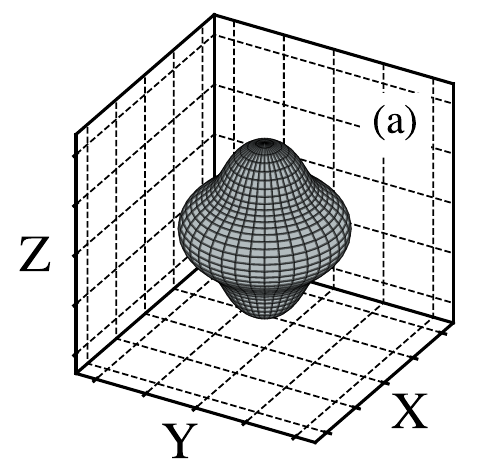}
\includegraphics[width=0.22\textwidth]{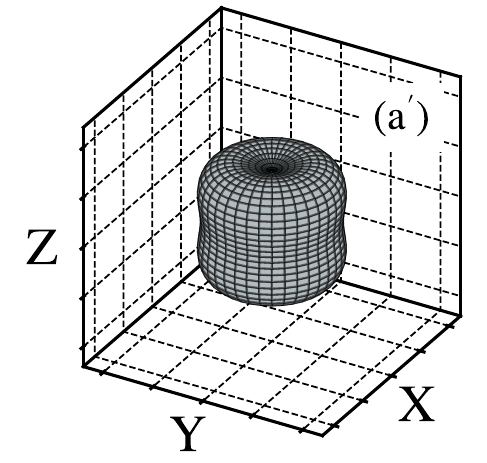}
\includegraphics[width=0.22\textwidth]{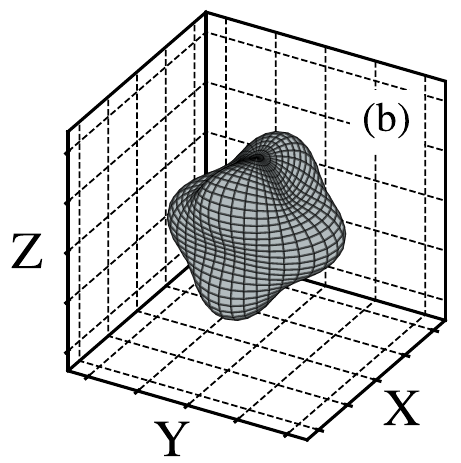}
\includegraphics[width=0.22\textwidth]{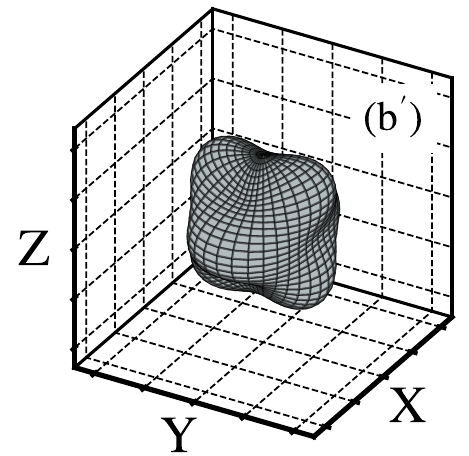}
\includegraphics[width=0.22\textwidth]{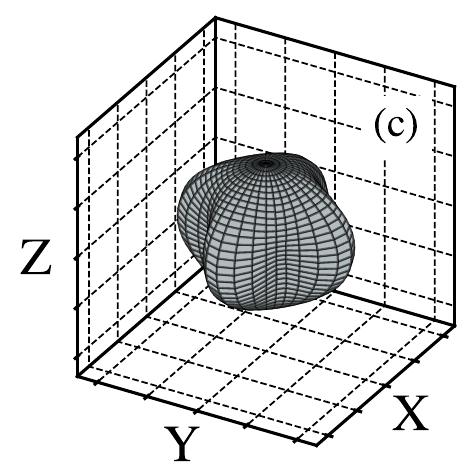}
\includegraphics[width=0.22\textwidth]{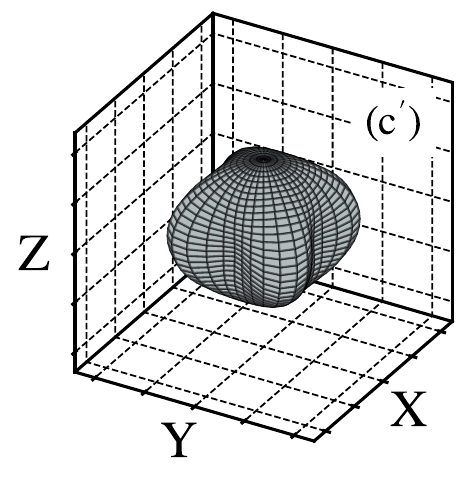}	
\caption{Similar to Fig.~\ref{Fig01} but for $\alpha_{4\mu}=+0.3$ (left) and $-0.3$ (right), $\mu=$0 (a and a$^{\prime}$), 2 (b and b$^{\prime}$), 4 (c and c$^{\prime}$).}
                                                                \label{Fig02}
\end{figure}
Let us start the presentation by visually illustrating the nuclear shapes with different quadrupole deformations and hexadecapole deformations involved in the present investigation. Figure~\ref{Fig01} illustrates the shapes for $\alpha_{2\mu=0,2}$ with arbitrarily selected values $\pm 0.3$. One can see that the shapes with positive and negative $\alpha_{20}$ respectively correspond to prolate and oblate shapes, breaking the spherical symmetry and prolongating along the $Z$ axis; and the positive and negative $\alpha_{22}$ values will respectively mean that the shape is compressed along the direction of $Y$ and $X$ axes, corresponding to the breaking of axial symmetry around $Z$ axis. Similarly, the shapes with $\alpha_{4\mu=0,2,4} = \pm 0.3$ are respectively displayed in Fig.~\ref{Fig02}. For the shape with $\alpha_{40}$, different signs correspond to different space distribution but other two hexadecapole deformations just have different orientations. Note that the relations between the collective coordinates $\alpha_{\lambda\mu}$ and Bohr deformation parameters $\beta_2$ and $\gamma$, cf. e.g., Eq.~(\ref{eqn.03}). One can easily understand that the nucleus with same shapes but different space orientations will obviously have the same mean-field potential, corresponding to the same model Hamiltonian and, certainly, the same single-particle levels. Practically, in the present calculations, nuclear shape are usually the combination of these separated shapes exhibited in Figs.~\ref{Fig01} and \ref{Fig02}. The detailed description of nuclear shape and its symmetry properties should follow the expression mentioned in Eq~(\ref{eqn.01}).

\begin{table*}
\centering
\begin{threeparttable}
\caption{Calculated ground-state equilibrium deformations $\beta_{2}$ and $\beta_{4}$ for even-even nuclei $^{180-184}$Yb, $^{182-184}$Hf and $^{184-188}$W in the present work, together with the theoretical results by the FY+FRDM~\cite{Moller1995}, HFBCS~\cite{Goriely2001} and ETFSI~\cite{Aboussir1995} calculations and part of experimental (Exp.) $\beta_{2}$ values~\cite{nndc_ensdf} for comparison. All the calculated equilibrium $\gamma$ deformations are almost zero and ignored here. See text for more explanations.}
\setlength{\tabcolsep}{4pt}  
\renewcommand{\arraystretch}{1.2}  
\begin{tabular}{lccccccccccl}
\toprule[1.0pt]
\multirow{2}{*}{Nuclei} & \multicolumn{5}{c}{$\beta_2$}& \multicolumn{4}{c}{$\beta_4$}\\    
\specialrule{0em}{2pt}{2pt} \cline{2-6}  \cline{8-12}
&Present & FY+FRDM & HFBCS & ETFSI & Exp. &
&Present & FY+FRDM & HFBCS & ETFSI \\ 
\midrule[0.5pt]
$^{180}_{70}$Yb$_{110}$& 0.267 &  0.250 & 0.260 & 0.310 &- && -0.071 &  -0.084 & -0.05 & -0.08 \\
$^{182}_{70}$Yb$_{112}$& 0.257 & 0.242 & 0.270 & 0.290 &- && -0.087  & -0.101 & -0.06 & -0.07 & \\
$^{184}_{70}$Yb$_{114}$& 0.240 & 0.233 & 0.240 & 0.280 &- && -0.084  & -0.104 & -0.05 & -0.09 & \\
$^{182}_{72}$Hf$_{110}$& 0.249 & 0.268 & 0.240 & 0.280 &0.274 && -0.079  & -0.099& -0.05 & -0.06  & \\
$^{184}_{72}$Hf$_{112}$&0.237 & 0.256 & 0.250 & 0.260 &- && -0.087 & -0.114 & -0.05 & -0.08 & \\
$^{186}_{72}$Hf$_{114}$& 0.220& 0.225 & 0.220 & 0.250 &- && -0.087 & -0.119& -0.05 & -0.07 & \\
$^{184}_{74}$W$_{110}$& 0.221 & 0.232 & 0.240 & 0.250 &0.234 && -0.070 & -0.093 & -0.05 & -0.07 & \\
$^{186}_{74}$W$_{112}$ &0.210 & 0.221 & 0.210 & 0.250 & 0.227&& -0.077 & -0.095 & -0.04 & -0.06 & \\
$^{188}_{74}$W$_{114}$& 0.194 & 0.220 & -0.210 & 0.200 & 0.198 && -0.077  & -0.109 & -0.05 & -0.08 & \\
\bottomrule[1.0pt]
\end{tabular}
                                                          \label{tab1}
\end{threeparttable}
\end{table*}

\begin{figure}[htbp]
\raggedright
\includegraphics[width=0.23\textwidth]{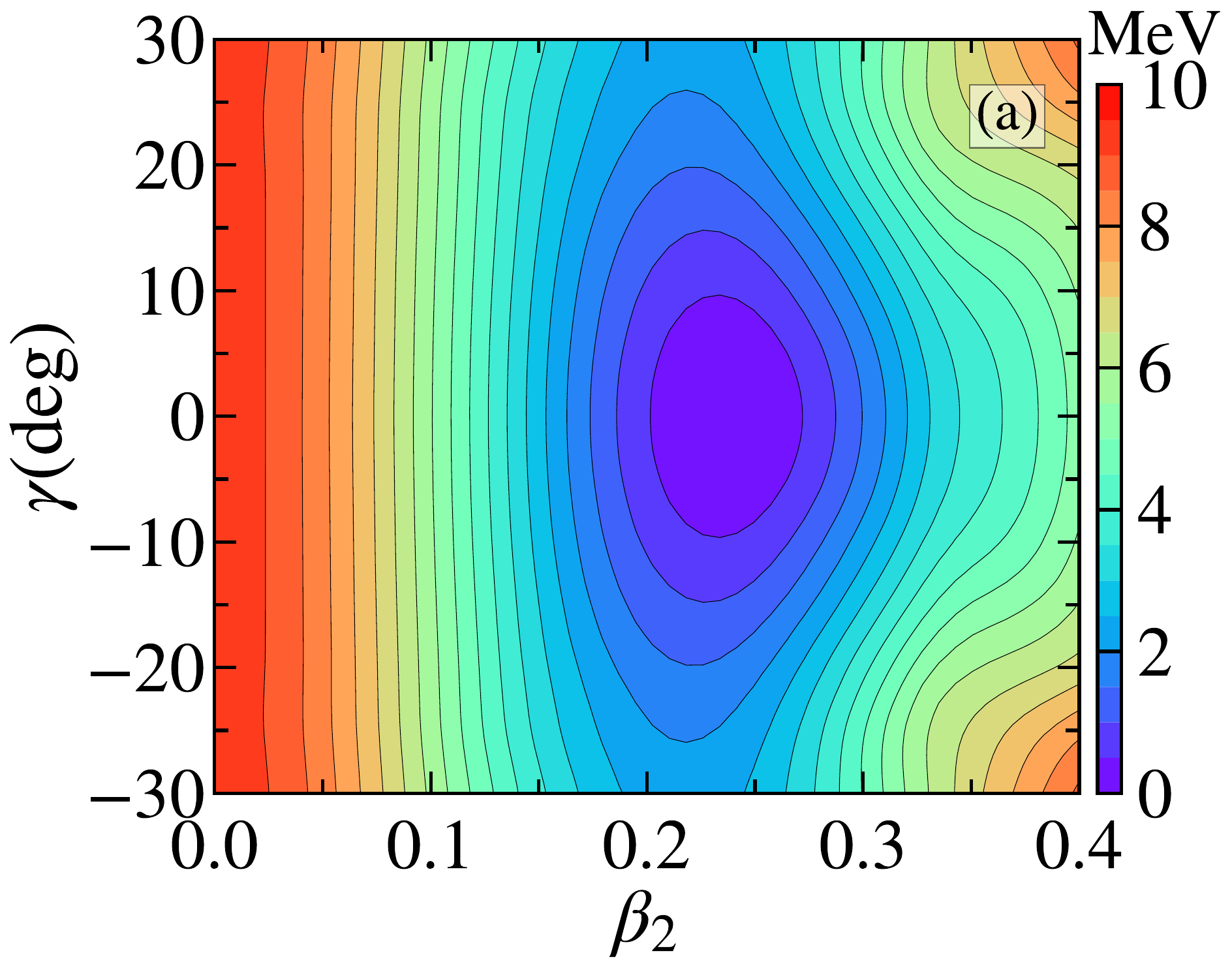}
\includegraphics[width=0.23\textwidth]{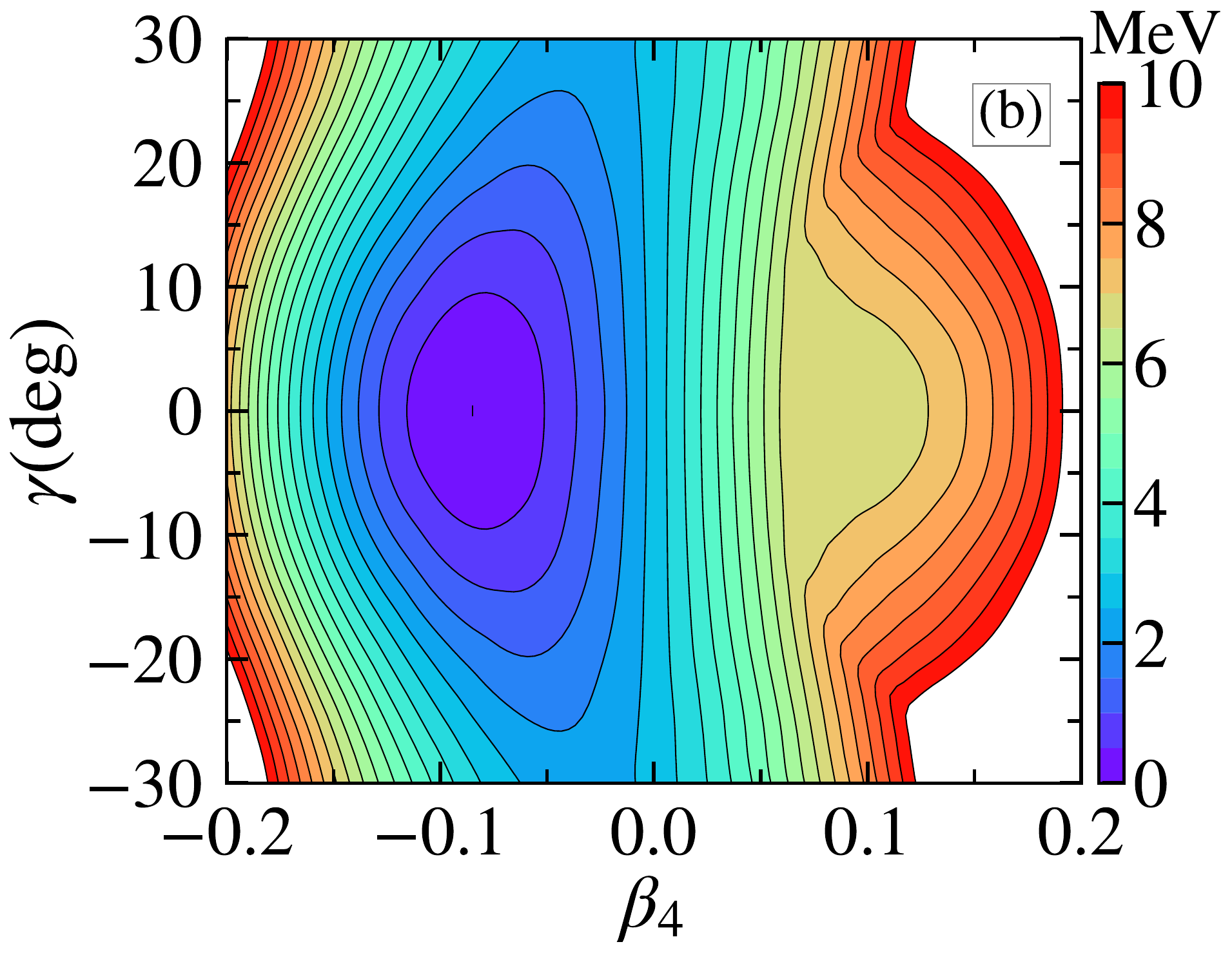}\\
\includegraphics[width=0.23\textwidth]{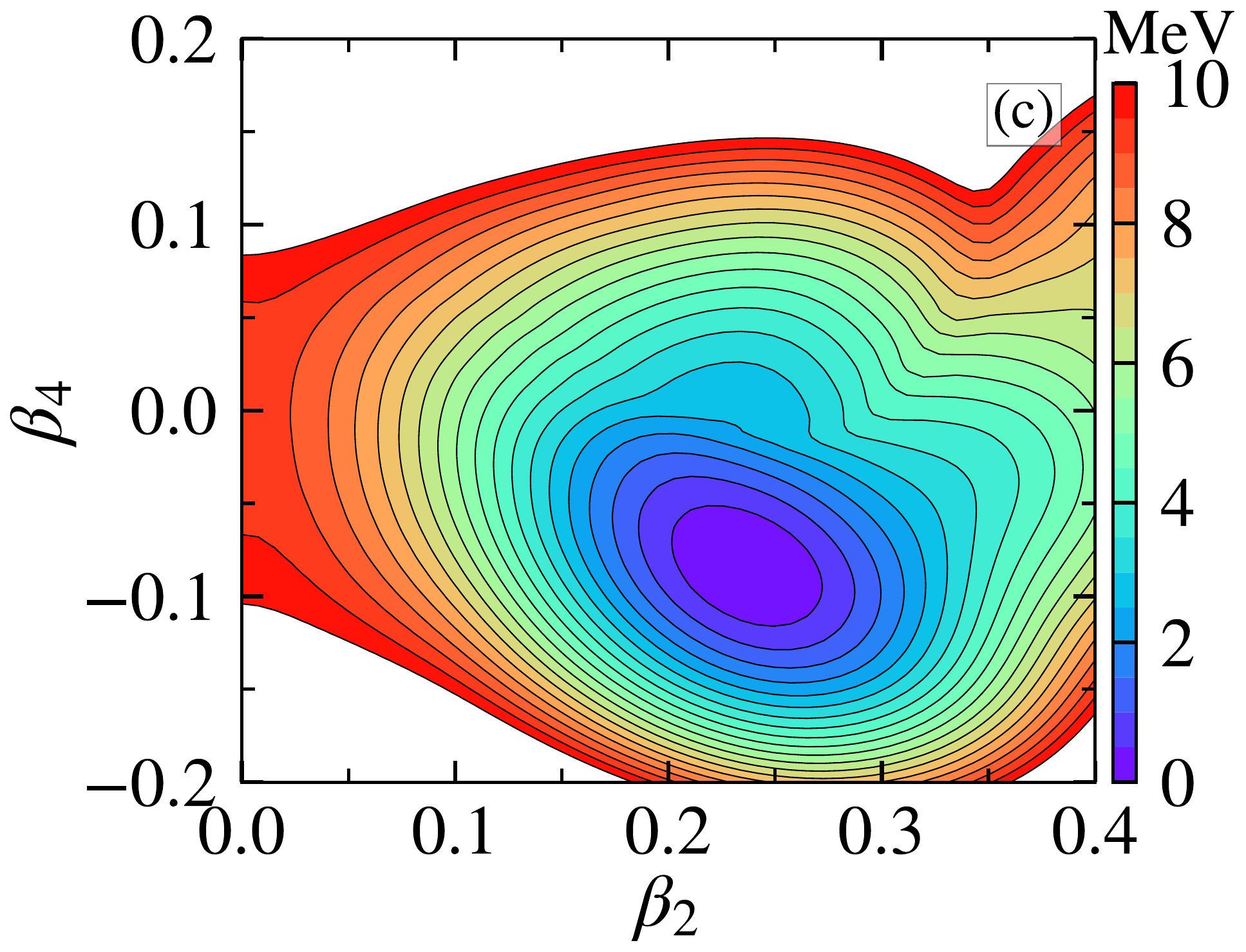}
\caption{ Projections of total energy on the ($\beta_2,\gamma$) (a), ($\beta_4,\gamma$) (b) and ($\beta_2,\beta_4$) (c) planes with contour-line separations of 0.5 MeV, minimized respectively at each deformation point over the remaining deformation, $\beta_4$, $\beta_2$ and $\gamma$, for the central nucleus $^{184}$Hf. Note that, for each subfigure, the energy normalization is specified by setting the minimum to zero at the equilibrium shape. See the text for more details. \\}
		                                                         \label{Fig03}
\end{figure}
\begin{figure}[htbp]	
\raggedright                            
\includegraphics[width=0.23\textwidth]{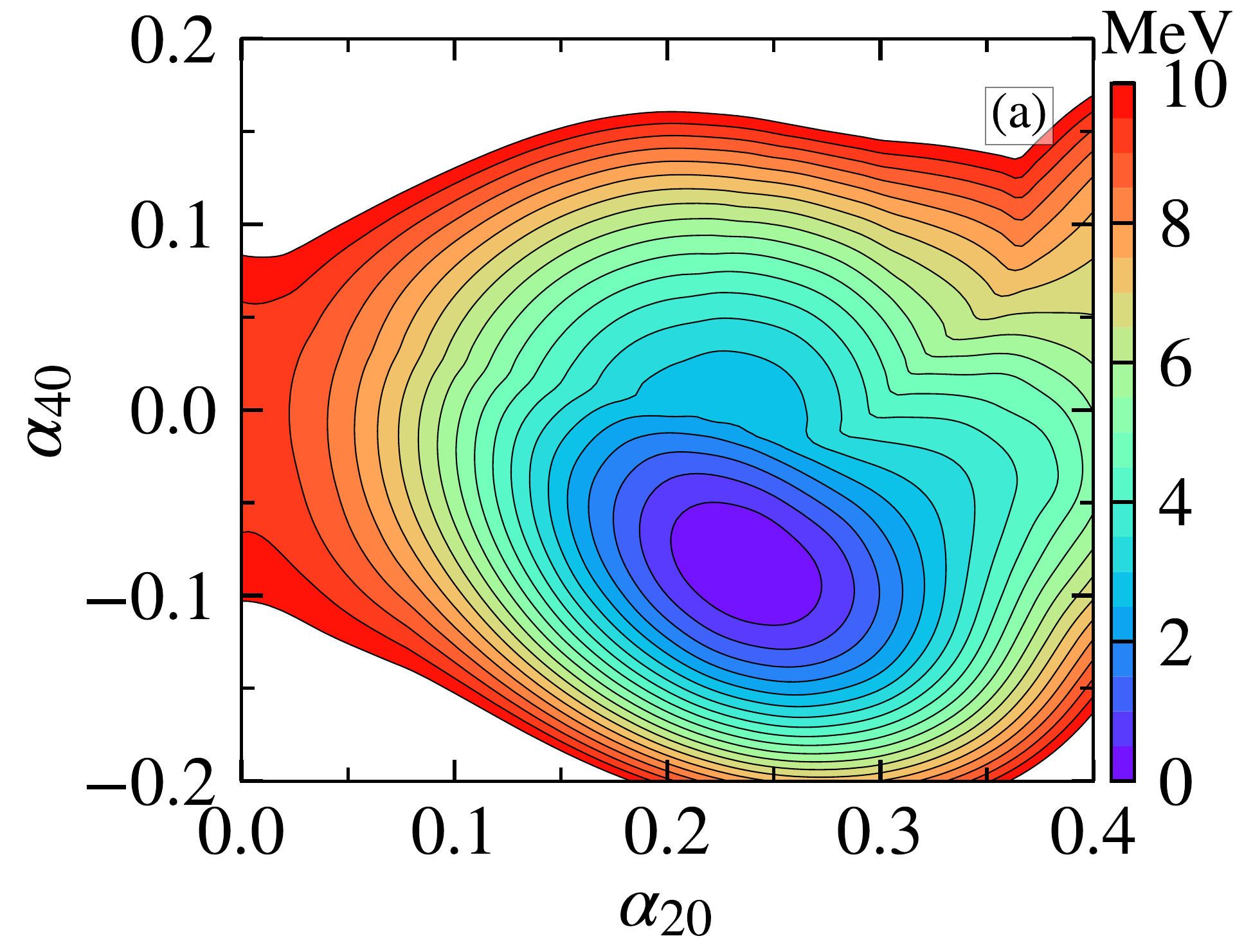}
\includegraphics[width=0.23\textwidth]{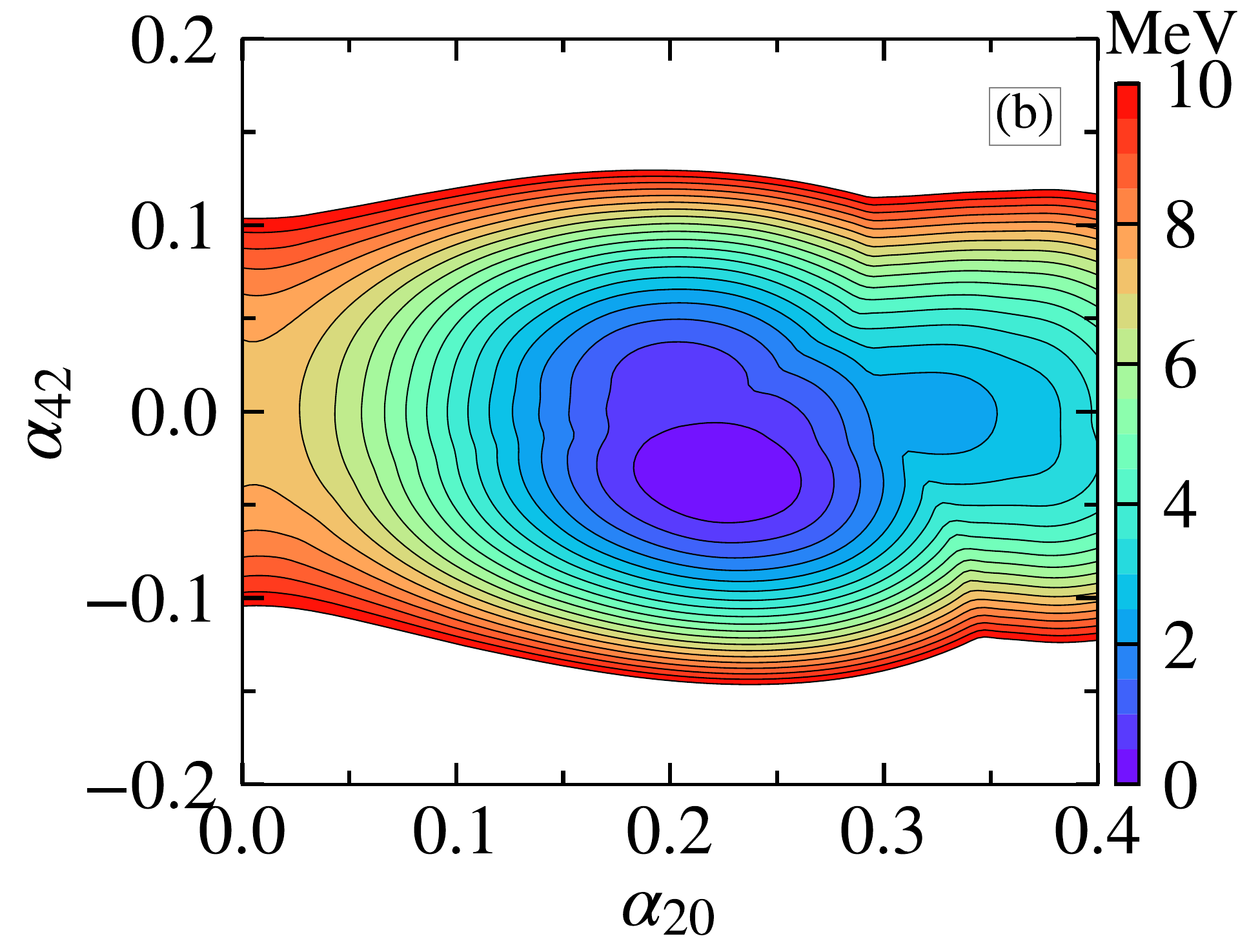}\\
\includegraphics[width=0.23\textwidth]{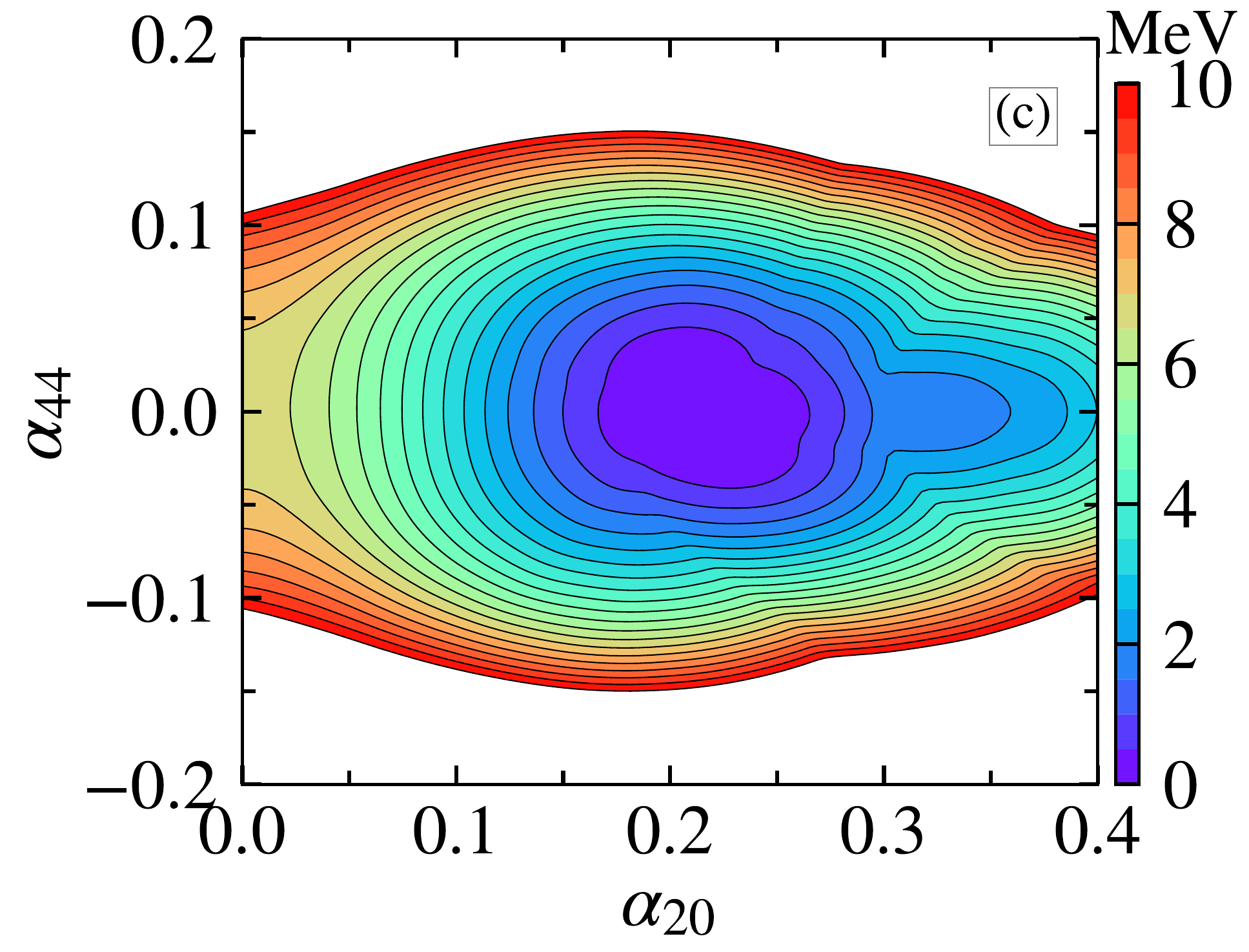}
\caption{ Similar to the preceding illustration in Fig.~\ref{Fig03}, but projected on the $(\alpha_{20},\alpha_{4\mu=0,2,4})$ planes for $^{184}$Hf.}
	                                                             \label{Fig04}
\end{figure}

As described in Section~\ref{introduction}, the hexadecapole deformation is still important in nuclear physics and attracts experimental and theoretical attention~\cite{Gupta2020,Han2021,Ryssens2023}. Within the framework of MM models, by using the PES method --- an important tool for studying the structural properties of atomic nuclei, we perform the energy surface calculations in the selected deformation spaces ($\beta_2$, $\gamma$, $\beta_4$), and ($\alpha_{20}$, $\alpha_{4\mu=0,2,4}$) for probing the different hexadecapole deformation effects. Taking the central nucleus $^{184}_{72}$Hf$_{112}$ as an example, the typical PESs are shown in Figs.~\ref{Fig03} and \ref{Fig04}. Using the frequently-used Bohr deformation parameters, Figure~\ref{Fig03} illustrates three energy maps projected on ($\beta_2$, $\gamma$), ($\beta_4$, $\gamma$) and ($\beta_2$, $\beta_4$) planes. In each plane, the energy is minimized over the remaining deformation degrees of freedom, e.g. minimized over $\beta_4$ in the ($\beta_2$, $\gamma$) plane.	Let us remind that in Fig.~\ref{Fig03}, the three minima in subfigures (a), (b) and (c) have the same energy and equilibrium shape. One can see that the triaxial $\gamma$ is zero and $\beta_2$ and $\beta_4$ deformations are obviously nonzero, agreeing with the theoretical values given by M\"oller et al~\cite{Moller2016} (experimental data is scarce in this nucleus). Similarly, Figure~\ref{Fig04} presents the energy projections on the ($\alpha_{20}$, $\alpha_{4\mu=0,2,4}$) planes. It is found that the axial hexadecapole deformation $\alpha_{40}$ has a relatively large value, $\sim 0.1$, but the non-axial $\alpha_{4\mu=2,4}$ deformations are weak though the $\alpha_{42}$ value seems to nonzero. Furthermore, the energy minimum ($\sim -5.44$ MeV) in Fig.~\ref{Fig04}(a) is about 2.5 MeV lower than, e.g., the minimum ($\sim -2.94$ MeV) in Fig.~\ref{Fig04}(c), indicating that the $\alpha_{40}$ deformation may cause a large reduction in energy.

To verify the effectiveness of the present calculation, e.g., the adopted StkI parameters~\cite{Meng2018,Zhang2021,Bhagwat2023}, it might be necessary to confront with experimental observations or other theoretical results. We show the calculated equilibrium deformations $\beta_{2}$ and  $\beta_{4}$ for present nine even-even nuclei on the selected hexadecapole deformation island in Table~\ref{tab1}, together with the theoretical results from the fold-Yukawa single-particle potential and the ﬁnite-range droplet model (FY+FRDM)~\cite{Moller1995}, Hartree-Fock-BCS (HFBCS)~\cite{Goriely2001} and the extended Thomas-Fermi plus Strutinsky integral (ETFSI) models~\cite{Aboussir1995}, as well as partial experimental (Exp.) values of $\beta_{2}$~\cite{nndc_ensdf} for comparison. From this table, it can be seen that all theoretical results, including phenomenological and self-consistent mean-field calculations, are somewhat smaller than the experimental data. However, according to the presently calculated results, one can see, as expected, that for each isotopic chain the quadrupole deformation $\beta_2$ decreases as the neutron number $N$ changes from 110 to 114, more and more approaching the magic number 126 from the mid-magic number 104; along each isotonic chain, $\beta_2$ decreases as the proton number $Z$ changes from 70 to 74, more and more approaching the magic number 82 from the mid-magic number 66. This trend is in good agreement the available data (e.g., see the $\beta_2$ evolution in even-even isotopes $^{184-188}$W) and also with those of the theoretical calculations in Refs.~\cite{Moller1995,Aboussir1995}, especially in Ref.~\cite{Moller1995} which also adopts the MM theoretical framework. The HFBC calculation seems to deviate from this general trend, particularly giving the oblate shape in $^{188}$W. As can be seen from the Table~\ref{tab1} , it can be concluded that the calculated results are model-dependent and no one can entirely agree with data. Nevertheless, all the theoretical calculations indicate the nonnegligible negative hexadecapole deformation $\beta_4$ though the magnitudes are somewhat different.


\begin{figure}[H]
\centering
\includegraphics[width=0.5\textwidth]{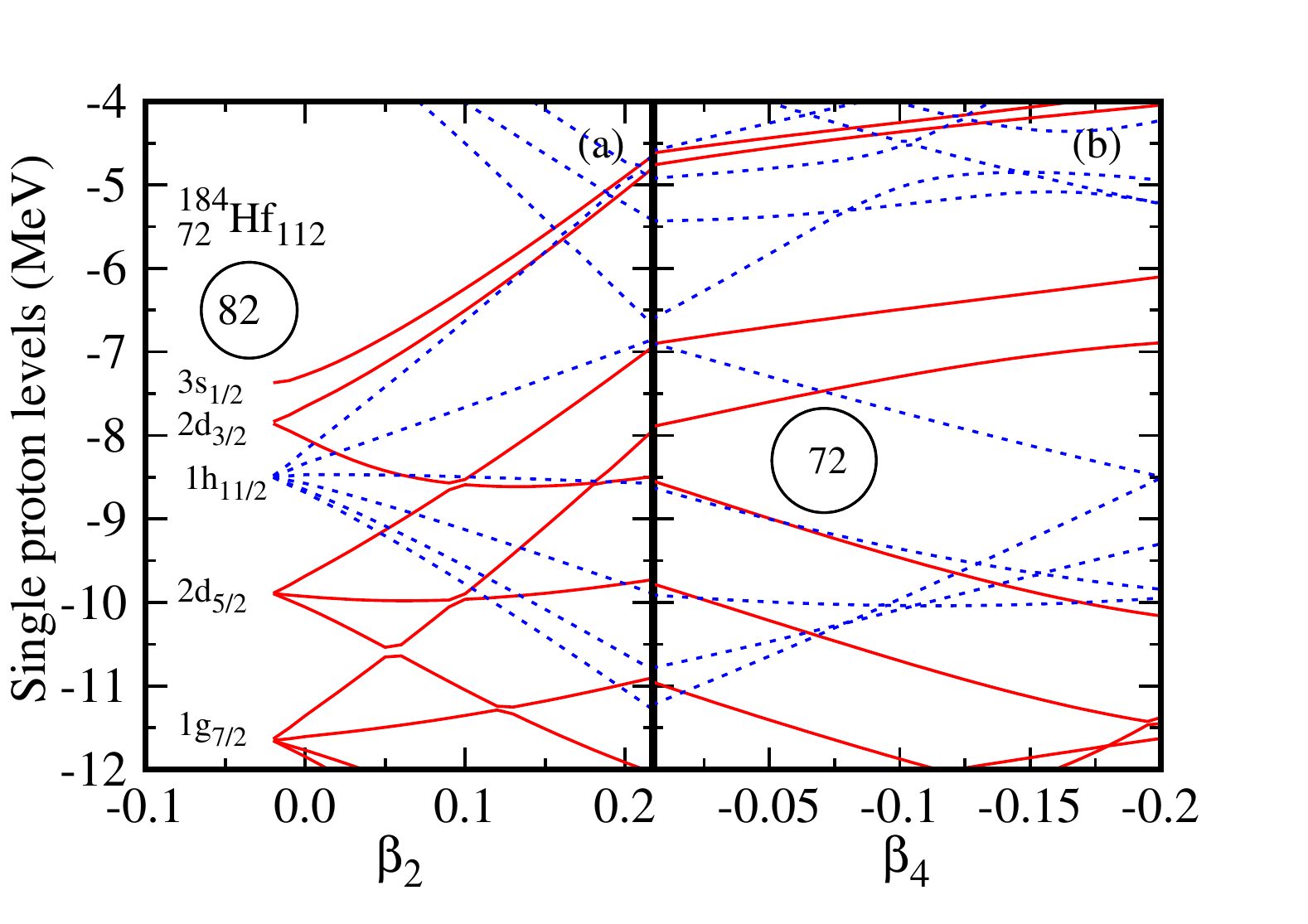} 
\includegraphics[width=0.5\textwidth]{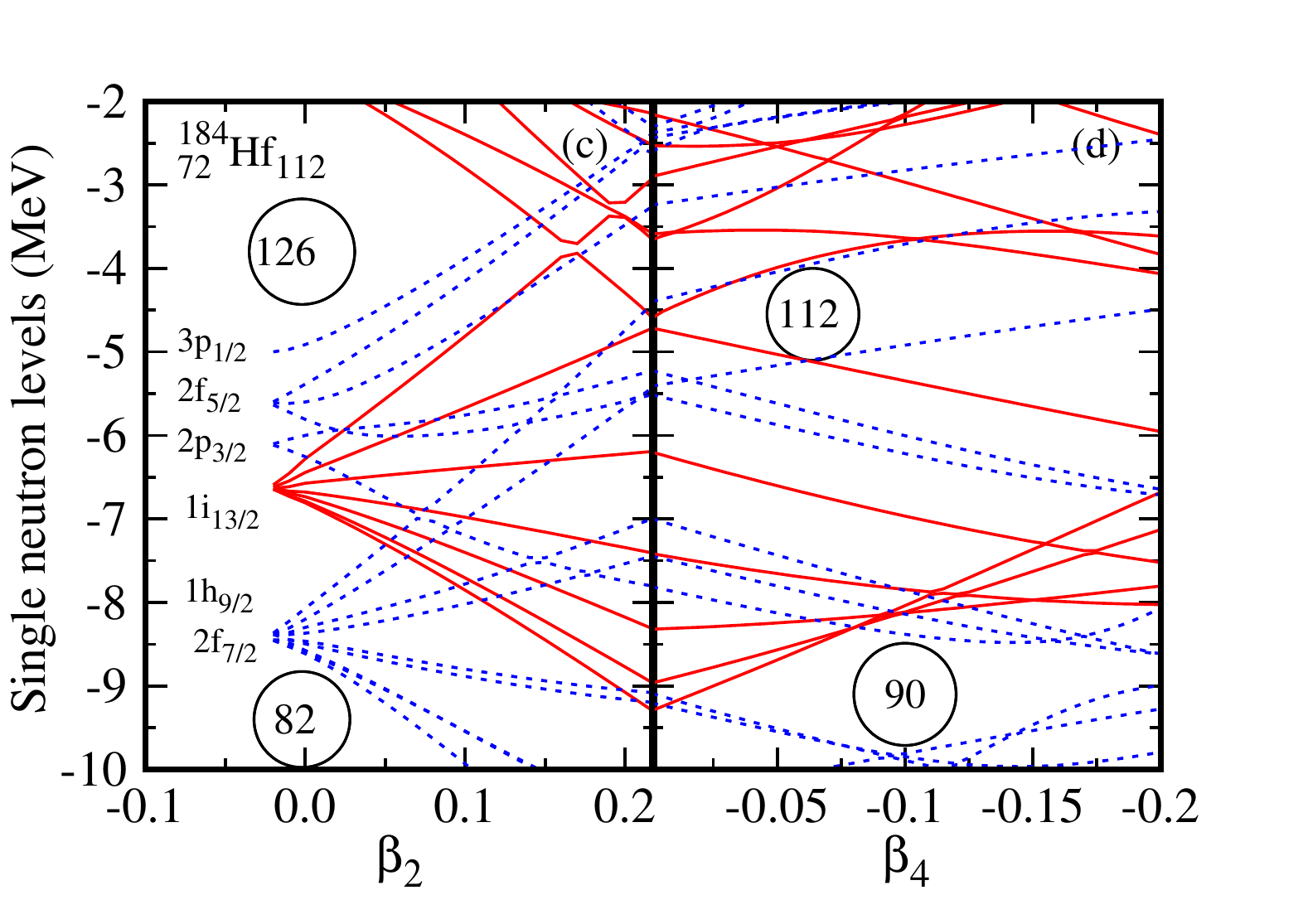}
\caption{Calculated proton (a, b) and neutron (c, d) single-particle energies as functions of the quadrupole deformation $\beta_2$ (a, c) and hexadecapole deformation $\beta_4$ (b, d) for the central nucleus $^{184}_{72}$Hf$_{112}$, focusing on the window of interest near the Fermi surface. Red solid (blue dotted) lines refer to positive and negative parity. In (a) and (c), the single-particle orbitals at $\beta_2 = 0.0$ are labelled by the spherical quantum numbers $nlj$ and the calculations extend to the equilibrium deformation ($\beta_2 = 0.237$), for further details, e.g. see Table~\ref{tab2}. In (b) and (d), the deformation $\beta_2$ is always set to the equilibrium value.}
                                                                   \label{Fig05}
\end{figure}

\begin{table*}
\centering
\caption{The calculated single-particle levels near the Fermi surface at $\beta_2= 0.237$,$\gamma= 0^o$ and $\beta_4 = 0.00$ for protons and neutrons in the selected nucleus $^{184}_{72}$Hf$_{112}$,together with their wave-function components expanded in the cylindrical basis $|Nn_z\Omega\rangle$ and spherical basis $|Nlj\Omega\rangle$.The calculations are performed using the WS Hamiltonian with the cranking parameters.The proton and neutron Fermi levels correspond to the energies -7.95 MeV and -5.43 MeV, respectively.}
\setlength{\tabcolsep}{5pt}  
\renewcommand{\arraystretch}{1.2}  
\begin{tabular}{lll}
\toprule[1.0pt]
&\textbf{$\varepsilon$}(MeV) & The first six main-components in terms of $|Nn_z\Lambda\Omega\rangle$ (upper) and $|Nlj\Omega\rangle$ (lower) \\ 
\midrule[0.5pt]
			Proton&-8.58 & $87.6\%|523\frac{7}{2}\rangle + 
			7.7\% |514\frac{7}{2}\rangle + 
			2.5 \% |503\frac{7}{2}\rangle + 
			0.6\% |743\frac{7}{2}\rangle + 
			0.5\% |303\frac{7}{2}\rangle+
			0.3\% |943\frac{7}{2}\rangle$\\
			& & 88.4\% $|5h_{\frac{11}{2}}\frac{7}{2}\rangle + 
			3.1\% |7j_{\frac{15}{2}}\frac{7}{2}\rangle + 
			2.7\% |5f_{\frac{7}{2}}\frac{7}{2}\rangle + 
			1.9\% |5h_{\frac{9}{2}}\frac{7}{2}\rangle + 
			0.6\% |3f_{\frac{7}{2}}\frac{7}{2}\rangle + 
			0.6\% |9h_{\frac{11}{2}}\frac{7}{2}
			\rangle$\\
			&-8.49 & $76.3\%|411\frac{1}{2}\rangle + 
			6.4\% |420 \frac{1}{2}\rangle + 
			6.0\% |431 \frac{1}{2}\rangle + 
			5.0\% |211\frac{1}{2}\rangle +
			1.5\% |631 \frac{1}{2}\rangle + 
			1.3\% |440 \frac{1}{2}\rangle $\\
			&&45.1\%$|4d_{\frac{3}{2}}\frac{1}{2}\rangle + 
			20.4\%|4d_{\frac{5}{2}}\frac{1}{2}\rangle + 
			15.0\% |4g _{\frac{7}{2}}\frac{1}{2}\rangle +
			3.0\% |4s_{\frac{1}{2}}\frac{1}{2}\rangle + 
			2.4\%|4g_{\frac{9}{2}}\frac{1}{2}\rangle + 
			1.9\% |2d_{\frac{3}{2}}\frac{1}{2}\rangle $ \\
			&-7.95 &  $96.1\%|404\frac{7}{2}\rangle + 
			3.2\% |413\frac{7}{2}\rangle+
			0.3\% |804\frac{7}{2}\rangle + 
			0.2\% |624 \frac{7}{2}\rangle + 
			0.0\% |824 \frac{7}{2}\rangle + 
			0.0\% |613 \frac{7}{2}\rangle$ \\
			& & 95.0\%$|4g_{\frac{7}{2}}\frac{7}{2}\rangle + 
			2.5\% |4g_{\frac{9}{2}}\frac{7}{2}\rangle + 
			1.3\%|6i_{\frac{11}{2}}\frac{7}{2}\rangle + 
			0.5\%|6g_{\frac{7}{2}}\frac{7}{2}\rangle + 
			0.2\% |10g_{\frac{7}{2}}\frac{7}{2}\rangle + 
			0.2\% |8g_{\frac{7}{2}}\frac{7}{2}\rangle$\\
			&-6.94& 87.9\%$|402\frac{5}{2}\rangle + 
			6.2\%|202\frac{5}{2}\rangle +
			1.7\%|602\frac{5}{2}\rangle + 
			1.5\%|422\frac{5}{2}\rangle + 
			1.0\%|622\frac{5}{2}\rangle +
			0.7\%|802\frac{5}{2}\rangle$\\
			&&86.0\%$|4d_{\frac{5}{2}}\frac{5}{2}\rangle + 
			21.2\% |2d_{\frac{5}{2}}\frac{5}{2}\rangle + 
			15.9\% |4g_{\frac{7}{2}}\frac{5}{2}\rangle + 
			5.3\% |4g_{\frac{9}{2}}\frac{5}{2}\rangle + 
			3.4\% |6d_{\frac{5}{2}}	\frac{5}{2}\rangle + 
			2.0\% |8d_{\frac{5}{2}}\frac{5}{2}\rangle$\\
			&-6.85 & 95.7\%$|514\frac{9}{2}\rangle + 
			3.0\% |505\frac{9}{2}\rangle+ 
			0.4\% |914\frac{9}{2}\rangle + 
			0.4\% |734\frac{9}{2}\rangle + 
			0.4\% |934\frac{9}{2}\rangle+
			0.0\%|954\frac{9}{2}\rangle$ \\
			&&94.3\%$|5h_{\frac{11}{2}}\frac{9}{2}\rangle + 
			2.7\%|7j_{\frac{15}{2}}\frac{9}{2}\rangle + 
			1.6\% |5h_{\frac{9}{2}}\frac{9}{2}\rangle + 
			0.8\% |9h_{\frac{11}{2}}\frac{9}{2}\rangle+
			0.2\% |7j_{\frac{13}{2}}\frac{9}{2}\rangle +
			0.2\% |11h_{\frac{11}{2}}\frac{9}{2}\rangle$\\
			\bottomrule[0.5pt]
			Neutron& -6.19& $86.5\%|624\frac{9}{2}\rangle + 
			7.1\% |615 \frac{9}{2}\rangle + 
			2.1 \% |604 \frac{9}{2}\rangle + 
			1.8\% |824 \frac{9}{2}\rangle + 
			1.7\% |844 \frac{9}{2}\rangle+
			0.4\% |404\frac{9}{2}\rangle$\\
			& &88.3\%$|6i_{\frac{13}{2}}\frac{9}{2}\rangle + 
			3.2\%|8i_{\frac{13}{2}}\frac{9}{2}\rangle + 
			2.6\% |8k _{\frac{17}{2}}\frac{9}{2}\rangle +
			2.0\% |6g_{\frac{9}{2}}\frac{9}{2}\rangle + 
			1.5\%|4g_{\frac{9}{2}}\frac{9}{2}\rangle + 
			1.2\% |6i_{\frac{11}{2}}\frac{9}{2}\rangle $ \\
			&-5.51 & $65.3\%|510\frac{1}{2}\rangle + 
			10.4\% |521 \frac{1}{2}	\rangle +
			6.2\% |310 \frac{1}{2}\rangle + 
			5.3\% |710\frac{1}{2}\rangle +
			3.5\% |730 \frac{1}{2}\rangle + 
			2.0\% |530 \frac{1}{2}	\rangle $\\
			& &
			29.7\%$|5p_{\frac{3}{2}}\frac{1}{2}\rangle + 
			25.5\%|5f_{\frac{5}{2}}\frac{1}{2}\rangle + 
			14.9\%|5f_{\frac{7}{2}}\frac{1}{2}\rangle + 
			10.0\%|5h_{\frac{9}{2}}\frac{1}{2}\rangle + 
			6.6\% |7p_{\frac{3}{2}}\frac{1}{2}\rangle + 
			3.5\% |3p_{\frac{3}{2}}\frac{1}{2}\rangle$\\
			& -5.43&  $78.1\%|503\frac{7}{2}\rangle + 
			7.8\% |703\frac{7}{2}\rangle+
			7.3\% |303\frac{7}{2}\rangle + 
			4.7\% |514 \frac{7}{2}\rangle + 
			1.5\% |723 \frac{7}{2}\rangle + 
			0.3\% | 523 \frac{7}{2}\rangle
			$ \\
			& &
			76.8\%$|5f_{\frac{7}{2}}\frac{7}{2}\rangle + 
			9.8\% |5h_{\frac{9}{2}}\frac{7}{2}\rangle + 
			5.9\% |7f_{\frac{7}{2}}\frac{7}{2}\rangle + 
			4.0\% |3f_{\frac{7}{2}}\frac{7}{2}\rangle + 
			2.1\% |5h_{\frac{11}{2}}\frac{7}{2}\rangle + 
			0.5\% |11f_{\frac{7}{2}}\frac{7}{2}\rangle$\\
			&-5.22& 70.1\%$|512\frac{3}{2}\rangle + 
			9.0\%|512\frac{3}{2}\rangle +
			4.1\%|521\frac{3}{2}\rangle + 
			3.3\%|712\frac{3}{2}\rangle + 
			2.8\%|501\frac{3}{2}\rangle +
			2.5\%|532\frac{3}{2}\rangle$\\
			&&48.2\%$|5f_{\frac{5}{2}}\frac{3}{2}\rangle + 
			13.4\% |5h_{\frac{9}{2}}\frac{3}{2}\rangle + 
			11.6\% |5f_{\frac{7}{2}}\frac{3}{2}\rangle + 
			11.2\% |5p_{\frac{3}{2}}\frac{3}{2}\rangle + 
			5.1\% |7f_{\frac{5}{2}}\frac{3}{2}\rangle + 
			1.9\% |3p_{\frac{3}{2}}\frac{3}{2}\rangle$\\
			&-4.72 & 93.6\%$|615\frac{11}{2}\rangle + 
			2.9\% |606\frac{11}{2}\rangle+ 
			2.3\% |815\frac{11}{2}\rangle + 
			1.0\% |835 \frac{11}{2}\rangle + 
			0.1\%|806\frac{11}{2}\rangle+ 
			0.0\% |1035 \frac{11}{2}\rangle$ \\
			&&94.7\%$|6i_{\frac{13}{2}}\frac{11}{2}\rangle + 1.8\%|8k_{\frac{17}{2}}\frac{11}{2}\rangle + 
			1.7\% |8i_{\frac{13}{2}}\frac{11}{2}\rangle + 
			1.0\% |6i_{\frac{11}{2}}\frac{11}{2}\rangle+
			0.3\% |8k_{\frac{15}{2}}\frac{11}{2}\rangle + 
			0.2\% |12i_{\frac{13}{2}}\frac{11}{2}\rangle$\\
\bottomrule[1.0pt]
		                                                            \label{tab2}
\end{tabular}\\
\vspace{0.6cm}
\centering
\caption{The same as table \ref{tab2},but  $\beta_4=-0.087$ and the proton and neutron Fermi levels correspond to the energies -7.62 MeV and -5.91 MeV, respectively.}
\setlength{\tabcolsep}{5pt}  
\renewcommand{\arraystretch}{1.2}  
\begin{tabular}{lll}
\toprule[1.0pt]
&\textbf{$\varepsilon$}(MeV) & The first six main-components in terms of $Nn_z\Lambda\Omega\rangle$	(upper) and $|Nlj\Omega\rangle$ (lower) \\ 
\midrule[0.5pt]
			Proton&-9.35 & $72.4\% |411 \frac{1}{2}\rangle + 
			7.7\% |420 \frac{1}{2}\rangle + 
			5.3 \% |211 \frac{1}{2}\rangle + 
			3.8\% |631 \frac{1}{2}	\rangle +  
			3.8\% |431\frac{1}{2}\rangle+ 
			3.0\% |440 \frac{1}{2}\rangle$\\
			& &
			44.9\%$|4d_{\frac{3}{2}}\frac{1}{2}\rangle + 
			15.6\% |4g_{\frac{7}{2}}\frac{1}{2}\rangle + 
			15.4\% |4d_{\frac{5}{2}}\frac{1}{2}\rangle + 
			7.1 \% |4s_{\frac{1}{2}}\frac{1}{2}\rangle + 
			3.3\% |4g_{\frac{9}{2}}\frac{1}{2}\rangle + 
			2.7\% |2d_{\frac{3}{2}}\frac{1}{2}\rangle$\\
			&-9.27& 85.8\%$|523\frac{7}{2}\rangle + 
			7.8\%|514\frac{7}{2}\rangle +
			2.6\%|303\frac{7}{2}\rangle + 
			1.3\%|743\frac{7}{2}\rangle + 
			1.0\%|503\frac{7}{2}\rangle+
			0.5\%|963\frac{7}{2}\rangle$\\
			&&82.9\%$|5h_{\frac{11}{2}}\frac{7}{2}\rangle + 
			4.7\%|5f_{\frac{7}{2}}\frac{7}{2}\rangle + 
			4.5\% |3f_{\frac{7}{2}}\frac{7}{2}\rangle + 
			3.6\% |7j_{\frac{15}{2}}\frac{7}{2}\rangle+
			1.5\% |5h_{\frac{9}{2}}\frac{7}{2}\rangle + 
			1.2\% |7h_{\frac{11}{2}}\frac{7}{2}\rangle$\\
			&-7.62 & 95.8\%$|514\frac{9}{2}\rangle + 
			1.8\% | 734\frac{9}{2}\rangle+ 
			1.5\% |505\frac{9}{2}\rangle + 
			0.5\% |914 \frac{9}{2}\rangle + 
			0.2\%|954\frac{9}{2}\rangle+ 
			0.1\% |934 \frac{9}{2}\rangle$ \\
			&&91.2\%$|5h_{\frac{11}{2}}\frac{9}{2}\rangle + 
			4.0\%|7j_{\frac{15}{2}}\frac{9}{2}\rangle + 
			3.0\% |5h_{\frac{9}{2}}\frac{9}{2}\rangle + 
			0.7\% |7j_{\frac{13}{2}}\frac{9}{2}\rangle+
			0.5\% |9h_{\frac{11}{2}}\frac{9}{2}\rangle + 
			0.2\% |7h_{\frac{11}{2}}\frac{9}{2}\rangle$\\
			&-7.39 & 94.5\% $ |404 \frac{7}{2}\rangle + 
			2.6\%|624 \frac{7}{2}\rangle + 
			1.9\% |413 \frac{7}{2}\rangle + 
			0.3\% |633 \frac{7}{2}\rangle +
			0.3\% |804\frac{7}{2}\rangle +
			0.2\% |604 \frac{7}{2}\rangle $\\
			&&$90.0\%|4g_{\frac{7}{2}}\frac{7}{2}\rangle + 
			3.7\% |6i_{\frac{11}{2}}\frac{7}{2}\rangle + 
			3.4\% |4g_{\frac{9}{2}}\frac{7}{2}\rangle + 
			1.9\% |6i_{\frac{13}{2}}\frac{7}{2}\rangle + 
			0.6\% |6g_{\frac{7}{2}}	\frac{7}{2}\rangle + 
			0.2\% |10g_{\frac{7}{2}}\frac{7}{2}\rangle$\\
			&-6.56 & 87.3\% $|402\frac{5}{2}\rangle +
			4.9\% |202 \frac{5}{2}\rangle+   
			4.6\% |622\frac{5}{2}\rangle+
			0.9\% |413\frac{5}{2}\rangle + 
			0.8\%|602\frac{5}{2}\rangle+ 
			0.7\% |802\frac{5}{2}\rangle$\\
			&&$79.5\%|4d_{\frac{5}{2}}\frac{5}{2}\rangle + 
			5.0\% |4g_{\frac{7}{2}}	\frac{5}{2}\rangle +
			4.7\% |4g_{\frac{9}{2}}\frac{5}{2}\rangle +
			3.8\%|6i_{\frac{13}{2}}\frac{5}{2}\rangle + 
			2.8\% |2d_{\frac{5}{2}}\frac{5}{2}\rangle +
			1.0\%|6i_{\frac{11}{2}}\frac{5}{2}\rangle$ \\
			\bottomrule[0.5pt]
			Neutron&-6.88& $84.6\% |624\frac{9}{2}\rangle + 
			6.6\% |615 \frac{9}{2}	\rangle + 
			3.2 \% |844\frac{9}{2}\rangle + 
			2.2\% |824\frac{9}{2}\rangle + 
			2.2\% |404\frac{9}{2}\rangle+
			0.9\% |604\frac{9}{2}\rangle$\\
			& &82.0\%$|6i_{\frac{13}{2}}\frac{9}{2}\rangle + 
			4.4\%|8i_{\frac{13}{2}}\frac{9}{2}\rangle + 
			3.9\% |4g_{\frac{9}{2}}\frac{9}{2}\rangle +
			3.5\% |6g_{\frac{9}{2}}\frac{9}{2}\rangle + 
			3.5\%|8k_{\frac{17}{2}}\frac{9}{2}\rangle + 
			1.2\% |6i_{\frac{11}{2}}\frac{9}{2}\rangle $ \\
			&-6.14 & $65.2\%|510\frac{1}{2}\rangle + 
			10.7\% |521\frac{1}{2}	\rangle + 
			6.1\% |310\frac{1}{2}\rangle + 
			5.9\% |730\frac{1}{2}\rangle +
			4.9\% |710 \frac{1}{2}\rangle + 
			1.5\% |301 \frac{1}{2}\rangle $\\
			&&27.1\%$|5f_{\frac{5}{2}}\frac{1}{2}\rangle + 
			23.6\% |5p_{\frac{3}{2}}\frac{1}{2}\rangle + 
			13.7\% |5f_{\frac{7}{2}}\frac{1}{2}\rangle + 
			12.5\% |5h_{\frac{9}{2}}\frac{1}{2}\rangle + 
			5.7\% |7p_{\frac{3}{2}}	\frac{1}{2}\rangle + 
			3.5\% |3p_{\frac{3}{2}}\frac{1}{2}\rangle$\\
			&-5.91 & 71.6\%$|512\frac{3}{2}\rangle + 
			8.4\% |521\frac{3}{2}\rangle+ 
			5.3\% |732\frac{3}{2}\rangle + 
			4.3\% |312\frac{3}{2}\rangle + 
			3.0\%|712\frac{3}{2}\rangle+ 
			1.6\% |501 \frac{3}{2}\rangle$ \\
			&&46.9\%$|5f_{\frac{5}{2}}\frac{3}{2}\rangle +
			17.0\%|5h_{\frac{9}{2}}	\frac{3}{2}\rangle + 
			9.0\% |5f_{\frac{7}{2}}\frac{3}{2}\rangle + 
			8.0\% |5p_{\frac{3}{2}}\frac{3}{2}\rangle+
			6.1\% |7f_{\frac{5}{2}}\frac{3}{2}\rangle + 
			2.2\% |3p_{\frac{3}{2}}\frac{3}{2}\rangle$\\
			&-5.27 & 93.2\% $ |615 \frac{11}{2}\rangle + 
			3.0\%|835\frac{11}{2}\rangle + 
			2.1\% |815 \frac{11}{2}\rangle + 
			1.5\% |606 \frac{11}{2}\rangle + 
			0.4\% |1055\frac{11}{2}\rangle + 
			0.4\% |806 \frac{11}{2}\rangle $\\
			&&$90.7\%|6i_{\frac{13}{2}}\frac{11}{2}\rangle + 
			3.6\% |8k_{\frac{17}{2}}\frac{11}{2}\rangle + 
			2.4\% |8i_{\frac{13}{2}}\frac{11}{2}\rangle + 
			2.1\% |6i_{\frac{11}{2}}\frac{11}{2}\rangle + 
			0.6\% |8k_{\frac{15}{2}}\frac{11}{2}\rangle + 
			0.1\% |12i_{\frac{13}{2}}\frac{11}{2}\rangle$\\
			&-4.99 & 80.4\% $|503\frac{7}{2}\rangle + 
			6.4\% |703\frac{7}{2}\rangle+ 
			5.9\% |303\frac{7}{2}\rangle +
			4.2\% |723\frac{7}{2}\rangle + 
			2.8\%|514\frac{7}{2}\rangle+ 
			1.6\% |923\frac{7}{2}\rangle$\\
			&&$74.9\%|5f_{\frac{7}{2}}\frac{7}{2}\rangle + 
			7.9\% |5h_{\frac{9}{2}}\frac{7}{2}\rangle +
			5.9\% |7f_{\frac{7}{2}}\frac{7}{2}\rangle +
			4.1\%|5h_{\frac{11}{2}}\rangle + 
			3.1\% |3f_{\frac{7}{2}}\frac{7}{2}\rangle +
			1.7\%|7j_{\frac{15}{2}}\frac{7}{2}\rangle$ \\
\bottomrule[1.0pt]
                                                                   \label{tab3}
\end{tabular}
\end{table*}	

For understanding the impact of the hexadecapole deformation on single-particle spectrum which corresponds to the eigenstates of the one-body WS Hamiltonian, Figure~\ref{Fig05} shows the single-particle diagrams for protons and neutrons in functions of the quadrupole deformation $\beta_2$ and hexadecapole deformation $\beta_4$ for $^{184}_{72}$Hf$_{112}$. One can notice that the hexadecapole deformation $\beta_4$ plays a critical role during the shell evolution. For instance, the shell gap is not prominent at $\beta_2\approx0.237$ (corresponding to the equilibrium deformation, e.g., see Table~\ref{tab1}) near the Fermi surface for both protons and neutrons but distinctly appears at $\beta_4\approx-0.09$, agreeing with the calculated $\beta_4$ deformation in Table \ref{tab1}). The appearances of the shell gaps, as see in Fig.~\ref{Fig05}, can be usually traced back to the strong coupling (via hexadecapole operator $\hat{Q}_{4\mu} \propto r^4Y_{4\mu=0,1,2,3,4}$ entering the hexadecapole-hexadecapole residual interaction Hamiltonian) between the $bra$ and the $ket$ states originating from the orbitals of $\Delta l= \lambda = 4$, e.g., see Table~\ref{tab3}, reflected in the structure of the corresponding wave functions. Note that in the spherical case, e.g., at $\beta_2=0.0$ in Figs. \ref{Fig05}(a) and \ref{Fig05}(c), the single-particle states are labeled by the principal quantum number $\mathit{n}$, the orbital angular momentum number $\mathit{l}$ and the total angular momentum number $\mathit{j}$. Similar to the standard notations in atomic spectroscopy, the $\mathit{s}$, $\mathit{p}$, $\mathit{d}$, $\mathit{f}$, $\cdots$ respectively correspond to the orbital angular momentum quantum numbers $\mathit{l}$ = 0, 1, 2, 3, $\cdots$. Owing to the strong spin-orbit coupling, the single-particle energy level labeled by $nl$, e.g., $2d$ in Fig.~\ref{Fig05}(a), will split into two partners $\mathit{j}$ = $\mathit{l}$ $\pm  1/2$, e.g., $2d_{3/2}$ and $2d_{5/2}$, and each $\mathit{j}$ level contains $2\mathit{j} + 1$  degenerate states. Of course, due to the large energy splitting for the high-$\mathit{j}$ orbitals, e.g., the $1h_{11/2}$ and $1i_{13/2}$, they often appears as the intruder states in the neighboring lower-$\mathit{N}$ shell. From Fig.~\ref{Fig05}, it can be seen that not only the spherical magic numbers 82 and 126 but also the expected single-particle properties at deformed case can be well reproduced. 

According to whether the hexadecapole deformation $\beta_4$ is included, namely, at two typical deformation points ($\beta_{2}$=0.237, $\gamma$=$0^\circ$, $\beta_{4}$=0.000) and ($\beta_{2}$=0.237, $\gamma$=$0^\circ$, $\beta_{4}$= -0.087), Tables~\ref{tab2} and \ref{tab3} respectively illustrate the first six wave-function components of proton and neutron single-particle levels near the Fermi surfaces for $^{184}_{72}$Hf$_{112}$. Through influencing the shell gaps and pairing interactions, the single-particle energy levels can determine nuclear internal structure as well as microscopic shell and pairing correction energies\cite{Goriely1996}. As it is well known, the spherical HO bases $|Nlj\Omega  \rangle$ and deformed HO bases (cf. Section~\ref{method}), especially characterized by the asymptotic Nilsson quantum numbers  $\Omega$[N$n_z\Lambda$], are usually used to understand, e.g., multipole interactions related to the deformation mechanism and to block the high-K configurations. In general, the single-particle levels are labeled by the maximum component of these wave functins. In these two tables, we show these two kinds of wave functions. Certainly, the spherical HO wave functions are indirectly obtained from the Moshinsky transformation mentioned above rather than the direct Hamiltonian diagonalization. It might be helpful to point out that the so-called wave-function label will be somewhat meaningless when the mixing of the corresponding wave function is too serious (e.g., the spherical label $|5p_{\frac{3}{2}}\frac{1}{2}\rangle$ for the neutron -5.51-MeV level in Table~\ref{tab2}). Due to the axial symmetry (triaxial $\gamma$ deformation is zero), the quantum number $\Omega$ (the angular momentum projection on the symmetry axis) is conserved in the two adopted bases. We can see that the largest components of the WS states are more prominent when adopting the deformed-basis expansion than that using the spherical ones since the similarity between the deformed HO and WS potentials is larger. Moreover, the high-$j$ high-$\Omega$ single-particle state is relatively pure due to the difficulties of the mixings for such high-$\Omega$ states with large energy difference. We note that similar to the case of the octupole deformation which leads to the mixing of the states with $\Delta l = \Delta j =3$, the hexadecapole deformation is expected to mix the states with $\Delta l = \Delta j =4$. Indeed, it seems that such phenomena occur though we cannot exclude the quadrupole-quadrupole coupling. For instance, by a comparison of the wave-function components of the proton single-particle states, we can observe that the coexisting partners $4.7\%|5f_{\frac{7}{2}}\frac{7}{2}\rangle \leftrightarrow 3.6\% |7j_{\frac{15}{2}}\frac{7}{2}\rangle$ in Table~\ref{tab3} is more prominent than the partners $3.1\% |7j_{\frac{15}{2}}\frac{7}{2}\rangle\leftrightarrow 2.7\% |5f_{\frac{7}{2}}\frac{7}{2}\rangle$ in Table~\ref{tab2} for the $|5h_{\frac{11}{2}}\frac{7}{2}\rangle$ state (note that the state is labelled by the largest component) and for the $|4d_{\frac{5}{2}}\frac{5}{2}\rangle$ state, the partner $|6i_{\frac{13}{2}}\frac{5}{2}\rangle$, apparently coexisted for the partners  $79.5\%|4d_{\frac{5}{2}}\frac{5}{2}\rangle \leftrightarrow 3.8\% |6i_{\frac{13}{2}}\frac{5}{2}\rangle$ at non-zero $\beta_4$ (see Table~\ref{tab3}), does not appear in the wave function, at least, in the first six components at $\beta_4 = 0.0$, indicating the enhanced mixing of the states with $\Delta l = \Delta j =4$ due to the hexadecapole-hexadecapole interactions.         

\begin{figure}
\centering
\includegraphics[width=0.53\textwidth]{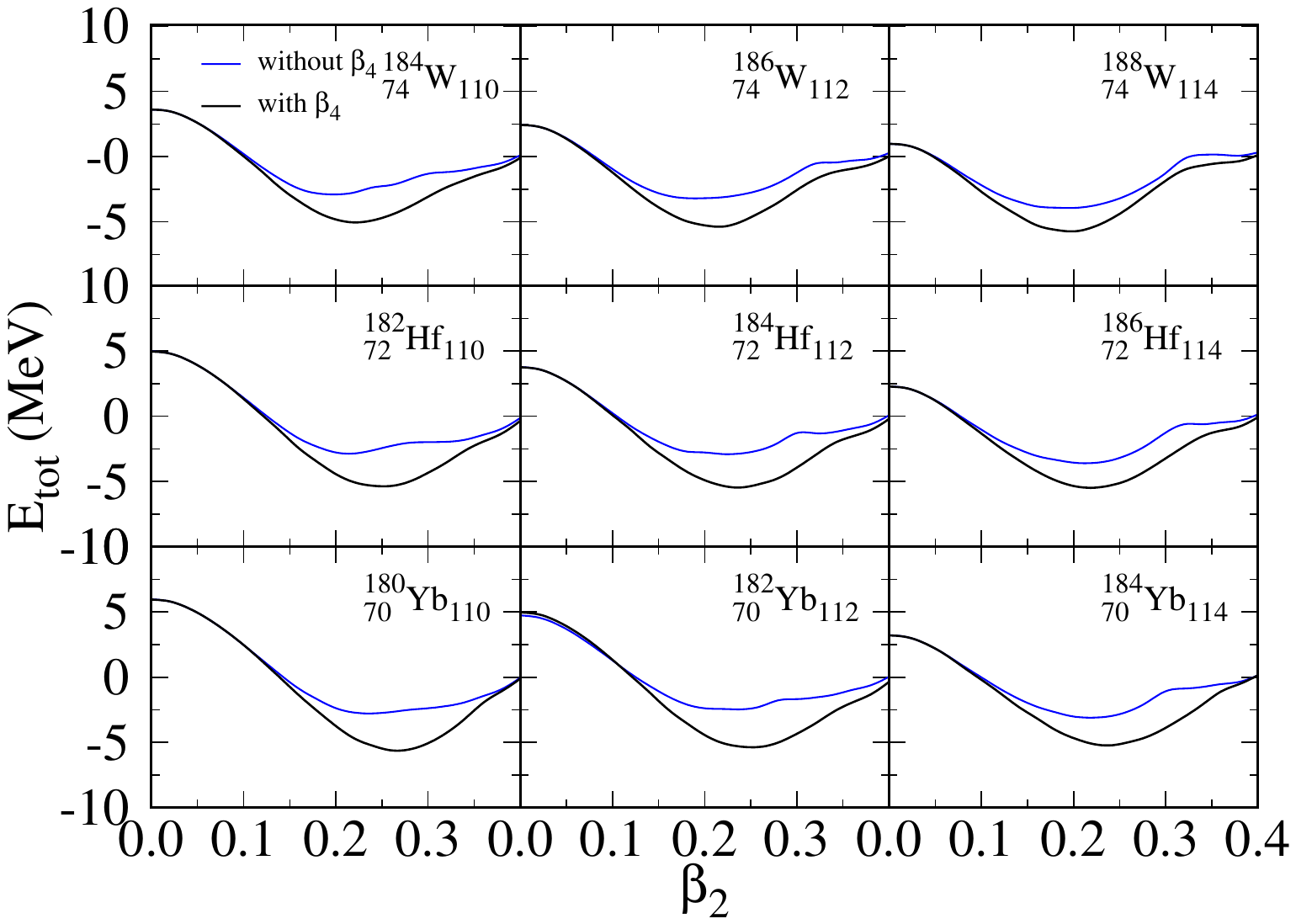}
\caption{Calculated total energy curves as function of the quadrupole deformation $\beta_2$ for nine selected even-even nuclei $^{180-184}$Yb, $^{182-186}$Hf and $^{184-188}$W. Note that, at each deformation point $\beta_2$, the energy is minimized over the triaxial deformation $\gamma$ and the hexadecapole deformation $\beta_4$ if $\beta_4$ is included (e.g., the green line). For more details, see the text.}
	                                                           \label{Fig06}
\end{figure}

Based on the MM method of Strutinsky~\cite{Strutinsky1975}, Figure~\ref{Fig06} exhibits two types of calculated potential-energy curves, focusing on the region around the equilibrium shapes, to denote the effects of $\beta_4$ on the total potential energy (binding energy) of the selected nuclei $^{180,182,184}$Yb, $^{182,184,186}$Hf, and $^{184,186,188}$W. From this figure, we can evaluate the nuclear stiffness and binding properties near the equilibrium deformations at the selected space ($\beta_2$, $\gamma$, $\beta_4$). It can be seen that the inclusion of the hexadecapole deformation degree of freedom can enhance the quadrupole deformations of the minima, describing the experimental data better, and lead to the energy reduction (about $1-3$ MeV) , helping to reproduce the nuclei masss in experiment (see, e.g., Refs.~\cite{Wei2024,Song2023}). Taking the hexadecapole deformation into account, we here note that all the nuclei will become stiffer near their equilibrium shapes. Far from these minima, one can see from Fig.~\ref{Fig06} that the green and blue curves overlap, indicating that the influence of the hexadecapole deformation $\beta_4$ cannot be neglected at that moment.      

\begin{figure}[htpb]
\includegraphics[width=0.234\textwidth]{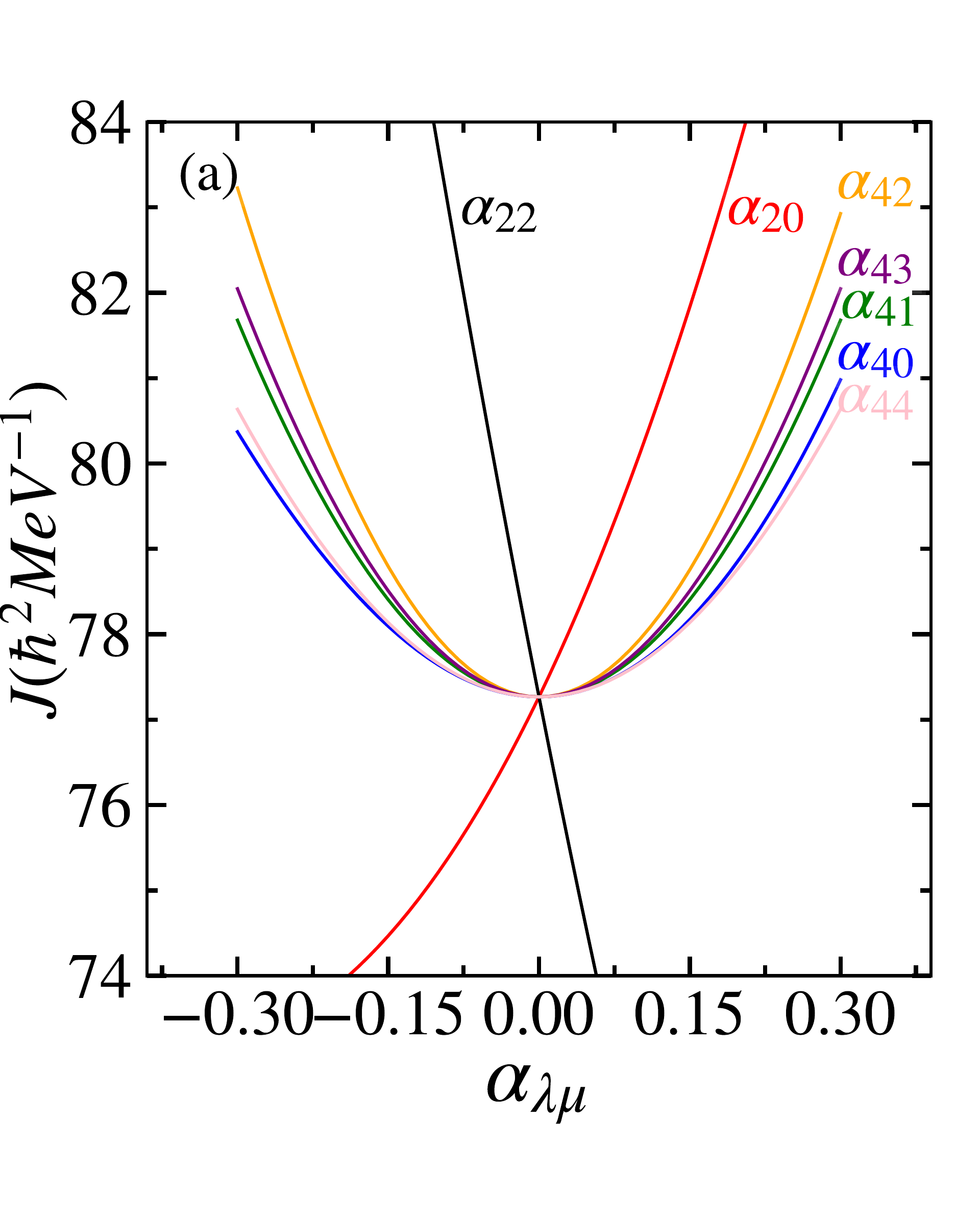} 
\includegraphics[width=0.234\textwidth]{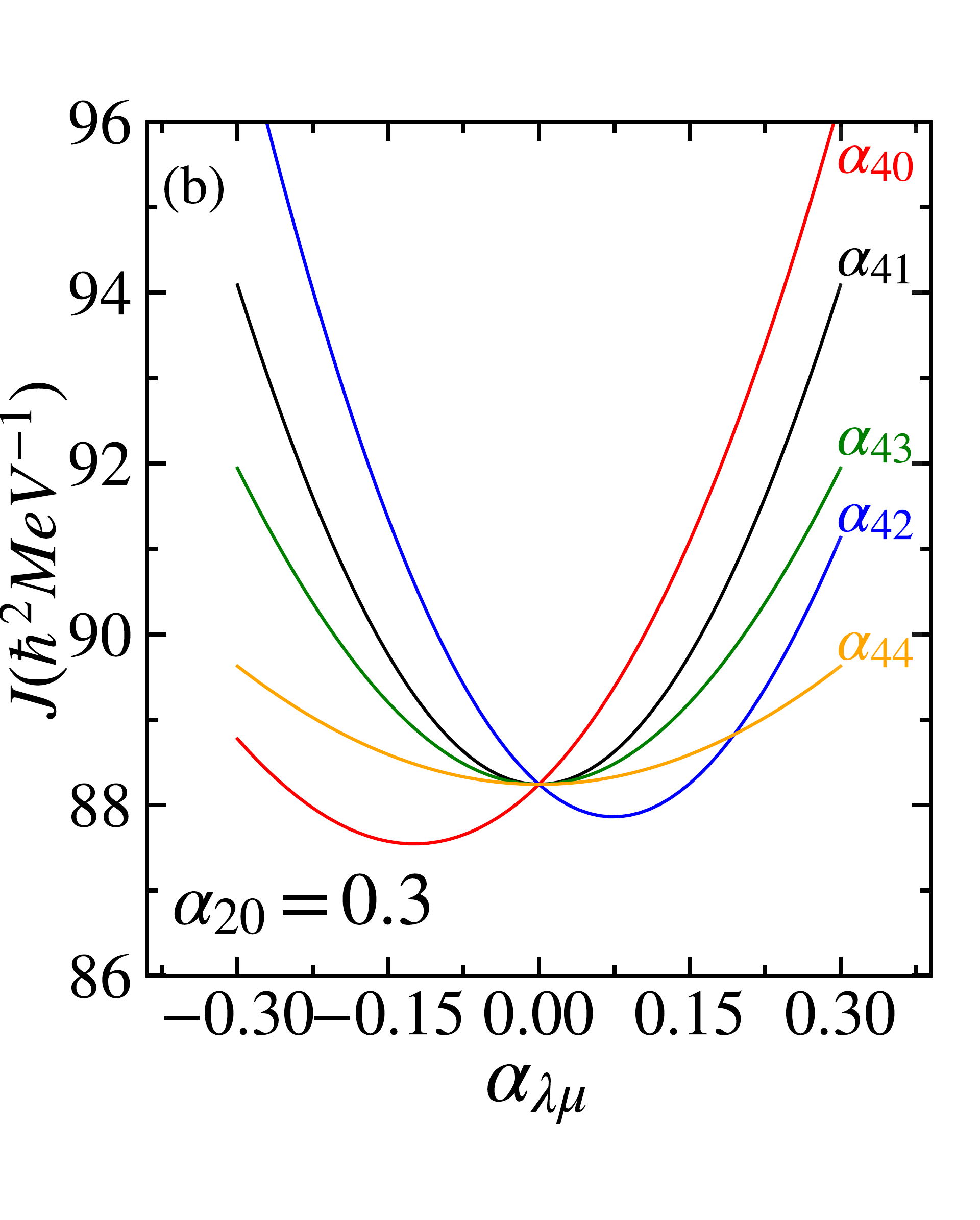}
\vspace{-0.51cm}
\caption{Calculated rigid-body MoIs around the $x$ axis as functions of the deformations $\alpha$$_{2\mu=0, 2}$ and $\alpha$$_{4\mu=0, 1, 2, 3, 4}$ for the central nucleus $^{178}$Hf. To see the coupling effect, the deformation $\alpha_{20}$​ is set to 0.3 in (b).}
		                                                           \label{Fig07}
\end{figure}

The impact of different deformations on the nuclear MoIs is one of our primary concerns in the present project. To date, it is known that a large number of ground-state nuclear electric quadrupole moments had been measured in atomic nuclei~\cite{Stone2005}. These nuclei are deformed and display a low-energy rotational level structure. The corresponding cranking approximations are adopted to explain the rotational spectra, generally by a comparison between theoretical and experimental MoIs. Before starting the microscopic discussion, let us calculate the MOI for a rigid body having the same volume, shape, and density of the nucleus. We know that it has been usual to consider the axially symmetric or asymmetric equilibrium shape~\cite{Allmond2017}. Presently, we perform a generalized treatment for a rigid body which can, in principle, involve any deformation degrees of freedom. As seen in Fig.~\ref{Fig07}, the calculated rigid-body MoIs in functions of different deformation parameters are illustrated for the central nuclei $^{178}$Hf. It is expected that the microscopic HFBC calculations will give the similar trends, even if considering the pairing effects. From Fig.~\ref{Fig07}, one can see that the MoI will rapidly increase (decrease) with increasing quadrupole $\alpha_{20}$ ($\alpha_{22}$) deformation parameter. Near the spherical shape, the hexadecapole deformations have a relatively small impact on the MoIs. At the large $\alpha_{4\mu}$ (e.g., see $\alpha_{4\mu} = \pm 0.3$), $\alpha_{42}$ has the largest influence on the nuclear MoI and the deformation parameters $\alpha_{44}$ ($\alpha_{40}$) respectively leads the smallest change in the MoI at positive and negative 0.3 positions. Nevertheless, at the elongated shape, e.g., cf. Fig.~\ref{Fig07}(b), the effects of different $\alpha_{4\mu}$ parameters on MoI obviously increase. From $-0.1$ to $+0.1$, the MoI will increase (decrease) with the increasing $\alpha_{40}$ ($\alpha_{42}$). For convenience, let us introduce the sensitivity coeffencient of MoI to deformation $\alpha_{\lambda\mu}$, e.g., defined by $S_{\lambda\mu} \equiv |\partial J^{(1)}/ \partial \alpha_{\lambda\mu}|$, to measure to what extent the $\alpha_{\lambda\mu}$ deformation will affect the MoI (e.g., in Fig.~\ref{Fig07}(a), the sensitivity coefficients satisfy $S_{42} > S_{43} > S_{41} > S_{40} > S_{44}$ at the side of $\alpha_{4\mu} >0$).

\begin{figure}[htbp]
\includegraphics[width=0.23\textwidth]{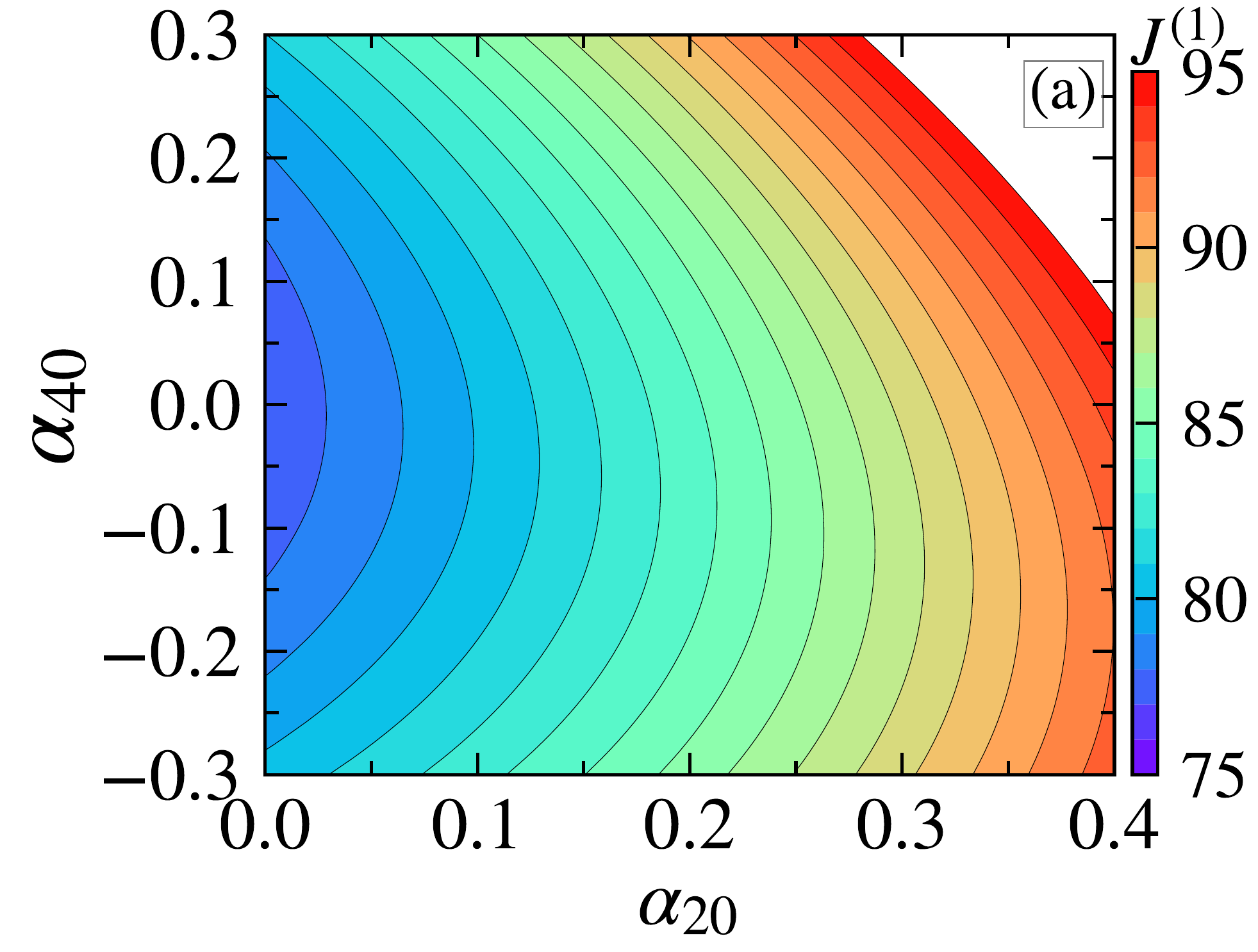}
\includegraphics[width=0.23\textwidth]{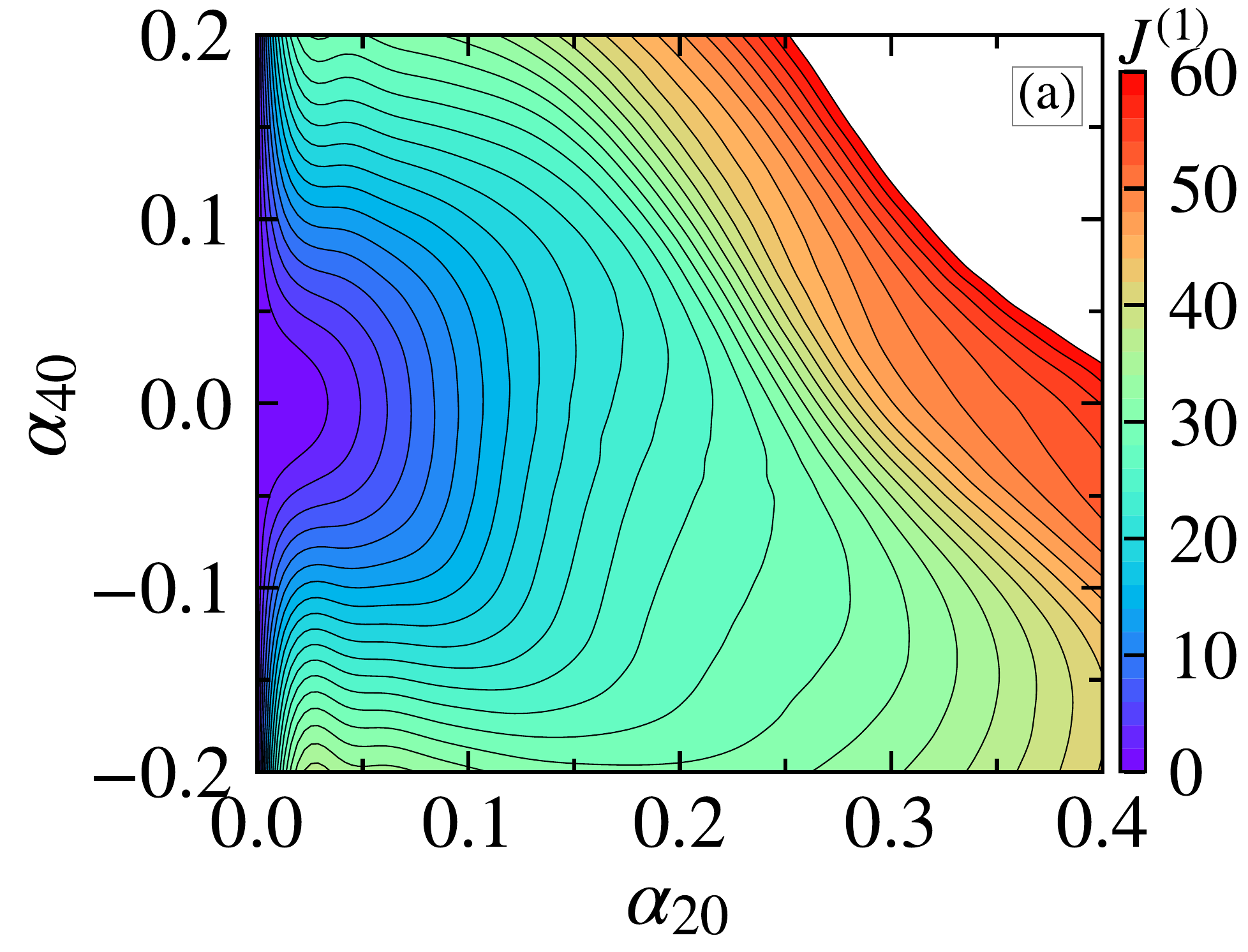}\\
\includegraphics[width=0.23\textwidth]{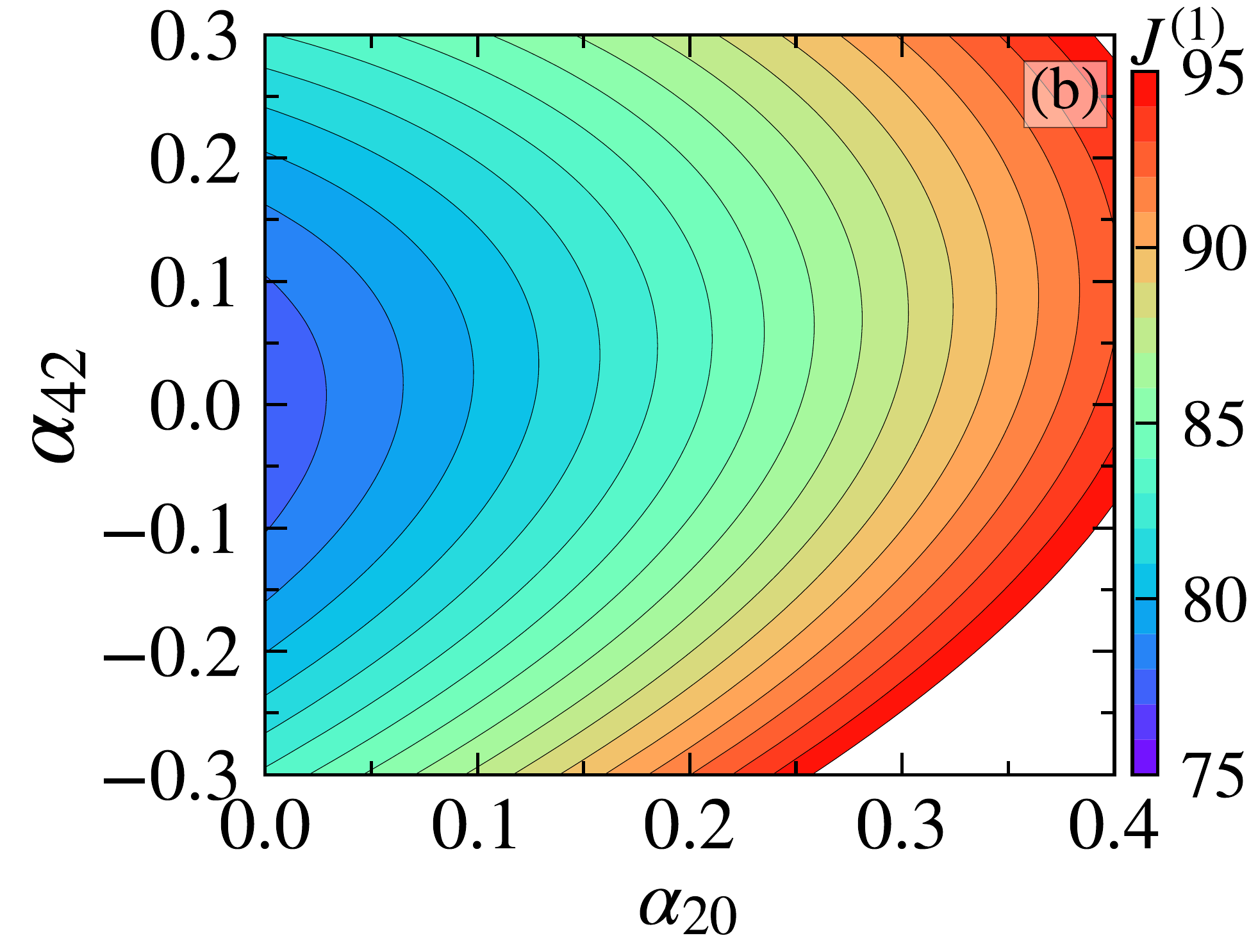}
\includegraphics[width=0.23\textwidth]{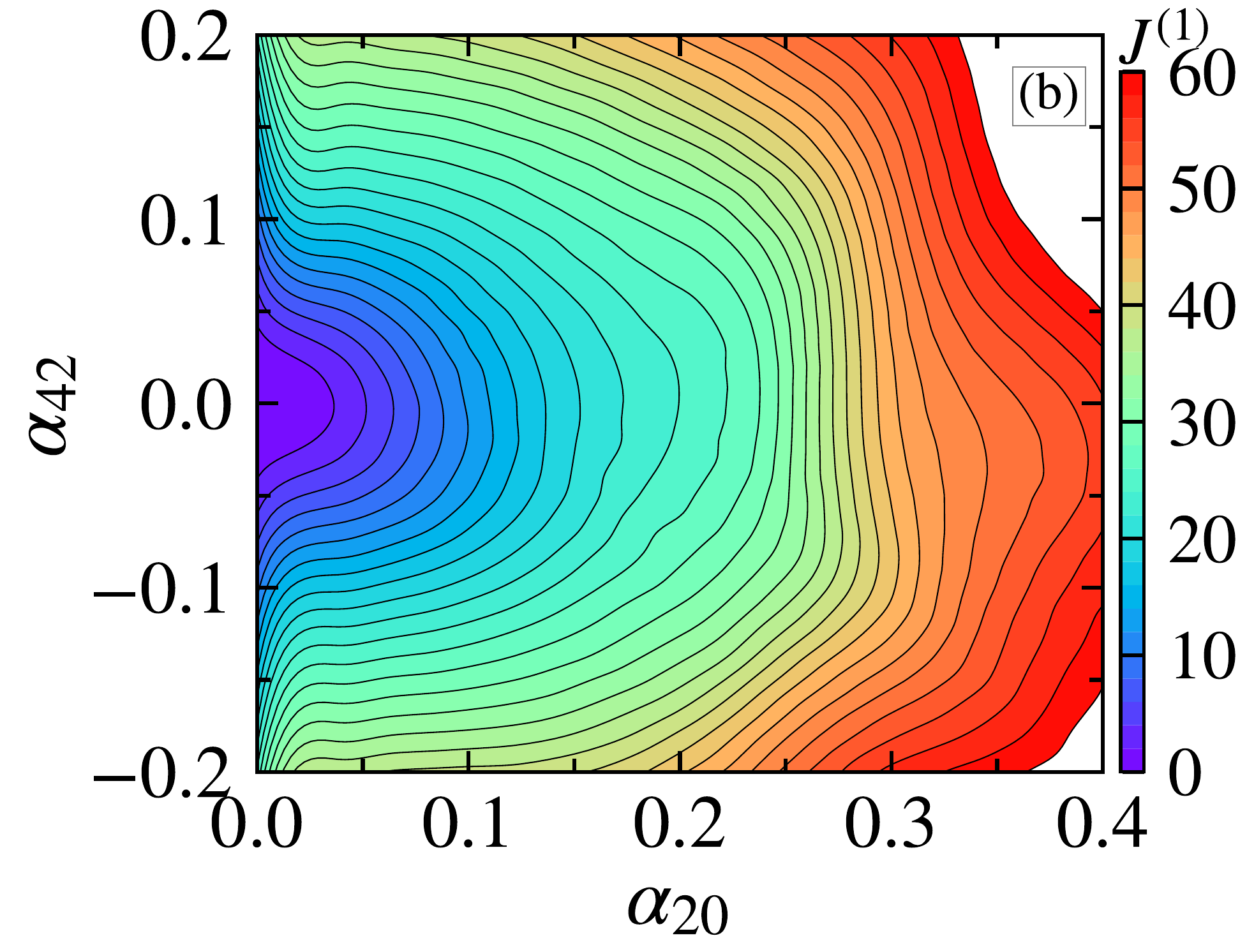}\\
\includegraphics[width=0.23\textwidth]{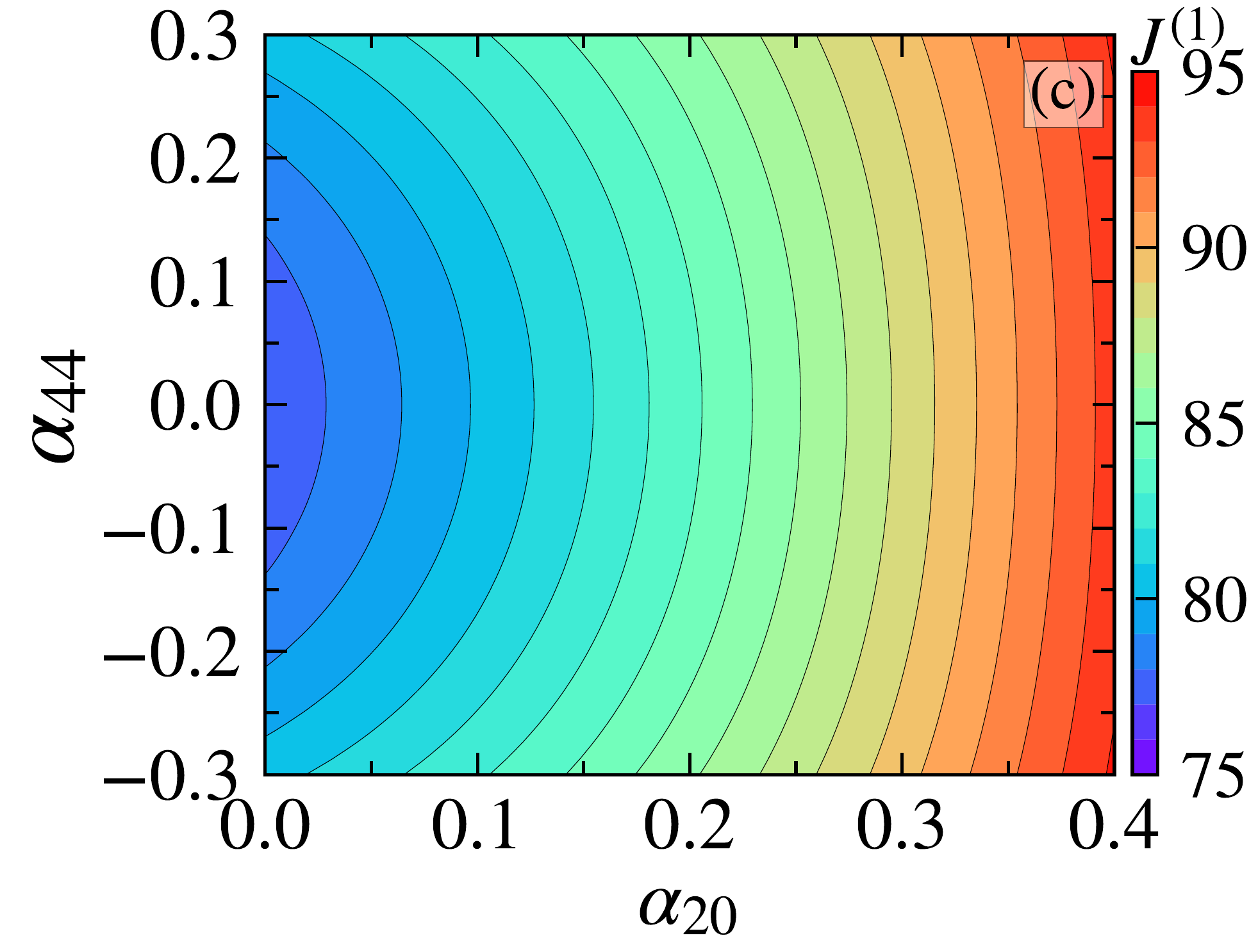}
\includegraphics[width=0.23\textwidth]{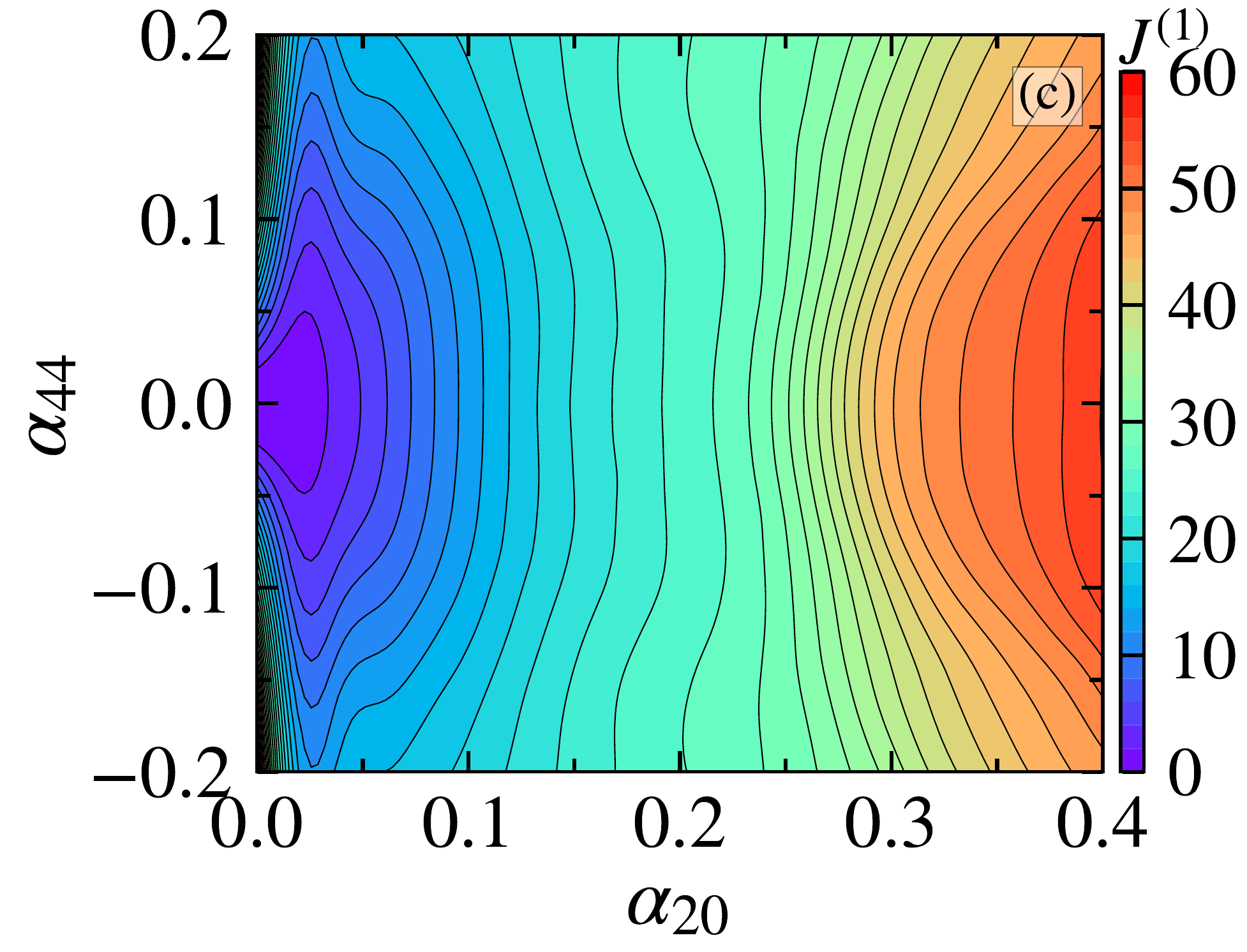}
\caption{Projections of calculated MoIs on  the ($\beta_2,\alpha_{40}$) (top), ($\beta_2,\alpha_{42}$) (middle) and ($\beta_2,\alpha_{44}$) (bottom) planes for the nucleus $^{178}$Hf. The maps in the left (a, b, c) and right (a$\prime$, b$\prime$, c$\prime$) sides are respectively obtained by the rigid-body and HFBC calculations (at rotational frequency $\hbar\omega =0.1$ MeV). See text for more details. \\}
                                                                \label{Fig08}
\end{figure}
	
From Fig.~\ref{Fig07}, one can notice that the hexadecapole deformations may have different impacts on the MoIs when the quadrupole deformation is different.  To observe the coupling effect, e.g., between $\alpha_{20}$ and $\alpha_{4\mu=0,2,4}$ degrees of freedom, Figure~\ref{Fig08} illustrates the calculated rigid-body MoIs in the ($\alpha_{20}$, $\alpha_{4\mu=0,2,4}$) planes. Relative to the MoI curves in Fig.~\ref{Fig07}, one can easily evaluate the MoI at the different combinations of the quadrupole and hexadexapole deformations by such 2D MoI maps. For instance, at $\alpha_{20} \sim 0.3$, a slightly negative $\alpha_{40}$ or positive $\alpha_{42}$ deformation will lead to a MoI reduction, agreeing with that in Fig.~\ref{Fig07}. Similarly, the MoIs based on the HFBC calculations are also presented in Fig.~\ref{Fig08} (see the corresponding planes on the rightside). Comparing with the contour maps between rigid-body and HFBC calculations, it can be indeed concluded that the MoIs seem to have the similar evolution trends no matter whether the pairig effects are included or not. This provides us a way to evaluate the effect of e.g., the exotic deformation degrees of freedom on the nuclear MoI before performing the corresponding microscopic calculations which may be relatively difficult.

\begin{figure}[htbp]
\centering
\includegraphics[width=0.48\textwidth]{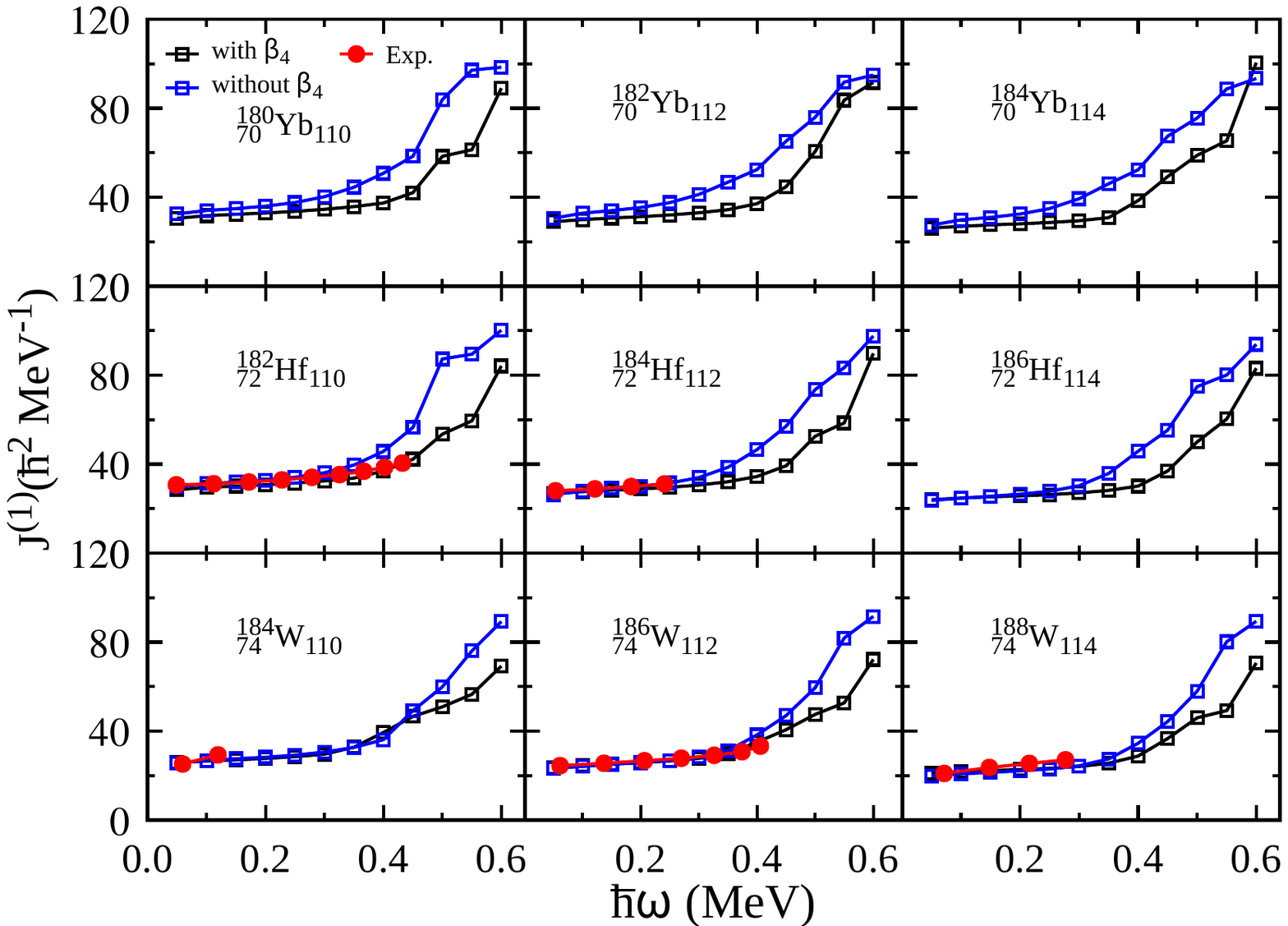}
\caption{Calculated MoIs, with (black square) and without (blue square) the inclusion of $\beta_4$ deformations, as functions of rotational frequency for nine selected even-even nuclei $^{180-184}$Yb, $^{182-186}$Hf and $^{184-188}$W, along with the experimental data (red circle). The experimental data are taken from \cite{nndc_ensdf}. See text for more explanations.}
		                                                         \label{Fig09}
\includegraphics[width=0.48\textwidth]{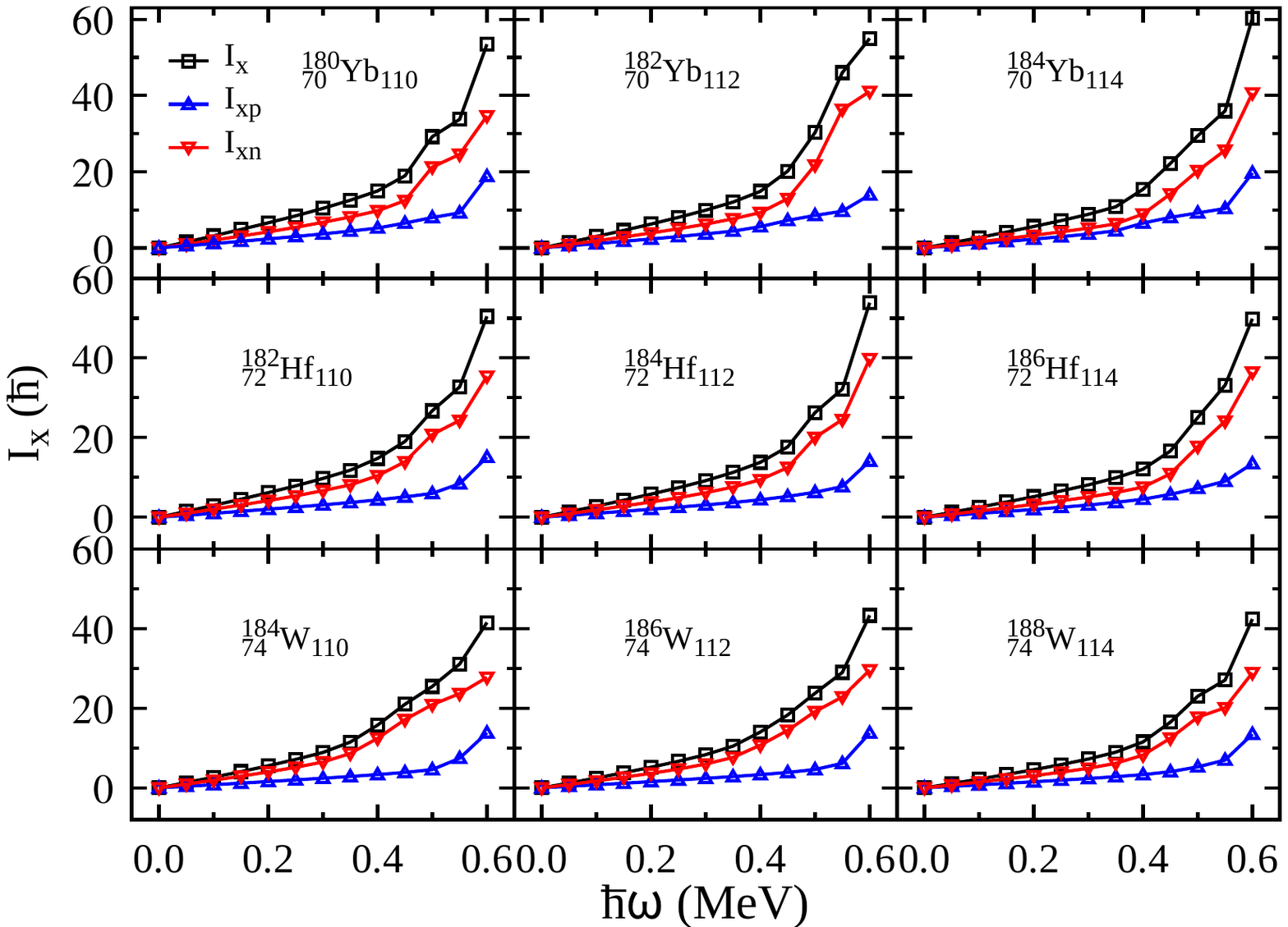}
\caption{Calculated aligned angular momenta $I_x$ as functions of rotational frequency $\hbar\omega$ for nine selected even-even nuclei $^{180-184}$Yb, $^{182-186}$Hf and $^{184-188}$W, together with the proton $I_{xp}$ and neutron $I_{xn}$ components.}
                                                                  \label{Fig10}
\end{figure}

For a nuclear state, the excitation energy and angular momentum are usually the observable attributes. In particular, the rotational bands can help to extract the MoI information in experiment. To investigate the rotational damping in the selected $^{180-184}$Yb, $^{182-186}$Hf and $^{184-188}$W nuclei, as seen in Fig.~\ref{Fig09}, we illustrate the calculated kinematic MoIs (by the HFBC method mentioned above) of the yrast bands in functions of rotational frequency, along with the experimental data for comparison. During the calculation proccess, we fixed the shapes of the ground-state values at two cases (with or without the inclusion of $\beta_4$, cf. e.g., Table~\ref{tab1}), ignoring the shape evolution. Such an approximation seems to be reasonable in these selected nuclei. One can see that the inclusion of the hexadecapole deformation $\beta_4$ does not affect most of the low-spin and low-lying states qualitatively, but can significantly improves the description of high-spin states of the ground-state bands. For instance, though the data are somewhat scarce, it can still be clearly noticed that the calculated MoIs reproduce the experimental trends, typically, in $^{182}$Hf and $^{186}$W. In other nuclei, there is no data or too few data points to observe this effect at present. The MoI decrease owing to the negative $\beta_4$ is in good agreement with our associated discussion of Figs.~\ref{Fig07} and \ref{Fig08}. In addition, the current conclusion also agrees with that recently obtained in Ref.~\cite{Lotina2024}.

From Fig.~\ref{Fig10}, it is found that the MoI undergoes a sudden increase during the process of the rotational frequency increase. Such a rapid change in MoI is generally caused by rotation alignments of a pair of nucleons (particularly those occupying high-$\mathit{j}$ low-$\Omega$ orbitals near the Fermi surface and usually accompanying shape transition, especially in the soft nucleus). To understand the mechanism of backbending or upbending, we show the aligned angular momenta by the HFBC calculations with fixed ground-state deformations in Fig.~\ref{Fig10}, including the proton and neutron contributions, for the selected nine nuclei. One can see that neutrons not only provide more angular alignments but also will be responsible for the MoI anomalies, which indicate that the rapid increase in the MoI is caused by neutron alignment, at least, at $\hbar \omega \le 0.5$ MeV. When $\hbar \omega > 0.5$ MeV, it seems that the second pair of nucleons (a pair of protons) begins to be broken and aligns along the rotational axis.

\begin{figure}[htbp]
\includegraphics[width=0.48\textwidth]{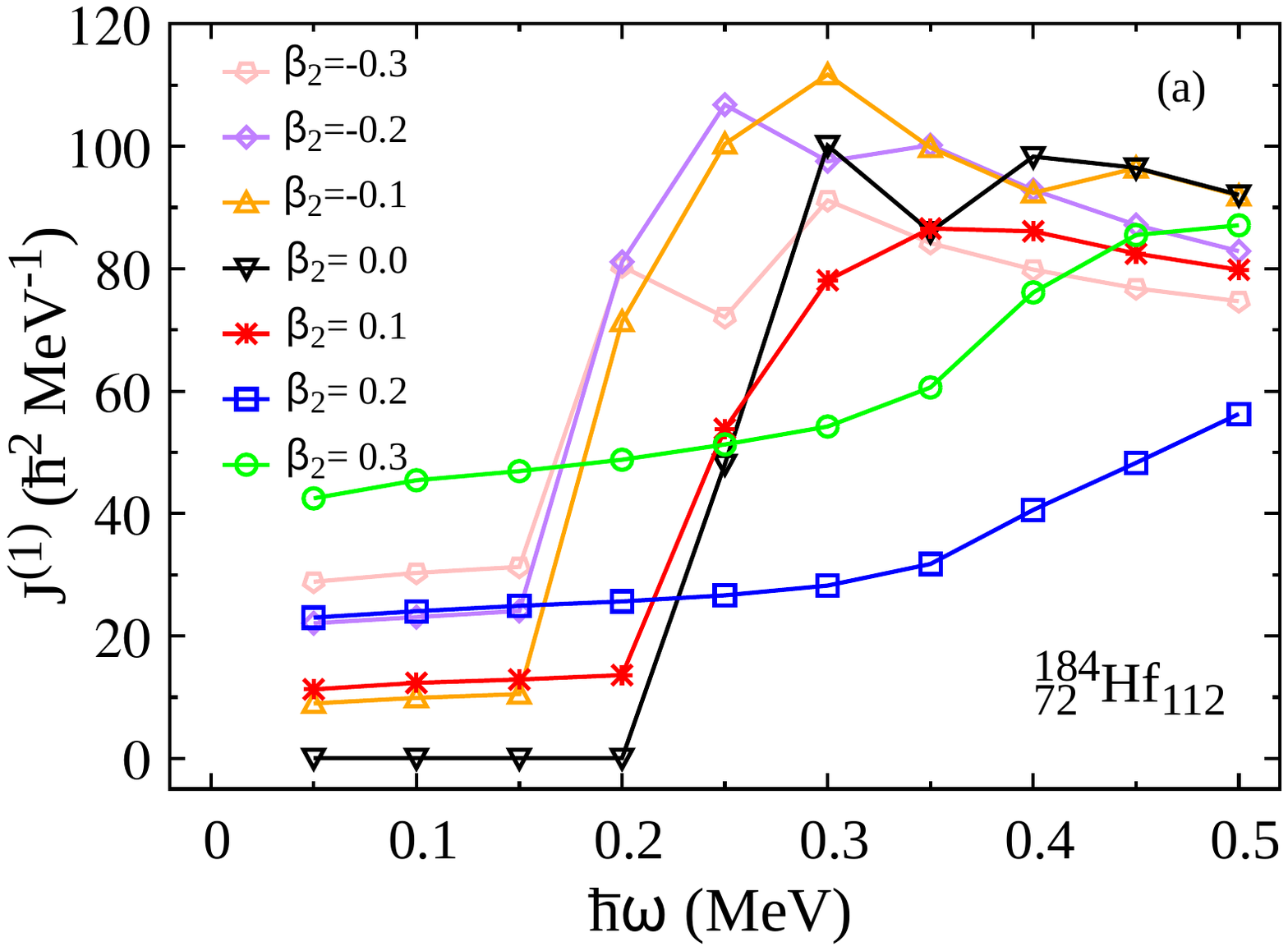}
\includegraphics[width=0.48\textwidth]{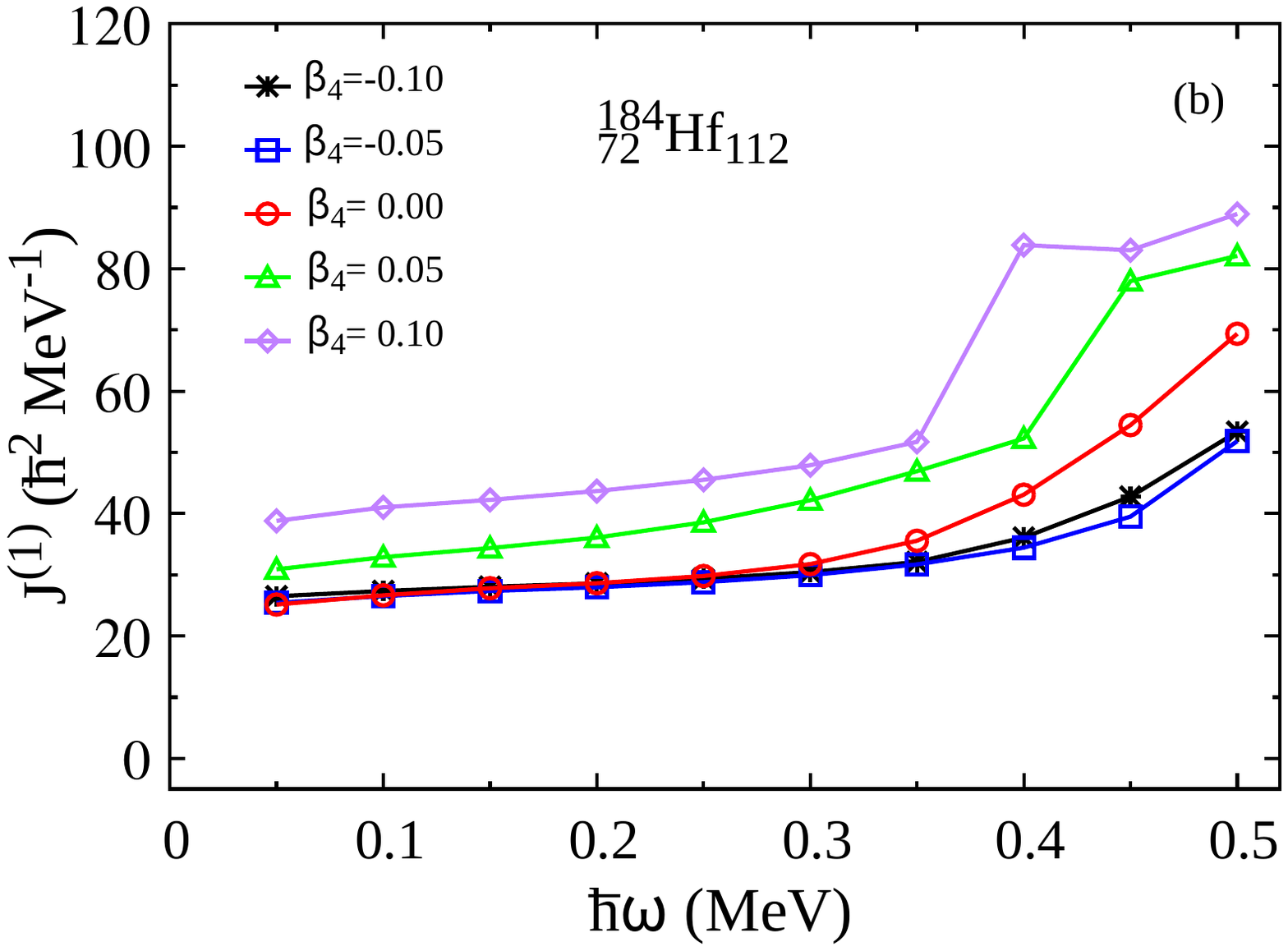}
\caption{Calculated kinematic MoIs in functions of $\hbar\omega$ at different $\beta_2$ (a) and $\beta_4$ (b) deformations for the central nucleus $^{184}$Hf. Note that, in subfigure (b), for each $\beta_4$, the $\beta_2$ is set to the equilibrium deformation 0.237, e.g., see Table~\ref{tab1}.}
                                                                  \label{Fig11}
\end{figure}

\begin{figure}[htbp]
\centering
\vspace{0cm}
\includegraphics[width=0.48\textwidth]{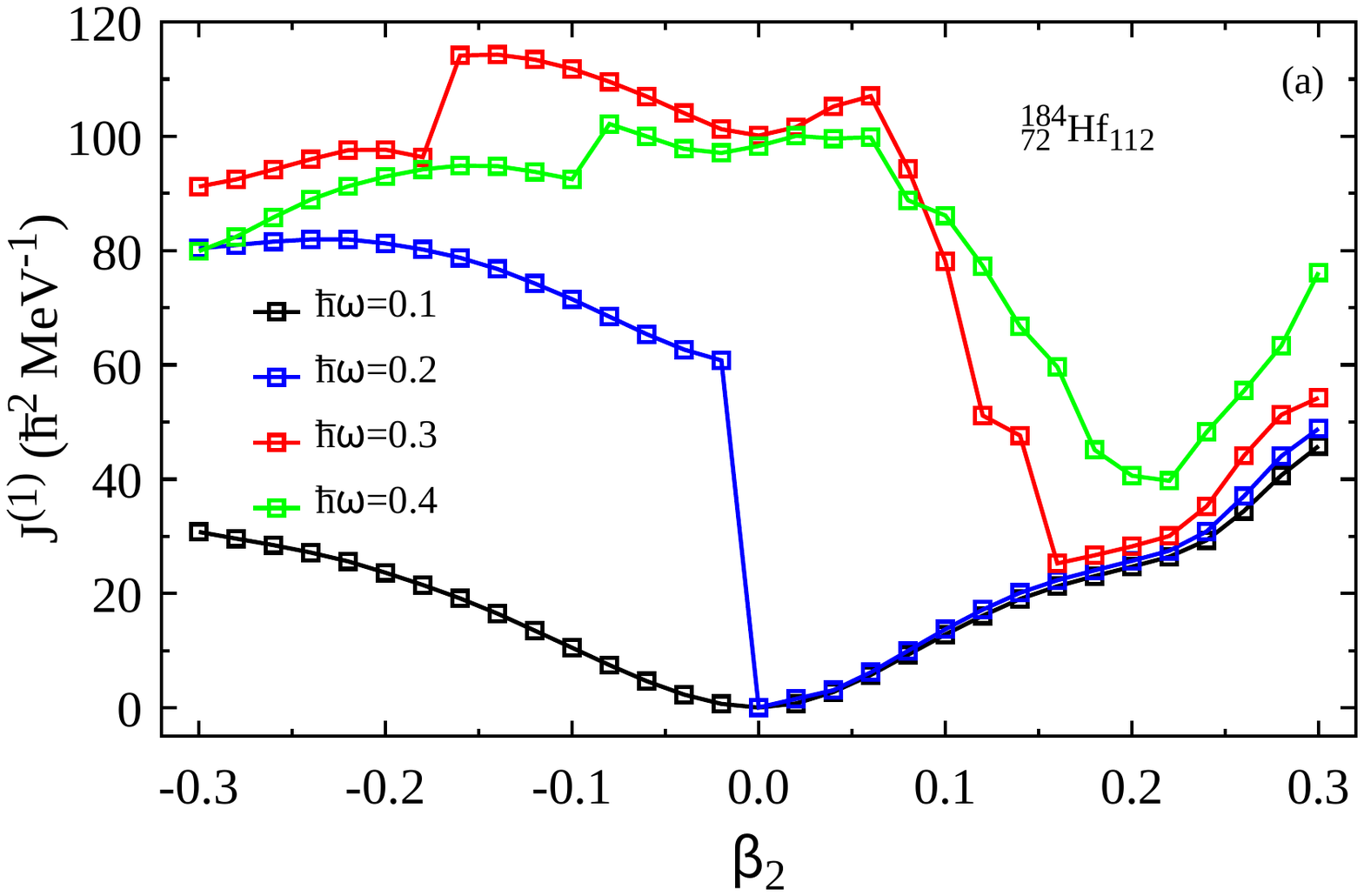}\\
\vspace{-0.7cm}
\includegraphics[width=0.48\textwidth]{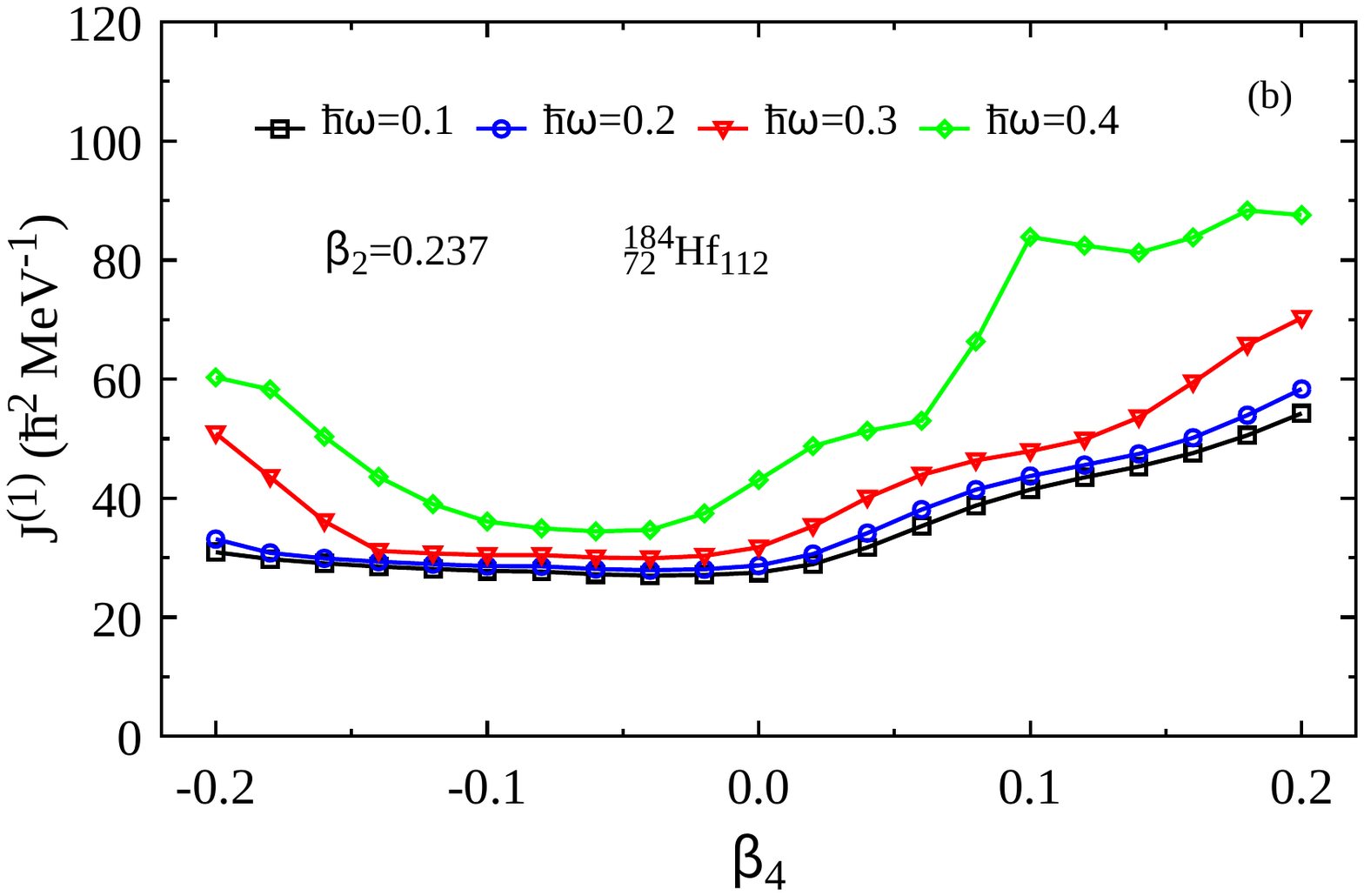}
\vspace{-0.9cm}
\caption{Calculated kinematic MoIs against $\beta_2$ (a) and $\beta_4$ 
(b) at different rotational frequencies $\hbar\omega$ for the central nucleus $^{184}$Hf. Similar to Fig.~\ref{Fig11}(b), the $\beta_2$ is fixed to 0.237 in (b).}
                                                                   \label{Fig12}
\end{figure}
How does the nuclear MoI vary with different deformation (rotational frequency) for fixed frequency $\omega$ (shape)? Taking the central nucleus $^{178}$Hf as an example, we intend to perform a detailed investigation and answer this question below. For simplicity and clarity, we merely allow one degree of freedom (deformation or rotational frequency) to change during the calculations. 
Figure~\ref{Fig11} presents the kinematic MoIs given by the HFBC calculations in functions of $\hbar\omega$ at different $\beta_2$ and $\beta_4$ deformations. In Fig.~\ref{Fig11}(a), we show the MoIs $J^{(1)}(\omega)$, fixing the $\beta_2$ at different values and setting other deformations to zero. One can see that, for the ground-state case (e.g., at $\hbar\omega < 0.2$ MeV) in the example nucleus $^{184}$Hf, the MoI will increase as the quadrupole deformation $\beta_2$ deviates from spherical shape (namely, $\beta_2 = 0.0$). For a spherical nucleus (a spherical quantum many-body system), as expected, the MoI is zero. The nucleus with the prolate shape has the larger MoI value than that of the oblate one at the same deformation amplitudes (e.g., the same $|\beta_2|$ values). The rapid increase of the MoI for a $\beta_2$-fixed shape orginates from the band crossing (e.g., see the following discussion). We can notice that the band-crossing frequency are somewhat different when fixing different deformation. Generally speaking, the oblate shape has the relatively smaller band-crossing frequency. Similarly, Figure~\ref{Fig11}(b) shows the MoIs in function of $\hbar\omega$ for the cases at different hexadecapole deformation $\beta_4$, always keeping the equilibrium $\beta_2$ deformation 0.237 unchanged. From this subplot, one can see that the negative $\beta_4$ deformation does not change the MoIs at low rotational frequency (e.g., at $\hbar\omega \le 0.25$ MeV) but decreases the MoIs at high rotational frequency, while the positive hexadecapole deformation always increases the MoIs, agreeing with our above results (e.g., see Figs.~\ref{Fig08}). 

Similar to Fig.~\ref{Fig11}, from the other perspectives, we present the MoI variations at different rotational frequencies in functions of $\beta_{2}$ and $\beta_{4}$ in Fig.~\ref{Fig12}. As can be seen from the figure, at low rotational frequency (e.g. at $\hbar \omega$ = 0.1 MeV), the MoI increases with the increase of $|\beta_{2}|$. For the cases of $\hbar \omega \ge 0.2$ MeV, at the oblate side, the band crossings will always occur, even, at very weak deformation. At the weak prolate deformation, e.g., $\beta_2 = 0.1$, the band crossings occur at higher rotational frequencies, e.g., $\hbar\omega = 0.3$ and 0.4 MeV. The conclusion is that, at low rotational frequency, the increasing $|\beta_2|$ and positive $\beta_4$ will lead to the increase of MoI. At the normal deformation case, e.g., $\beta_2 \sim 0.2-0.3$, the slightly negative $\beta_4$ will contribute to the decline of the MoIs, at least, in this region. Note that, by the present analysis, we just intend to provide insights into the effect of deformatons (especially the hexadecapole one) on the MoI and the practical equilibrium shape should still be determined from the minimum of the potential energy landscape of the nucleon system.

\begin{figure}
\includegraphics[width=0.23\textwidth]{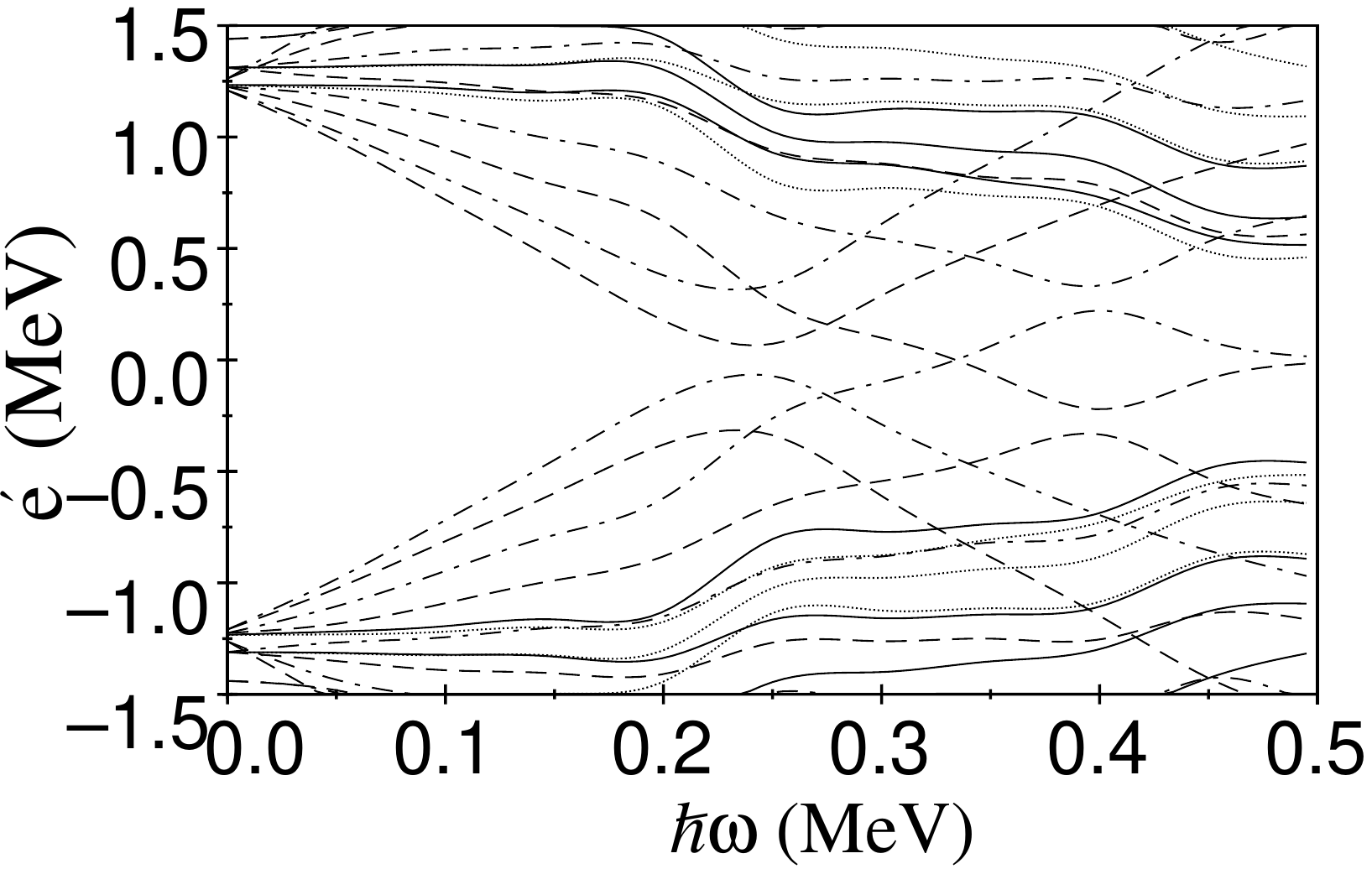}
\includegraphics[width=0.23\textwidth]{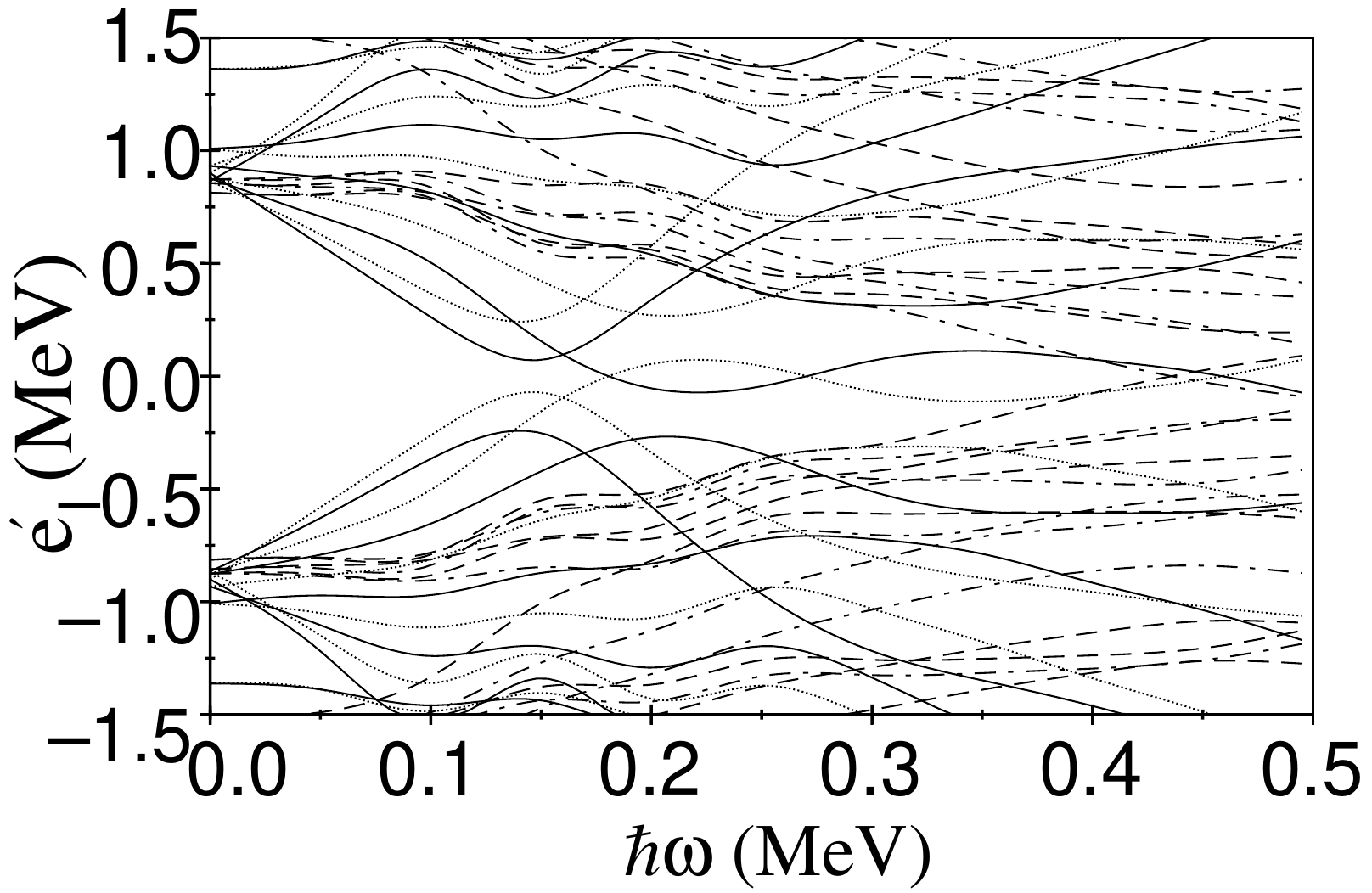} \\
\includegraphics[width=0.23\textwidth]{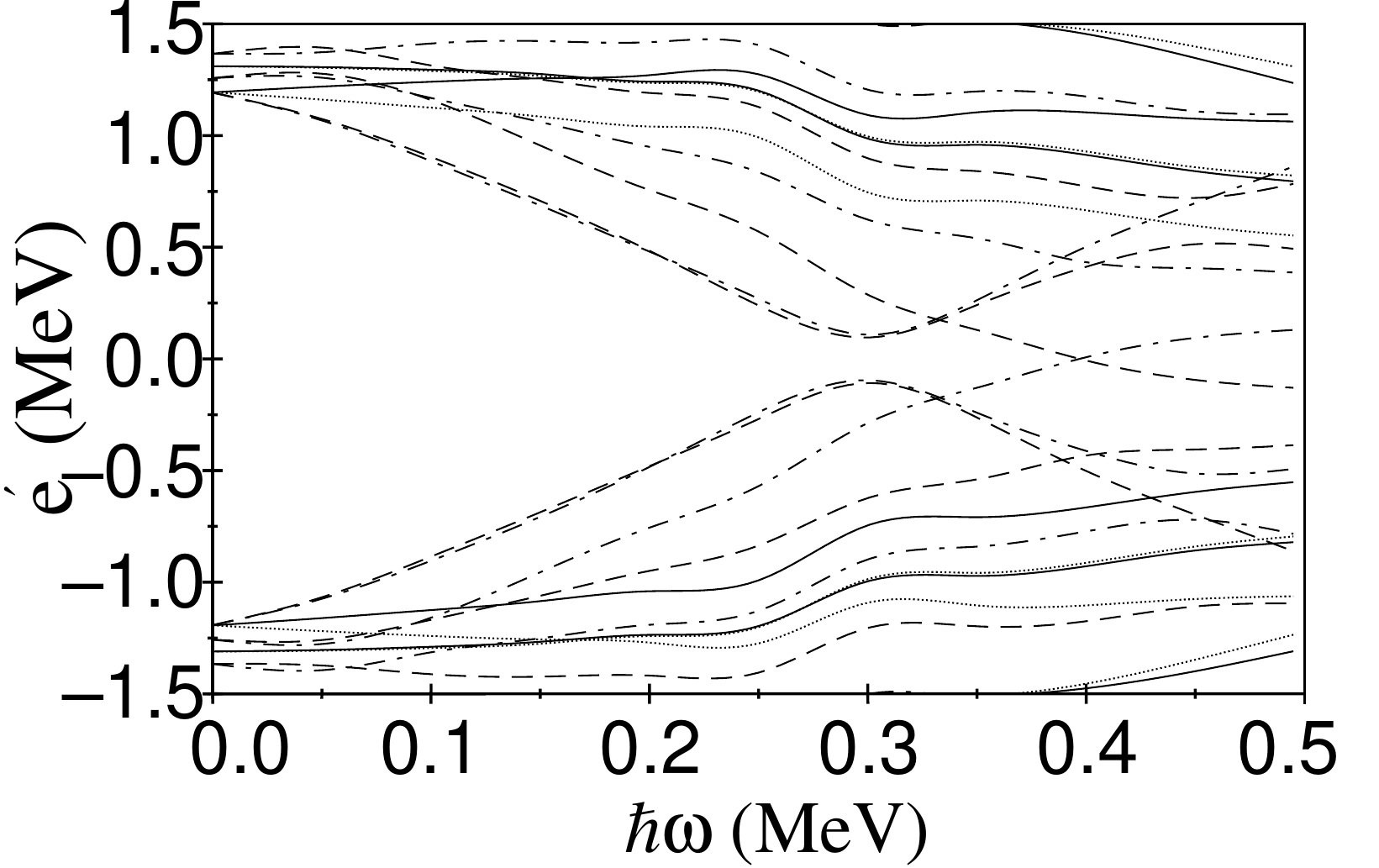}  
\includegraphics[width=0.235\textwidth]{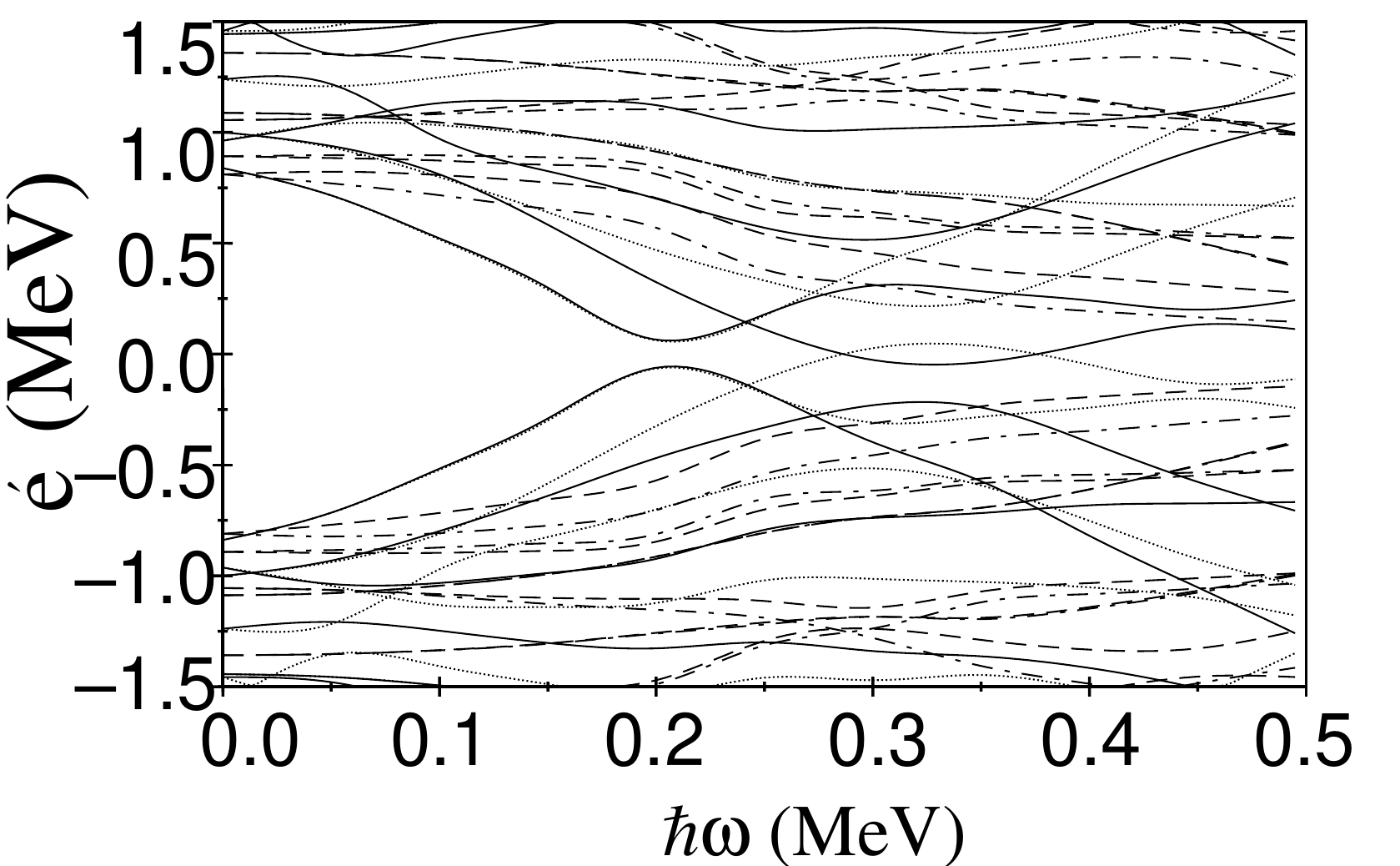}\\
\includegraphics[width=0.23\textwidth]{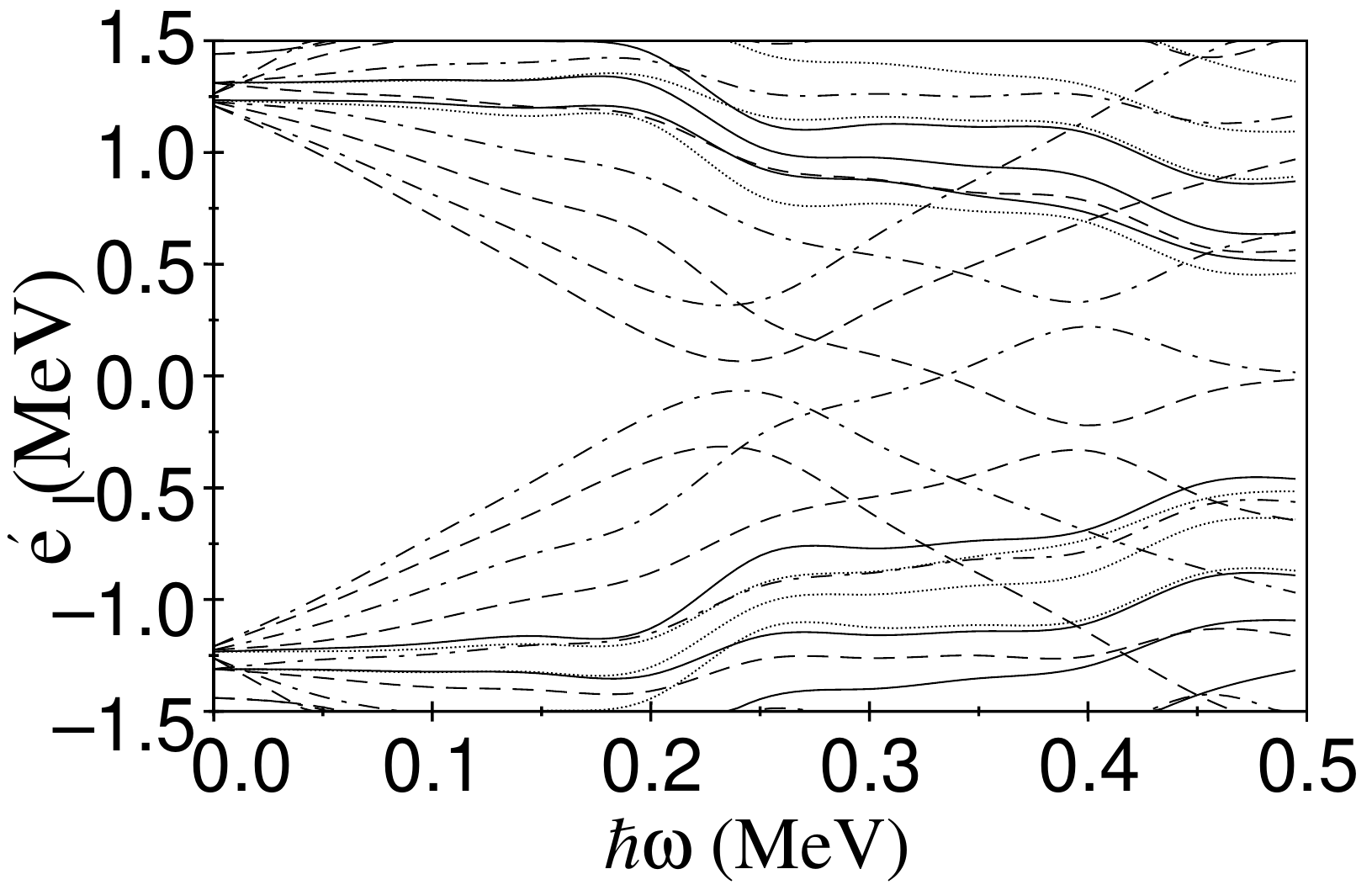}    
\includegraphics[width=0.23\textwidth]{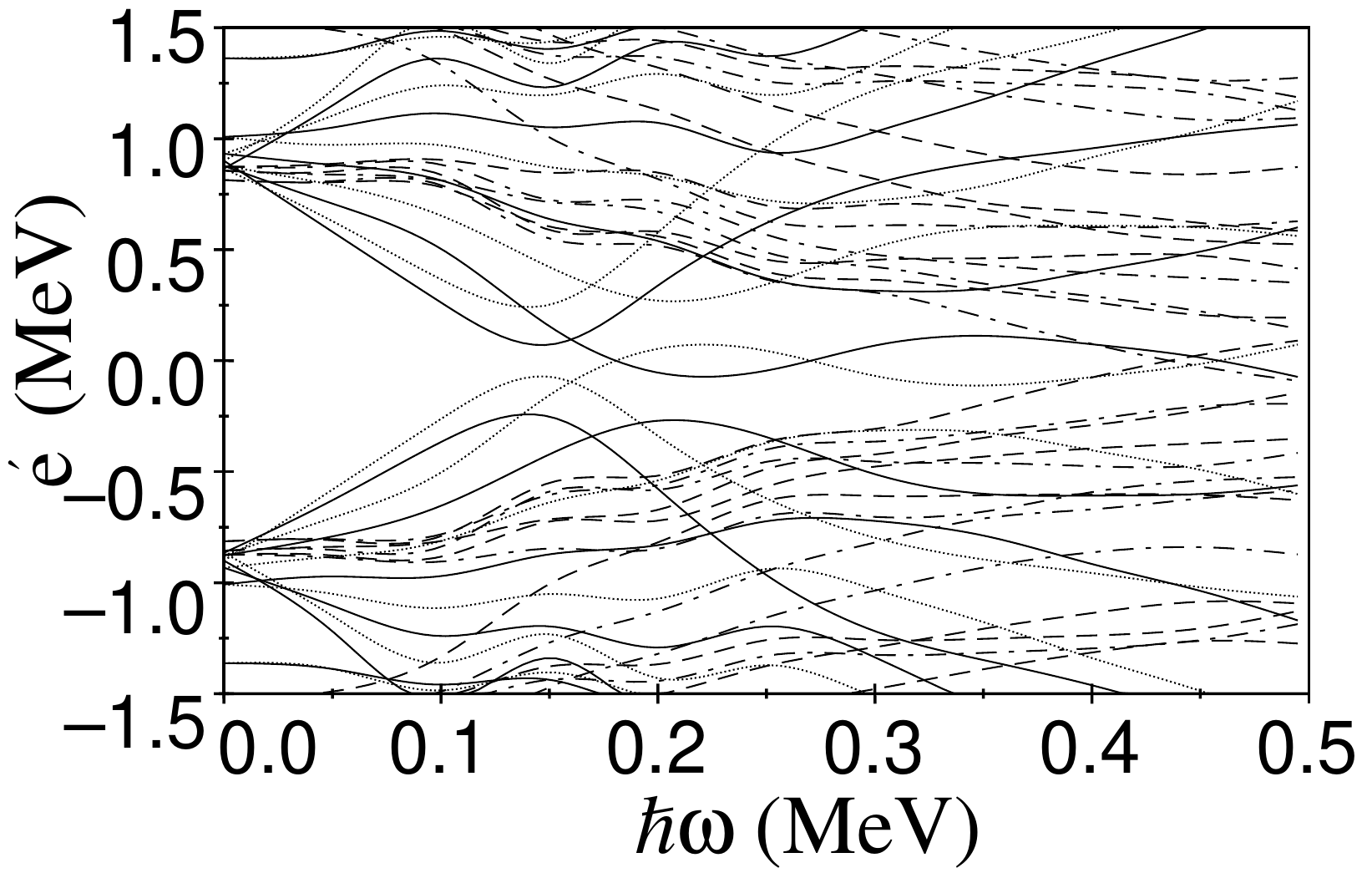} 
\caption{Proton (left) and neutron (right) quasi-particle diagrams at three typical deformation points, namely, $\beta_2 =$ $-0.1$ (top), $ +0.1$ (middle) and $0.2$ (botton), for $^{184}$Hf. Note that the oblate deformation $\beta_2 = -0.1$ is equivalent to the deformation point ($\beta_2 = +0.1, \gamma= -60$) in the deformation space ($\beta_2, \gamma$). }
                                                                    \label{Fig13}
\end{figure}

To reveal the details of rotation alignments and band crossings to some extent, as seen in Fig.~\ref{Fig13}, we show the calculated quasi-particle diagrams for protons and neutrons at three typical deformation points, namely, at $\beta_2 = -0.1, 0.1$ and 0.2, for the example nucleus $^{184}$Hf. The phenomena of low-frequency band crossing appear at, e.g., $\beta_2 = -0.1$ and 0.1. One can notice that the alignment of proton pair at the positive-parity orbital firstly occurs and then the neutron pair at the negative-parity orbital. Let us remind that the collective angular alignment $I_x$ is the summation of the minus values of the slopes of quasi-particle orbitals~\cite{Voigt1983}. All the quasi-particle orbitals contribute to the collective angular alignment which will change smoothly. However, once the band crossing occurs, the angular alignment will have a rapid increase and, consequently, the MoI will follow the change.

\begin{figure}
\centering
\includegraphics[width=0.23\textwidth]{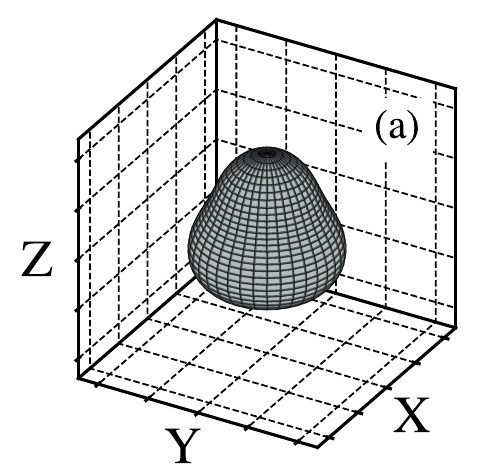}
\includegraphics[width=0.23\textwidth]{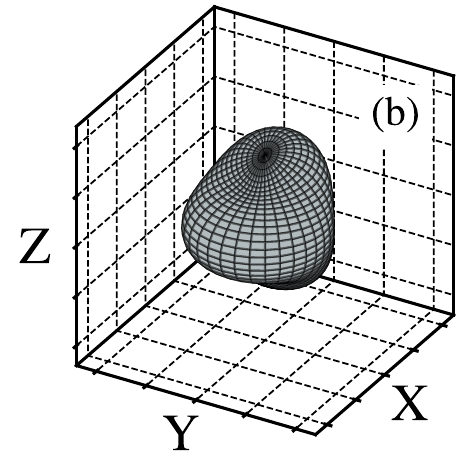}
\includegraphics[width=0.23\textwidth]{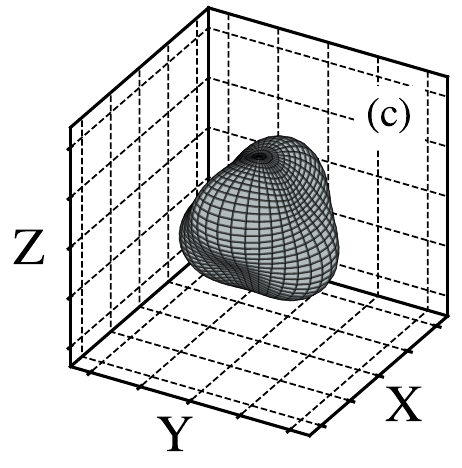}
\includegraphics[width=0.23\textwidth]{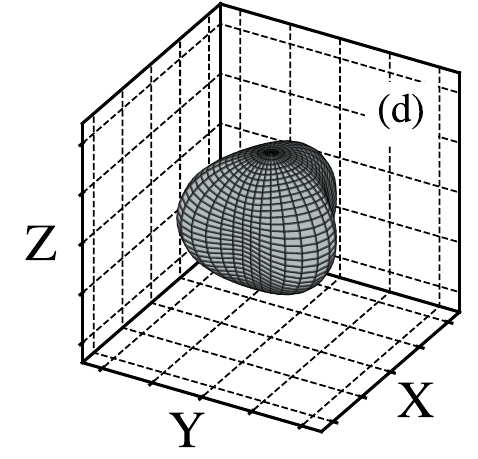}
\caption{Illustrations of nuclear shapes at single deformation parameter $\alpha_{3\mu}$ = +0.3, $\mu$ = 0 (a), 1 (b), 2 (c), and 3 (d).}
	                                                             \label{Fig14}
\centering
\includegraphics[width=0.235\textwidth]{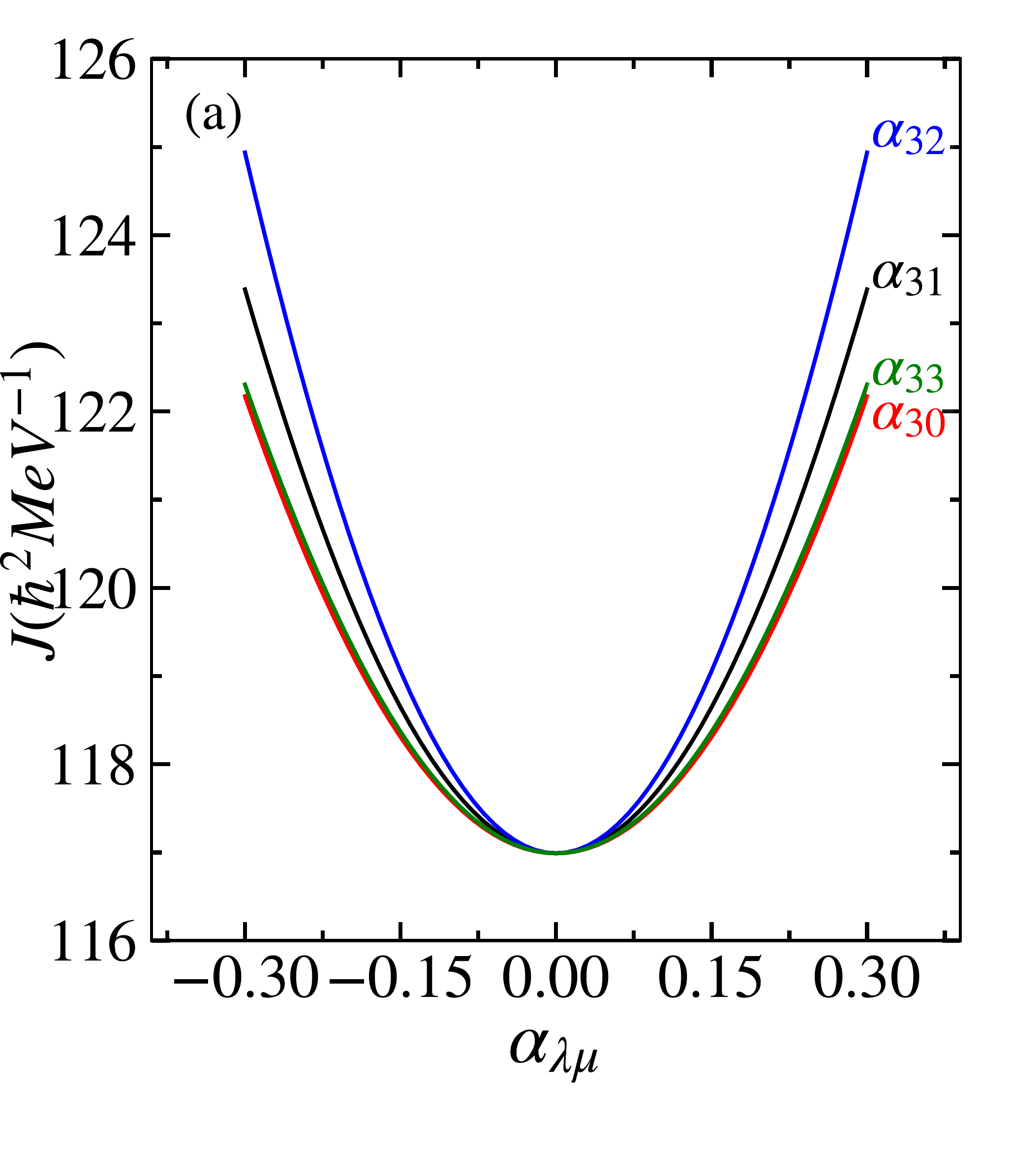}
\includegraphics[width=0.235\textwidth]{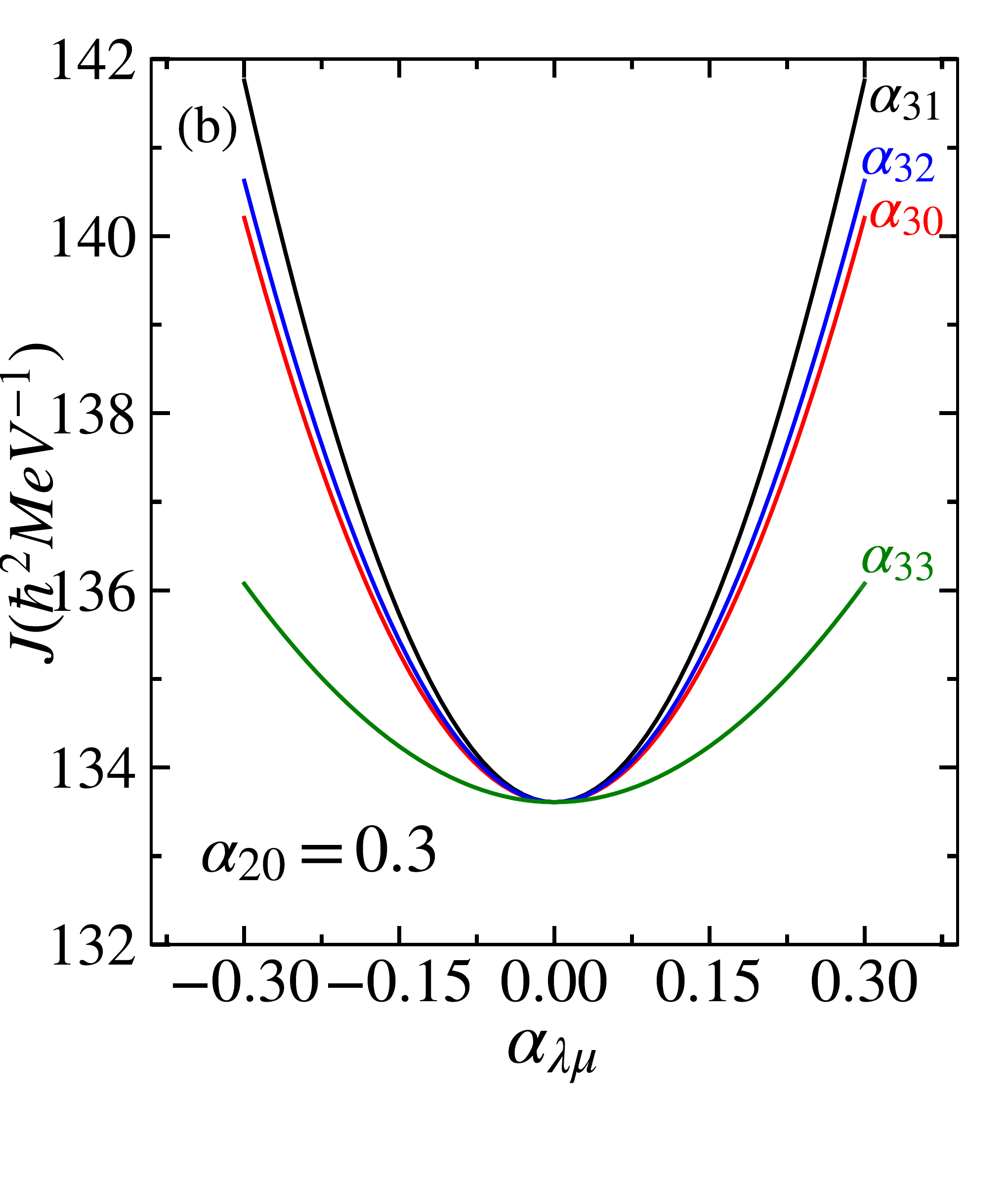}
\vspace{-0.5cm}
\caption{Calculated rigid-body MoIs around the $x$ axis in functions of the octupole deformation $\alpha_{3\mu=0,1,2,3}$ for $^{236}$U. Note that, in (b), the quadrupole deformation $\alpha_{20}$ is set to 0.3 during the MoI calculations with the changing $\alpha_{3\mu=0,1,2,3}$.}
                                                                 \label{Fig15}
\end{figure}

Very recently, the STAR Collaboration provided the first experimental evidence from the high-energy regime for octupole deformation of $^{238}$U~\cite{Star2025}. Dedes et al~\cite{Irene2025} discussed the unprecedented experimental identification and properties of the molecular $H_2O$ $C_{2v}$-symmetry rotational band in $^{236}$U involved the exotic $\alpha_{3\mu=0,1,2,3}$ deformations. Similar to the cases of the hexadecapole deformations, naturally, it is of interest to understand how these deformations $\alpha_{3\mu=0,1,2,3}$ theoretically affect the MoI in the nucleus. The computing time was quanlitatively estimated in typical potential energy calculations, e.g., see Ref.~\cite{Irene2025}. It was illustrated that the high-dimensional potential-energy-surface calculations, due to the huge consumption of CPU time, are still difficult for general computing conditions so far. Before extending the deformation space to include these deformation parameters, it may be meaningful and possible to give a qualitative prediction (even if not quite correctly) based on the rigid-body approximation by considering the corresponding degree(s) of freedom. In Fig.~\ref{Fig14}, we firstly present the geometry shapes with different $\alpha_{3\mu}$ deformations at an arbitrary deformation amplitude $\alpha_{3\mu} = 0.3$. Similar to Fig.~\ref{Fig07}, the rigid-body MoIs around the $x$ axis are illustrated in functions of $\alpha_{3\mu=0,1,2,3}$ in Fig.~\ref{Fig15}. Note that, for clarity, only one deformation $\alpha_{3\mu}$ is remained to vary, setting others to zero, e.g., cf. Fig.~\ref{Fig15}(a). To display the quadrupole-octupole coupling, in Fig.~\ref{Fig15}(b), the quadrupole deformation $\alpha_{20}$ is set to 0.3 during the MoI calculation. Indeed, one can find that the sensitity coefficient $S_{\lambda\mu}$ of MoI to deformation $\alpha_{\lambda\mu}$ as defined above (namely, the slope of MoI along the corresponding deformation), are rather different in Fig.~\ref{Fig15} (a) and (b), indicating the coupling effect between quadruple and octupole deformations. We can see, e.g., at $\alpha_{\lambda\mu} > 0$, that the sensitivity coefficients respectively satisfy $S_{32} > S_{31} > S_{33} >S_{30}$ and $S_{31} > S_{32} > S_{30} >S_{33}$ in Fig.~\ref{Fig15}(a) and Fig.~\ref{Fig15}(b). That is to say, in the present case, the $\alpha_{32}$ and $\alpha_{30}$ deformations will respectively have the largest and the smallest influences on the nuclear MoI near the spherical shape. Correspondingly, the largest and the smallest influences will respectively come from the deformations $\alpha_{31}$ and $\alpha_{33}$ at the elongated shape, and, simultaneously, the sensitivity coefficient $S_{33}$ significantly reduces. Further, the microscopic HFBC calculations are of interest and will be done in our future work.

\section{Summary}
\label{summary}
	
In summary, focusing on one of hexadecapole-deformation islands, we have performed an investigation on nuclear ground-state and rotational properties in $A \approx 180$ mass region within the theoretical framework of the MM model and cranking shell model. The impact of axial and nonaxial quadrupole and hexadecapole deformation degrees of freedom on nuclear structure (e.g., single-particle levels, equilibrium deformations, moment of inertia, etc.) is analyzed for $^{184}$Hf and its eight even-even neighbours $^{180-184}$Yb, $^{182,186}$Hf and $^{184-188}$W. It is found that the inclusion of hexadecapole deformations, especially the axial one $\alpha_{40}$, can not only enhance nuclear binding and lead to suitable quadrupole deformation $\beta_2$, but also well reproduce the experimental moments of inertia. With the HFBC calculation with fixed deformation(s), taking the nucleus $^{184}$Hf as an example, we present the evolution properties of moment of inertia with different deformations and rotational frequency. Understanding the effect of hexadecapole deformation and quadrupole-hexadecapole coupling will be of importance for both nuclear structure and nuclear reaction (e.g., the driving-potential calculation in dinuclear-system model). Comparing with the calculation of the corresponding deformed rigid body by considering the constant nuclear-density distribution, the present investigation reveals the similarity of the effects different quadrupole and hexadecapole deformations on nuclear moment of inertia. Combining such a similarity and a recent study, we predict the effects of different exotic deformations on nuclear rotational properties, e.g., the moment of inertia, and point out the sensitivity of menment of inertia to different octupole deformation degrees of freedom. Of course, whether such an exotic deformation will occur depends on the potential-energy-surface calculation rather than that of the moment of inertia. However, this prediction based on the simple rigid-body approximation might still provide us some important information on nuclear level structure once such an exotic deformation occurs.             
	
\section*{Conflict of Interest}
	
The authors declare that they have no known competing financial interests or personal relationships that could have appeared to influence the work reported in this paper.

\section*{Acknowledgement}
This work was supported by the Natural Science Foundation of Henan Province (No. 252300421478) and the National Natural Science Foundation of China (No.11975209, No. U2032211, and No. 12075287). Some of the calculations were conducted at the National Supercomputing Center in Zhengzhou.
%

\end{document}